# When Better Opportunities Backfire: Service Worker Responses to Parallel AI Deployment in Customer Acquisition

**Xinyu Fu**
J. Mack Robinson College of Business, Georgia State University
55 Park Place NE, Atlanta, GA 30303, USA
Email: xinyufu@gsu.edu

**Shulong Luo**
School of Applied Economics, Renmin University of China
59 Zhongguancun St, Haidian District, Beijing 100086, China
Email: Shulongluoecon@gmail.com

**Narayan Ramasubbu**
School of Business, University of Pittsburgh
3959 Roberto and Vera Clemente Dr, Pittsburgh, PA 15260, USA
Email: narayanr@pitt.edu

**Peng Shen**
Waterdrop Research Institute
Building B, WaterDrop Building, Lize Middle 2nd Road
Chaoyang District, Beijing 100102, China
Email: shenpeng@shuidi-inc.com

**Xincheng Wang**
School of Accounting and Finance, The Research Centre for ESG Advancement (RCESGA), The Hong Kong Polytechnic University
8 Hung Lok Road, Hung Hom, Kowloon, Hong Kong
Email: xincheng.wang@polyu.edu.hk

**Yucheng Wang**
School of Economics, The University of Sydney
Social Science Building (A02), Camperdown, NSW 2006, Australia
Email: yuchengwangecon@gmail.com

**When Better Opportunities Backfire:**

**Service Worker Responses to Parallel AI Deployment in Customer Acquisition**

## Abstract

As AI voice agents get deployed in high-volume contact centers, firms can route low-intent leads to AI and allocate high-intent leads to human workers. Drawing on sequential search theory, we posit that AI-induced task allocation reshapes work, enriching opportunities facing human workers but also raising their reservation value for continuing with the lead in hand, making them less willing to persist. We study the effects of an AI voice-agent deployment at a Chinese online health-insurance brokerage using more than a half million customer outreach records surrounding the rollout. Adopting a regression-discontinuity-in-time design, we trace how worker behavior changes across the workday. Results show that the enrichment of human-facing opportunities concentrates early in the workday, raising the probability that a call is a high-intent lead by about 23 percentage points on the first three calls and fading to zero later. However, customer acquisition by human workers declines precisely where the enrichment is greatest, falling about 5 percentage points on the first three calls and flattening over the day. Consistent with the theorized reservation value mechanism, workers reallocate their earliest leads about 8 percentage points more often and persist less with them, shortening those calls by about 13%. A calibrated decomposition analysis further shows that the modest gain from improved lead ordering is outweighed by a much larger negative behavioral response. By reshaping the quality and within-workday distribution of opportunities, parallel AI deployment alters how workers pursue tasks, highlighting opportunity distribution as an important service operations state variable beyond workload and congestion.

## 1. Introduction

Allocating work is a longstanding problem in service operations, where staffing and routing decisions are used to match capacity with fluctuating demand (Wallace and Whitt 2005; Akşin et al. 2007; Mehrotra et al. 2012). Behavior-aware operations research further shows that service workers adjust their pace, effort, persistence, and task choices in response to operating conditions (Tan and Netessine 2014; Cho et al. 2019; Kostami 2024). Prior research has accordingly examined workload and congestion, scheduling and task ordering, and the value and discretion associated with the task currently in hand. But work allocation also shapes what workers expect to receive next. We refer to the quality composition and within-shift sequencing of these opportunities as the worker-facing opportunity distribution.

AI makes this aspect of work allocation especially important because it allows firms to reshape the work left for human workers at scale. A growing literature examines how AI automates service tasks, augments worker decisions, changes task assignments, and affects worker productivity and behavior (Tong et al. 2021; Jia et al. 2024; Brynjolfsson et al. 2025; Fügener et al. 2026; Snyder et al. 2026). Much of this work examines AI's direct involvement in the focal task, decision, or assignment. Yet AI can also affect human work without directly participating in the focal human task. When AI operates in parallel with workers and selectively handles part of a shared task stream, it changes both the opportunities left for humans and those they expect to receive next.

At first glance, such a reallocation should only help. If AI absorbs the least promising opportunities, the average quality of the work reaching humans mechanically rises and, holding worker behavior fixed, realized human performance should rise with it. Extant accounts of how AI affects work reinforce this expectation: substitution perspectives emphasize gains from allocating tasks according to the comparative advantages of humans and technology (Acemoglu and Restrepo 2019), whereas augmentation studies document productivity improvements when AI enhances the focal task or the worker's capabilities (Brynjolfsson et al. 2025; Dell'Acqua et al. 2026b; Fügener et al. 2026). None of these perspectives predicts that improving the opportunities routed to humans would reduce their realized performance. They leave open, however, whether a mechanically better allocation can itself change worker behavior in ways that offset its expected benefit.

To address this question, we draw on sequential search theory (Stigler 1961; McCall 1970; Lippman and McCall 1976), which suggests that workers' willingness to persist with the opportunity in hand depends partly on the attractiveness of the opportunities they expect to encounter next. Applied to our setting, where an AI voice agent absorbs low-intent leads while human agents retain priority over high-intent leads, this logic implies that AI may weaken persistence on the very opportunities it leaves for humans, and it yields a position-graded prediction, developed in §2: the behavioral response should be strongest early in the workday when many future opportunities remain, and attenuate as the day's shift progresses.

We test this position-graded prediction using a large-scale AI voice-agent deployment at a Chinese online insurance brokerage. On August 28, 2024, the firm expanded the deployment of an AI voice agent that absorbed most low-intent ("cold") leads, concentrating the allocation of high-intent ("hot") leads to human agents. We examine the impact of this real-world deployment using a regression-discontinuity-in-time design applied to more than a half million customer outreach records over ±60 days around the AI expansion (409,622 human-handled and 157,468 AI-handled; about 13,000 agent-days). We find that AI improved the human-facing opportunity pool most strongly at the beginning of the daily shift and the improvement faded over the day. Because of the AI deployment, the probability that a human-handled call was a hot lead increased by 23.4 percentage points at the first three calls, attenuated over the shift, and reached zero at the end-of-the-shift tail. However, human agents' customer acquisition rate declined precisely where the opportunity improvement was greatest; it fell by 4.5 percentage points on the first three calls and by 2.5 points on the next three, remained unchanged at intermediate positions, and increased by 3.4 points among the end-of-the-shift calls. These patterns reflect changes in how agents worked each lead, along the two margins the reservation-value logic identifies. Consistent with a higher reservation value, agents became more willing to return the lead in hand to the shared opportunity queue for other agents to pursue. This reallocation rose by 8.1 percentage points on the first three calls. They also invested less effort in the calls they did pursue, and average call duration on the first three calls fell by roughly 13%. Both behavioral responses concentrated early in the day, where the improvement in the opportunity pool was largest, and were flat later on. A calibrated accounting decomposition analysis indicates that the small mechanical gain from the AI-induced improvement in the opportunity pool was more than offset by the behavioral responses, leaving the aggregate day-level performance impact close to zero.

These findings show that deploying AI alongside service workers is not behaviorally neutral. Even when AI and service workers handle different task streams, routing better work to humans can reshape how they pursue it, because service workers weigh each task against the opportunities they expect to come next. The value of AI-enabled specialization therefore depends not only on which opportunities humans receive, but on the expectations that AI-induced task routing creates about the opportunities still to come. More broadly, our results suggest that as firms delegate parts of a shared workflow to AI, the returns to that delegation will depend on how the remaining human work is composed and sequenced over the workday, a service operations design choice that critically shapes worker behavior.

## 2. Theoretical Background

Our study connects two streams of operations research: behavior-aware service operations and AI-enabled work allocation. The first examines how operating conditions shape worker behavior; the second examines how AI changes who performs which tasks. Their intersection raises a distinct question: *when AI selectively removes tasks from a shared stream of work, does it also change how humans approach the tasks left for them?* We use sequential search theory to explain why it might. Routing lower-value work to AI mechanically improves the human-facing opportunity pool, but the same improvement can make moving on more attractive and reduce workers' persistence and effort on the current task. This tension shapes a distinctive within-day work pattern that guides our empirical analysis.

### 2.1 Behavior-Aware Service Operations and Opportunity Distribution

Traditional service staffing and routing models often represent workers using fixed service-rate parameters (Wallace and Whitt 2005; Akşin et al. 2007; Mehrotra et al. 2012). Behavior-aware service operations research instead recognizes that workers adjust their pace, effort, persistence, and task choices in response to operating conditions (Donohue et al. 2020; Roels and Staats 2021). Prior work has shown that workload and congestion affect service speed and output (Tan and Netessine 2014; Cho et al. 2019) and that scheduling and task ordering shape worker execution and downstream performance (Ibanez and Toffel 2020; Kesavan et al. 2022). Moreover, task value and operating discretion influence which tasks workers pursue and how much effort they invest in them (Ibanez et al. 2018; KC et al. 2020; Beil et al. 2026). We extend this literature by focusing on worker-facing opportunity distribution, which we define as the distribution of opportunity quality across the stream of tasks that reaches a worker over a shift. It captures both the quality composition of the opportunities available to the worker and how that quality is distributed across positions in the workday. Opportunity distribution differs from (a) workload, which captures the quantity or rate of work; (b) focal-task value, which captures the value of the task currently in hand; and (c) scheduling and task ordering, which take the set of tasks as given and ask how their arrangement affects execution. Thus, two workers can face the same workload but different opportunity distributions if one receives a greater share of high-value opportunities, and the same set of opportunities can generate different opportunity distributions across the workday depending on where higher- and lower-value opportunities appear.

Opportunity distribution shapes workers' expectations about the opportunities they will encounter next and, consequently, the alternatives against which they evaluate the task currently in hand. A difficult task may remain worth persisting with when future opportunities are scarce or unattractive, but the same task may become less attractive when workers expect better opportunities ahead, even if its objective value has not changed. This distinction is essential because opportunity distribution concerns the opportunity environment surrounding the focal task, but not the value of the focal task itself. Related service operations research examines similar choices between continuing with the current task and moving work elsewhere. For example, gatekeepers may decide whether to retain or refer a customer (Shumsky and Pinker 2003), and workers in discretionary shared-pool systems may decide whether and when to participate (Hu et al. 2022; Liu et al. 2023). We extend this logic to parallel AI deployment by examining how the composition of opportunities changes workers' treatment of the task currently in hand. In summary, when AI selectively processes one portion of a shared opportunity pool, it can change the human-facing opportunity distribution even if worker capacity, the focal human task, and the underlying task-ordering rule remain unchanged.

### 2.2 AI-Enabled Work Allocation and the Human-Facing Opportunity Pool

AI can affect service operations through automation, augmentation, and task allocation (Caro et al. 2026;

Cohen et al. 2026; Dai and Swaminathan 2026). It may assist human workers during a focal interaction, interact with customers in place of a human worker, or absorb selected tasks before the remaining work reaches humans (Brynjolfsson et al. 2025; Cui et al. 2026; Fügener et al. 2026; Krakowski et al. 2026). Extant research also shows that workers may adjust their behavior in response to algorithmic systems and the operating conditions those systems create (Kawaguchi 2021; Bai et al. 2022), and their willingness to use, follow, or override an algorithmic recommendation itself depends on task and lead characteristics (Cao et al. 2026; Snyder et al. 2026), including strategic responses to algorithmic rules and market conditions (Wang Q et al. 2023; Benjaafar et al. 2025). Recent field-experimental evidence further documents how generative AI can reshape customer-service operations (Wang L et al. 2023; Ni et al. 2026; Zhang and Narayandas 2026), and process automation can produce behavioral externalities on human collaborators (Beer et al. 2026). Related research further shows that whether AI operates as a gatekeeper or as a second opinion changes the structure of the service workflow and the allocation of human attention (Dai and Singh 2025).

Research on human–algorithm collaboration examines how workers combine their own judgment with algorithmic recommendations and how system interventions can improve that collaboration (Ibrahim et al. 2021; DosSantos DiSorbo et al. 2025; Balakrishnan et al. 2026). These studies focus on AI's involvement in the focal decision. We examine a structurally different channel, in which AI affects workers without entering the focal human task at all. AI does not directly assist workers with the task currently in hand; it changes the opportunity environment surrounding that task by selectively processing other opportunities from a shared pool. We study a selective parallel-allocation setting in which AI and human workers perform the same customer acquisition activity but handle different portions of a shared opportunity pool. What changes is the quality composition of the opportunity stream routed to human agents and, consequently, what workers can expect to receive next. Thus, whereas AI-augmentation studies examine how AI changes human performance by intervening within the focal task or providing workers with information and decision support for that task, our opportunity-distribution channel operates outside the focal task and changes the alternatives against which workers evaluate it. By absorbing more opportunities with lower conversion probabilities, AI leaves humans with a richer opportunity pool. AI and humans therefore handle separate opportunities, yet their work remains behaviorally connected because allocating opportunities to AI changes the opportunity environment facing humans. Thus, parallel AI deployment can shape workers' behavior through allocation, without AI participating in the focal human task.

### 2.3 Theoretical Perspective: Opposing Forces and a Position-Graded Prediction

Sequential search theory provides an account of how a change in the opportunity pool can affect worker behavior (Stigler 1961; McCall 1970; Lippman and McCall 1976). In a standard search problem, a worker decides whether to continue with the current opportunity or move on to the next one. This decision depends partly on the value of the current opportunity and partly on how attractive future opportunities are expected to be. When the distribution of future opportunities improves, the expected value of continuing to search rises. The threshold for continuing with the current opportunity, what search theory calls the *reservation value*, therefore also rises.

This sequential search and reservation value logic maps naturally to our setting. During a customer acquisition task, each lead is a draw from a shared pool of opportunities. After contacting a potential customer, a human agent retains discretion over two margins: whether to keep working the lead or return it to the queue for another agent (reallocation), and how much effort to invest in the interaction. The setting therefore extends the standard search problem by letting service workers choose their effort before moving on. Manzoor et al. (2025) provide a related formulation of the salesperson's stopping decision in outbound sales, formalizing when to end a call as an optimal-stopping problem. When AI absorbs a larger share of cold leads, the expected quality of subsequent human assignments improves. The objective quality of the current hot lead is unchanged, but it becomes less attractive relative to the agent's expected next assignment. Agents should therefore become more willing to reallocate the current lead and less willing to invest costly effort in closing it. We use reservation value as an organizing theoretical mechanism; we do not derive behavior from a fully specified optimal-stopping model. The central prediction is straightforward: *improving future opportunities raises the opportunity cost of continuing to work on the current one*.

The AI expansion thus creates two opposing forces on human output. The first is a mechanical enrichment effect. By absorbing more low-conversion cold leads, AI increases the average quality of the human-facing opportunity pool. If worker behavior remained unchanged, this enrichment would mechanically raise the average conversion rate. The second is a countervailing reservation-value-based behavioral response, which makes future assignments more attractive, leading workers to release leads more readily and invest less effort in persisting with a lead. The two forces operate in opposite directions. The mechanical effect improves the mix of opportunities handled by humans, whereas the behavioral response may reduce workers' engagement with those opportunities. Which force dominates depends on how strongly worker behavior responds to the composition of the opportunity pool. Similar tensions, where operational improvements are eroded by endogenous behavioral responses, have been documented in other operational-design contexts (e.g., Sunar et al. 2021). We therefore treat the net effect of the AI-induced opportunity pool enrichment and countervailing reservation-value-related behavioral response as an empirical question.

The reservation-value mechanism does, however, generate a clear within-workday prediction. Early in an agent's daily shift, many future assignments remain, making it relatively attractive to move on from the current lead. Later in the shift, fewer same-day opportunities remain, reducing the value of continued search. The behavioral response should therefore be strongest at the earliest call positions and weaken as the sequence progresses. This position-graded pattern plays a dual role in our argument. First, it follows directly from the reservation-value mechanism: the relevant state variable is the remaining opportunity horizon, not position itself. As that horizon contracts over the workday, the continuation value generated by future opportunities declines, so an improvement in the future opportunity pool should exert progressively less influence on workers' treatment of the current lead. Second, this gradient provides a mechanism-discriminating implication. Broader explanations such as a general disruption to worker productivity, an agent-level shift in beliefs, a customer-side response, or a change in operating routines could generate an overall change in performance, but they do not by themselves imply an effect that is strongest when many future opportunities remain and attenuates as that horizon shrinks.

Our empirical strategy is designed to test this prediction by matching position-varying changes in opportunity exposure to position-varying changes in worker behavior. In addition, we examine whether conversion, reallocation, and call duration exhibit a corresponding gradient. Evidence that these outcomes respond most strongly at positions where AI most improves the human-facing opportunity pool, and weaken as both the enrichment and the remaining opportunity horizon diminish, would be more consistent with the reservation-value mechanism than with a conventional AI-induced level shift in worker performance.

## 3. Empirical Setting and Research Design

We study the operational records of a large online health-insurance brokerage in China that operates a high-volume outbound calling center and adopted an AI voice agent through a staged rollout beginning in late 2023. This section describes the firm and its workflow, the institutional details of the AI rollout that are central to our identification strategy, the call-level dataset, and the within-day profile design, based on a regression-discontinuity-in-time approach, which we use to assess how the rollout changed worker behavior.

### 3.1 Institutional Background

Platform X (pseudonym) is a licensed online insurance brokerage established in 2017 and authorized by China's National Financial Regulatory Administration. The firm distributes health, critical-illness, accident, and life insurance products on behalf of dozens of partnering insurers. Most products distributed through the platform are individual annual policies with first-year premiums in the low- to mid-thousand RMB range. Within the operations covered by our analysis sample, the firm places on the order of ten thousand outbound calls per workday. Prospective customers enter the platform through targeted online advertising and referrals from affiliated digital health platforms in the firm's broader ecosystem. We omit the firm's identity from the study to comply with the terms of our non-disclosure and data-sharing agreement.

#### 3.1.1 Lead Generation and Human Task Workflow

The firm operates a high-volume customer outreach service center that follows up with customers who left the online insurance-purchase process without completing a transaction. When such an exit occurs, the tasking system creates a follow-up lead, classified automatically by where the customer stopped. Customers who reached the payment interface but did not complete payment are hot leads (higher intent to purchase), whereas those who exited earlier in the order process are cold leads. Historically, both entered a single centralized queue ordered by lead freshness, and the task scheduling system assigned the next lead at the head of the queue whenever an agent became available. Agents could not themselves select leads or observe which lead would come next. After receiving a lead, an agent had discretion over three post-assignment actions:

retain the lead and continue working it, return it to the centralized queue for reassignment to another agent (reallocation), or discontinue further pursuit. A six-month no-recontact rule shapes the choice to discontinue: when a customer explicitly refuses further telephone contact, the lead becomes subject to the do-not-call restriction and cannot be contacted again for six months, whereas a reallocated lead remains available for other agents to pursue. Reallocation carried no direct penalty and was recorded but did not enter performance evaluations. Agents are compensated only on policies they personally close and receive no share when a reallocated lead later converts under another agent, so willingness to release a hot lead reflects a genuine forgone-value decision. Our analysis focuses on the post-assignment margins this workflow creates: reallocation and within-call effort, the latter observed through conversion and persistence measures.

### 3.1.2 Introduction and Expansion of the AI Voice Agent

The firm introduced an AI voice agent system to supplement human workers for customer acquisition. The AI system uses automated speech recognition and generation to run scripted calls to customers assigned to it by the scheduling system. As shown in Figure 1, the rollout of the AI voice agent proceeded in three phases. In the training phase (starting December 23, 2023), the AI placed evening-only calls, so daytime workflow was unaffected. In the trial phase (starting June 5, 2024), the AI began daytime operation under a human-first, AI-backup routing rule: leads reached the AI only when human capacity was exhausted, so the daytime pool facing human agents was largely unchanged. In the expansion phase (starting August 28, 2024), the firm reconfigured call routing to assign cold leads to the AI by default while preserving human priority over hot leads. Human agents received cold leads only when no hot leads remained available, and hot leads flowed to the AI only when all human agents were occupied. The AI agent rollout did not alter the basic customer acquisition task of the human agents, the assignment mechanism, or the post-assignment options available to the agents. What changed was the composition of the pool reaching them: before the AI expansion, agents worked a mix of hot and cold leads; after, the AI absorbed most cold leads, enriching the opportunity pool and concentrating human work on hot leads.

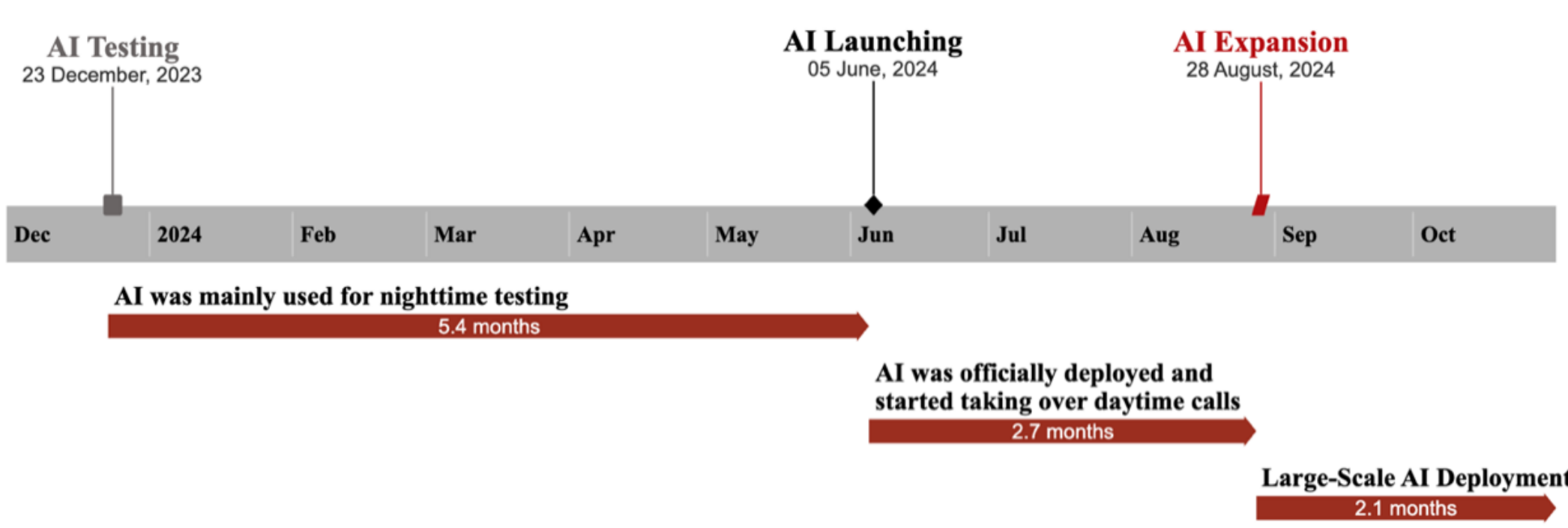

**Figure 1. AI Voice-Agent Deployment Timeline**

Figure 2 illustrates the post-expansion workflow. During the trial phase the AI handled under 25% of cold leads; at the August 28 expansion, its cold-lead share jumped to above 60% and remained there through the end of our observation period. Two features of the routing environment anchor our analysis and interpretation. First, the firm operates a single shared lead pool, not agent-, team-, or region-specific queues, so the opportunities each agent expects to draw are those in the common pool. Second, lead transfers between the AI and human agents are limited by design: once assigned, cold leads stay with the AI across subsequent attempts, and hot leads stay within the human operation. The reallocation margin we study is therefore a human-to-human return of a lead to the shared pool, not a transfer between the AI and human agents. Management adopted the routing rule for two reasons: cold leads had substantially lower conversion probabilities than hot leads, so the AI's near-zero marginal calling cost let it operate economically on tasks that were often unprofitable for humans; and transferring cold leads to the AI freed human capacity for hot leads. This strategic rationale is distinct from the behavioral mechanism we document. Prior work on voice-AI deployment in call centers has focused on customer-side outcomes (Wang L et al. 2023) or on how upstream AI-generated leads shape human salesperson creativity when AI performs an initial handling role (Jia et al. 2024); we focus instead on how such parallel AI deployment reshapes the reservation-value calculus and effort allocation of human agents.

### 3.1.3 Agent Incentives and Identification-Relevant Institutional Features

Human agents receive a modest base salary supplemented by commissions tied to the premium value of the policies they close, with commission income the majority of monthly take-home pay. This compensation structure, unchanged through the AI expansion, gives agents a direct interest in successful conversion, so the expansion of AI into daily operations altered neither commission rates nor the marginal reward from a sale. The August 28 change was scheduled as part of the firm's strategic deployment plan; the firm made no concurrent changes to compensation, training, scripts, evaluation, or hiring. Structured interviews with management confirmed the routing reconfiguration was the only major operational change in the surrounding months, and Appendix Tables C2 and C3 show that agent characteristics and the total number of calls handled per agent-day did not change discontinuously at the expansion date, although the composition of that volume shifted toward hot leads (about four more hot calls per agent-day), consistent with the routing change itself. Agents perceive the pool change through the composition of leads they are assigned, not through any formal announcement, and the within-day gradient in §4 (strongest at the earliest positions, fading through the day) is itself evidence that agents track and respond to the opportunities they expect still to receive.

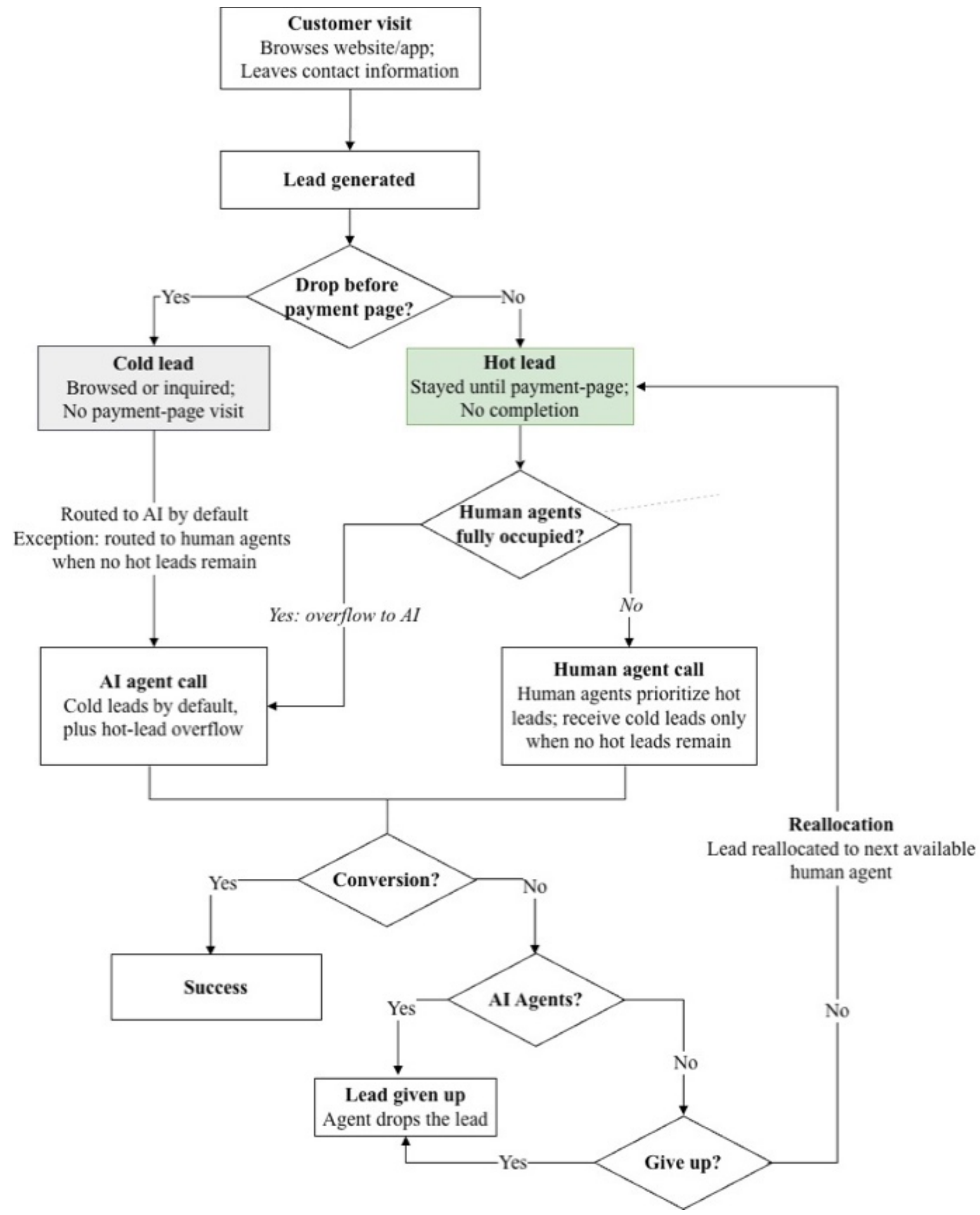

**Figure 2. Lead Generation, Classification, and Call Allocation under AI Expansion**

***Note.*** Post-expansion workflow. Hot leads route to human agents and cold leads to AI. Human agents receive cold leads only when they have no hot leads; hot leads overflow to AI only when all agents are occupied.

## 3.2 Data and Variable Construction

### 3.2.1 Data Source and Coverage

Our data consist of customer outreach records for both human-handled and AI-handled interactions covering January 1, 2023 through November 28, 2024. The records allow us to trace each lead from its creation through subsequent call attempts and eventual acquisition. Our primary analysis focuses on the August 28, 2024 routing expansion. Earlier observations are used only in analysis examining whether comparable changes in human-agent outcomes occurred during the AI system's earlier rollout stages (reported in Appendix). For each lead, we observe its creation time, first outbound call time, hot/cold classification, the handler identity (a specific human agent or the AI voice agent), and whether each call successfully connected. At the call level, we observe conversation duration in seconds, whether the call resulted in a completed transaction, the number of transactions, and the total payment amount in Chinese yuan. The data also include basic customer and human-agent characteristics. Customer characteristics include gender and whether the customer resides in a provincial capital or first-tier city. Agent characteristics include gender, tenure in months, shift starting time, working hours, and average daily task volume.

### 3.2.2 Outcome Variables

We examine three primary call-level outcomes that capture distinct dimensions of human-agent response to the change in the lead pool: realized customer acquisition performance, lead-allocation decisions, and effort within the current customer interaction. We also construct a supplementary measure of work pacing between successive calls.

The first outcome is *conversion*. The conversion indicator equals one when the focal call results in a completed insurance transaction and zero otherwise. It captures the immediate realized customer acquisition performance associated with the current call. This call-level measure is distinct from the downstream, lead-level conversion outcomes for leads that are reallocated and subsequently contacted by another agent.

The second outcome is *reallocation*. Following a call that does not result in a transaction, the focal agent may return the lead to the shared queue, after which the system can assign it to another agent. The reallocation indicator equals one when the focal agent returns the unconverted lead to the queue and zero otherwise. This variable captures the agent's decision to release the current opportunity. Reallocation is therefore a post-call allocation decision, not an ex ante choice of which customer to contact, because the initial lead assignment is determined by the centralized tasking system. Reallocation becomes observable only once the system's redistribution batch has executed. Lead creation ends 2024-10-31 whereas allocation events run to 2024-12-05, so the shortest follow-up window in the post-expansion sample is 39 days. Since 99.75% of observed redistributions complete within 35 days of the first call and 99.99% within 40 days, truncation can affect only a negligible share of post-period releases; any truncation attenuates the post-period rate, making the estimated increase conservative.

The third primary outcome is *call duration*, defined as the natural logarithm of the duration of the focal call in seconds. We use call duration as a proxy for effort within the customer interaction. Longer conversations generally indicate that the agent remains engaged with the customer for a longer period and devotes more time to explaining the product, addressing questions, or attempting to complete the transaction. Call duration is not a direct measure of effort, however, because it may also reflect customer responsiveness, call complexity, or other features of the interaction. We therefore interpret it as an observable proxy for within-call effort and report a robustness analysis restricted to successfully connected calls.

Finally, we construct the *inter-call gap*, measured as the elapsed time in minutes between the focal call and the agent's immediately preceding call. This variable captures the pace at which the agent moves through assigned leads. A longer gap may reflect time spent recording information, determining the disposition of the prior lead, preparing for the next call, or temporarily pausing between assignments. Because the measure may also include system idle time or other activities not directly related to selling effort, we treat it as a supplementary measure of between-call pacing, not as a primary measure of effort on the focal lead.

### 3.3 Empirical Strategy

We exploit the August 28, 2024 AI expansion, the sharpest change in AI exposure in the firm's rollout, when the cold-lead share handled by AI jumped from below 25% to above 60% (Figure 3), as a regression discontinuity in time (RDiT). Unlike a standard RDiT, our design does not target a discontinuity in the level of an outcome. We residualize each outcome on calendar-date fixed effects (FE) before estimating the discontinuity, so the object of interest is the discontinuous change in the within-day position profile of the outcome, measured relative to the same-date average, not an absolute level change at the cutoff. We refer to this residualized RDiT as a within-day profile design. Our master estimation sample covers at least ±65 days around the expansion, with a 10-day donut on each side of the cutoff to reduce contamination from implementation adjustments. The baseline specifications use mean square error (MSE)-optimal bandwidths (which vary by outcome and position bin); Appendix E reports results under a fixed ±60-day bandwidth for comparison. Following Hausman and Rapson (2018), we residualize each call-level outcome on agent, calendar-date, and hour-of-day fixed effects (first stage) and estimate a local linear RDiT with separate slopes on each side of the August 28 cutoff and a ±10-day donut on the residuals (second stage). Appendix H reports the estimating equations and the fixed-effects rationale. We estimate the local linear specification separately for five call-attempt-order bins within an agent's daily shift: positions 1–3, 4–6, 7–9, 10–12, and 81–120. The first four bins trace the beginning of the workday at fine resolution, where our theoretical framework predicts the strongest behavioral response: an agent at position 1 expects many future assignments, so a change in the expected quality of those assignments can meaningfully alter the reservation value for continuing to work the current lead. Positions 81–120 serve as an extreme late-position range (end of the shift) where substantially fewer same-day opportunities remain and the forward-looking response should be attenuated. Table 2 confirms this distinction empirically: the residualized AI-induced increase in the probability that a human-handled call is a hot lead is large at the earliest positions and statistically indistinguishable from zero at positions 81–120. We therefore treat positions 81–120 as a late-position benchmark, not an untreated control group; human agents continue to handle both hot and cold leads there.

The position-specific estimates in Table 2 order the bins in the opposite direction from the raw descriptives in Table 1, Panel D. In the descriptives, the pre–post increase in hot-lead share is largest at the tail (positions 81–120, from 0.379 to 0.558) and negligible at the earliest positions; in Table 2, the estimate is largest at positions 1–3 and indistinguishable from zero at the tail. The two sets of results are not in conflict, because they identify different estimands. Figure B3 plots them side by side. Panel A reports raw levels. Hot-lead share at early positions exceeds 0.86 before the expansion, so the shift in pool composition, which removes predominantly cold leads from the human queue, has little scope to raise the early-position share and is absorbed almost entirely by the tail. Panel B reports the estimates from Table 2, in which the dependent variable is the hot-lead indicator residualized on agent, hour-of-day, and calendar-date fixed effects. Panel A thus characterizes raw hot-lead exposure levels, whereas Panel B characterizes changes in the within-day allocation profile relative to the same-date average. So the contrasting cross-position patterns reflect the distinction between raw level changes and residualized position-specific discontinuities.

Because the first-stage residualization absorbs calendar-date fixed effects, we focus on how the outcome at a given position changes relative to the same-date average, not an absolute level change. Therefore, our primary estimand is a change in the shape of the within-day outcome profile, and differences in discontinuous change across position bins identify the position-graded response predicted by the reservation-value mechanism. Appendix H presents this interpretation in greater detail. Standard errors are clustered at the agent-day level. Because the AI expansion varies at the calendar-date level, we complement the call-level estimates with a date-level early-versus-late position contrast, reported alongside the main results. A simple date discontinuity at the August 28 expansion could reflect concurrent organizational, cohort, or marketing shifts. Our identification therefore does not rest on the RDiT discontinuity alone; it rests on the conjunction of two graded patterns. The first is a position-graded first stage. Table 2 shows an AI-induced increase in relative hot-lead exposure of +23.4 percentage points at positions 1–3, +15.7 percentage points at 4–6, +11.7 percentage points at 7–9, and statistically zero at 81–120. The human-facing opportunity pool improves most at the earliest positions and is unaffected at the late tail. The second is a matching position-graded response. In §§4.2–4.4 we document that conversion, reallocation, and per-lead effort each shift most sharply at positions 1–3 and progressively less at later positions, fading or reversing in the end-of-the-shift tail. The reservation-value mechanism developed in §2.3 predicts a response that inherits this gradient. Position-invariant confounders, such as organization-level, cohort, or marketing shifts, would produce a uniform shift across bins, not one that vanishes at the late tail. Because the specification absorbs agent, calendar-date, and hour-of-day fixed effects, cross-agent and cross-day selection into long workdays cannot mechanically generate the position gradient, and §4.4 provides evidence against fatigue-, learning-, and pace-based mechanisms that could otherwise track position monotonically. We therefore report position-specific RDiT estimates, not day-level averages, as the primary empirical results.

### 3.4 Analytic Sample and Descriptive Statistics

Table 1 reports descriptive statistics for human-handled calls in the 60-day windows around the August 28 expansion. The descriptive sample imposes the same 10-day donut as the primary estimation sample (§3.3), so the statistics characterize the raw before-after patterns on the same sample used in the RDiT estimates below. The human-handled sample contains 243,473 calls before the expansion and 166,149 after, placed by human agents on 7,492 and 5,561 agent-days respectively.

Panel A of Table 1 reports the three primary outcomes. The conversion rate is low in both periods, declining from 3.2% before the expansion to 2.9% afterward. This pattern is consistent with the high-volume and low-conversion nature of customer acquisition in the insurance context. The raw reallocation rate rises from 14.1% to 22.9%, an increase of 8.8 percentage points. This difference provides initial descriptive evidence consistent with the fading reallocation prediction, but it should not be interpreted as a causal estimate. It does not adjust for agent heterogeneity, date-specific conditions, intraday rhythms, or the position of the call within the daily shift. Average log call duration is nearly unchanged across the two windows, at 3.846 before the expansion and 3.848 afterward. The absence of an aggregate difference, however, conceals heterogeneous responses across within-day positions. The position-specific RDiT design is intended to reveal these offsetting changes.

Panel B reports workflow and lead-composition variables. The share of human-handled calls involving hot leads increases from 60.6% to 64.6%, consistent with the new routing rule concentrating human work on higher-intent customers. Average lead freshness also improves, with the elapsed time between lead creation and the first call falling from 56.3 to 30.8 seconds. The raw descriptive increase in the human-facing hot-lead share should be distinguished from the position-specific estimates in Table 2. The raw statistic measures the absolute composition of all human-handled calls in each window. Table 2 estimates measure how hot-lead exposure at a particular position changes relative to the overall composition on the same date.

Panel C reports observed characteristics of the human workforce. Mean tenure is 8.8 months before the expansion and 8.5 months afterward. Average daily working hours are 4.2 and 4.0 hours, respectively, and the average shift starting time remains at approximately 9:25 to 9:30 a.m. The share of female agents increases from 37.0% to 41.2% in the raw descriptive windows. Because these figures are unconditional before-after averages, they do not by themselves establish a discontinuous workforce-composition change. Appendix Table C3 formally examines agent presence and observed workforce characteristics around the cutoff and finds no corresponding discontinuity.

Panel D reports the four call-level measures across selected within-day position bins. The descriptive statistics reveal pronounced gradients across the daily shift. Conversion is substantially higher at the beginning of the sequence than in the late tail. Before the expansion, the conversion rate across all human-handled calls is 9.1% at positions 1–3 but only 1.8% at positions 81–120. Calls at earlier positions also tend to last longer than calls in the late-position range. Reallocation exhibits a different raw pattern, rising after the expansion across both early and late positions. These unconditional changes combine the effects of lead composition, agent behavior, daily conditions, and regular position gradients. They therefore motivate, but do not substitute for, the residualized position-specific RDiT estimates. The late-position descriptive statistics should not be interpreted as evidence that positions 81–120 contain only cold leads. The call-volume records show that human agents continue to handle both hot and cold leads at these positions. The role of this range is instead to provide an extreme late-position benchmark at which relatively few same-day assignments remain and no additional relative hot-lead discontinuity is detected.

**Table 1. Summary Statistics: Pre-Expansion and Post-Expansion Windows**

| Variable | Definition | Pre-Expansion | | Post-Expansion | | |
|---|---|---|---|---|---|---|
| | | Mean | SD | Mean | SD | Unit |
| ***Panel A. Outcomes*** | | | | | | |
| Conversion rate | =1 if call yields transaction | 0.032 | 0.176 | 0.029 | 0.168 | 0/1 |
| Reallocation | =1 if lead returned to pool | 0.141 | 0.348 | 0.229 | 0.420 | 0/1 |
| Log call duration | Natural log of call seconds | 3.846 | 1.246 | 3.848 | 1.197 | log sec |
| ***Panel B. Workflow and Lead Composition*** | | | | | | |
| Hot-lead share | =1 if call is on a hot lead | 0.606 | 0.489 | 0.646 | 0.478 | 0/1 |
| Lead freshness | Elapsed time, lead creation to first call | 56.294 | 209.339 | 30.798 | 107.289 | sec |
| Customer female | =1 if customer is female | 0.533 | 0.499 | 0.528 | 0.499 | 0/1 |
| Call gap | Minutes since agent's prior call | 6.583 | 22.465 | 5.993 | 20.601 | min |
| Agent-day level: | | | | | | |
| Hot calls per agent-day | Hot leads dialed per agent per day | 19.698 | 13.275 | 19.312 | 12.621 | calls |
| Cold calls per agent-day | Cold leads dialed per agent per day | 12.800 | 12.668 | 10.565 | 11.757 | calls |
| ***Panel C. Agent Characteristics*** | | | | | | |
| Tenure | Months since agent hire date | 8.768 | 9.042 | 8.533 | 9.517 | months |
| Agent female | =1 if agent is female | 0.370 | 0.483 | 0.412 | 0.492 | 0/1 |
| Working hours | Hours worked on call date | 4.246 | 3.049 | 3.990 | 3.008 | hours |
| Shift start | First-call hour of agent shift | 9.415 | 0.946 | 9.476 | 1.042 | hour |
| ***Panel D. Outcomes by Within-Day Call-Attempt Position*** | | | | | | |
| Hot-lead share | | | | | | |
| positions 1–3 | | 0.862 | 0.345 | 0.895 | 0.306 | 0/1 |
| positions 4–6 | | 0.796 | 0.403 | 0.779 | 0.415 | 0/1 |
| positions 7–9 | | 0.715 | 0.452 | 0.697 | 0.460 | 0/1 |
| positions 10–12 | | 0.657 | 0.475 | 0.650 | 0.477 | 0/1 |
| 81–120 | | 0.379 | 0.485 | 0.558 | 0.497 | 0/1 |
| Conversion rate | | | | | | |
| positions 1–3 | | 0.091 | 0.288 | 0.072 | 0.259 | 0/1 |
| positions 4–6 | | 0.051 | 0.221 | 0.044 | 0.205 | 0/1 |
| positions 7–9 | | 0.036 | 0.187 | 0.034 | 0.181 | 0/1 |
| positions 10–12 | | 0.032 | 0.176 | 0.028 | 0.165 | 0/1 |
| 81–120 | | 0.018 | 0.132 | 0.024 | 0.154 | 0/1 |
| Reallocation | | | | | | |
| positions 1–3 | | 0.245 | 0.430 | 0.381 | 0.486 | 0/1 |
| positions 4–6 | | 0.190 | 0.392 | 0.327 | 0.469 | 0/1 |
| positions 7–9 | | 0.168 | 0.374 | 0.297 | 0.457 | 0/1 |
| positions 10–12 | | 0.157 | 0.364 | 0.279 | 0.449 | 0/1 |
| 81–120 | | 0.164 | 0.370 | 0.297 | 0.457 | 0/1 |
| Log call duration | | | | | | |
| positions 1–3 | | 4.118 | 1.358 | 4.058 | 1.268 | log sec |
| positions 4–6 | | 3.970 | 1.278 | 3.982 | 1.243 | log sec |
| positions 7–9 | | 3.930 | 1.269 | 3.926 | 1.223 | log sec |
| positions 10–12 | | 3.912 | 1.244 | 3.914 | 1.206 | log sec |
| 81–120 | | 3.553 | 1.158 | 3.614 | 1.156 | log sec |
| Call gap | | | | | | |
| positions 1–3 | | 6.879 | 28.931 | 5.332 | 24.068 | min |
| positions 4–6 | | 6.471 | 24.635 | 5.580 | 22.055 | min |
| positions 7–9 | | 6.730 | 25.903 | 5.404 | 21.422 | min |

| | | | | | |
|---|---|---|---|---|---|
| positions 10–12 | 7.182 | 25.795 | 5.522 | 19.154 | min |
| 81–120 | 3.283 | 11.619 | 3.445 | 13.776 | min |
| Observations (calls) | 243,473 | | 166,149 | | |
| Observations (agent-days) | 7,492 | | 5,561 | | |

***Notes.*** Descriptive statistics for human-handled calls, ±60-day windows around August 28, 2024, ±10-day donut. Panels A–C: full human-handled sample. Panel D reports raw, unadjusted means by within-day position bin (late tail = 81–120). See Appendix Figure B3 for a comparison of raw levels and residualized within-day estimates.

We exclude calls missing either the human-agent identifier or the within-day call-attempt position. Missingness is below 1% and does not differ materially across the cutoff. We retain both connected and nonconnected call attempts, because reallocation decisions and within-day position are defined for both. Call duration, however, is defined only for connected calls, so the log-duration outcome is measured on the connected-call sample; we report this restriction explicitly and, for duration, present a robustness analysis on connected calls. Agent turnover is modest, and agent presence on the active roster remains balanced around the cutoff, as documented in Appendix Table C3. Together with the institutional evidence in §3.1, these patterns reduce concerns that the estimated discontinuities are driven by changes in sample construction, workforce composition, or other observed confounders.

## 4. Results

This section tests the position-graded theoretical predictions developed in §2. We organize the analysis to mirror the mechanism's structure: (§4.1) the position-graded first-stage change in the human-facing pool, (§4.2) the position-graded conversion response, (§4.3) the position-graded behavioral margins in reallocation and effort, and (§4.4) confounders that fail to track the same position gradient. For each outcome, we first present visual evidence comparing the within-day profile before and after the AI expansion, and then report RDiT estimates of the discontinuous change at each within-day position bin. Alignment between the position-graded first stage and the position-graded response identifies the reservation-value mechanism; failure to align at any margin would refute it. Two reading conventions guide this section. First, the daily average pools all within-day positions together. Because the behavioral channel and any compositional support are predicted to move in opposite directions across positions, the daily average is uninformative about the mechanism even when the mechanism is operative. We therefore report estimates position by position throughout the section and read the daily average only as the net of these countervailing effects. Second, the flattening of the within-day conversion profile predicted by the framework is visible directly in Figure 4 without modeling assumptions.

### 4.1 Position-Graded First Stage: The AI-Induced Opportunity Shift

Our empirical strategy requires a sharp first-stage discontinuity in AI exposure at the August 28, 2024 cutoff. Figure 3 reports the daily share of cold leads handled by AI in the sixty days before and after the cutoff. In the pre-cutoff period, the share rises gradually from below 10% to roughly 25%, reflecting the prior daytime deployment that began in June 2024. At the cutoff, the share jumps discontinuously from approximately 25% to above 60%, and then stabilizes between 55% and 75% over the subsequent two months. The corresponding figure for hot leads, reported in Appendix Figure B1, shows a much smaller change. The share of hot leads handled by AI remains below 10% throughout the daytime hours both before and after the cutoff, with majority AI handling concentrated during overflow periods, consistent with the routing rule described in §3.

The first-stage shift in AI exposure thus operates almost entirely through cold leads: AI takes over cold leads whereas the flow of hot leads is essentially unchanged. The composition of leads facing human agents during regular daytime working hours becomes mechanically richer in hot leads after the cutoff, whereas the human task itself, an outbound call to a previously interested customer, does not change. Table 2 reports the corresponding RDiT estimates and shows that the increase is concentrated in the early within-day call positions, where the reservation-value mechanism predicts the strongest behavioral response.
The aggregate shift in AI absorption translates into a position-graded shift in the composition of leads facing humans. Table 2 reports, by within-day call-attempt-order bin, the discontinuous change at the cutoff in the probability that a human-handled call is a hot lead. The position-specific shift is largest at the earliest within-day positions, where the framework predicts the strongest behavioral response: 23.4 percentage points at positions 1–3, 15.7 at positions 4–6, 11.7 at positions 7–9, and 10.9 at positions 10–12. The shift declines across the early positions, and at positions 81–120, which represent the extreme late portion of the within-day sequence, it is essentially zero (−0.8 percentage points, not significant). The position-graded first stage is the mirror of the asymmetric structure that the position-bin design is built to test, and it is what allows the later-position bin to serve as a comparison bin showing no detectable relative hot-lead discontinuity at the cutoff. This position gradient is the identifying variation exploited by the position-specific analyses (§§4.2–4.4).

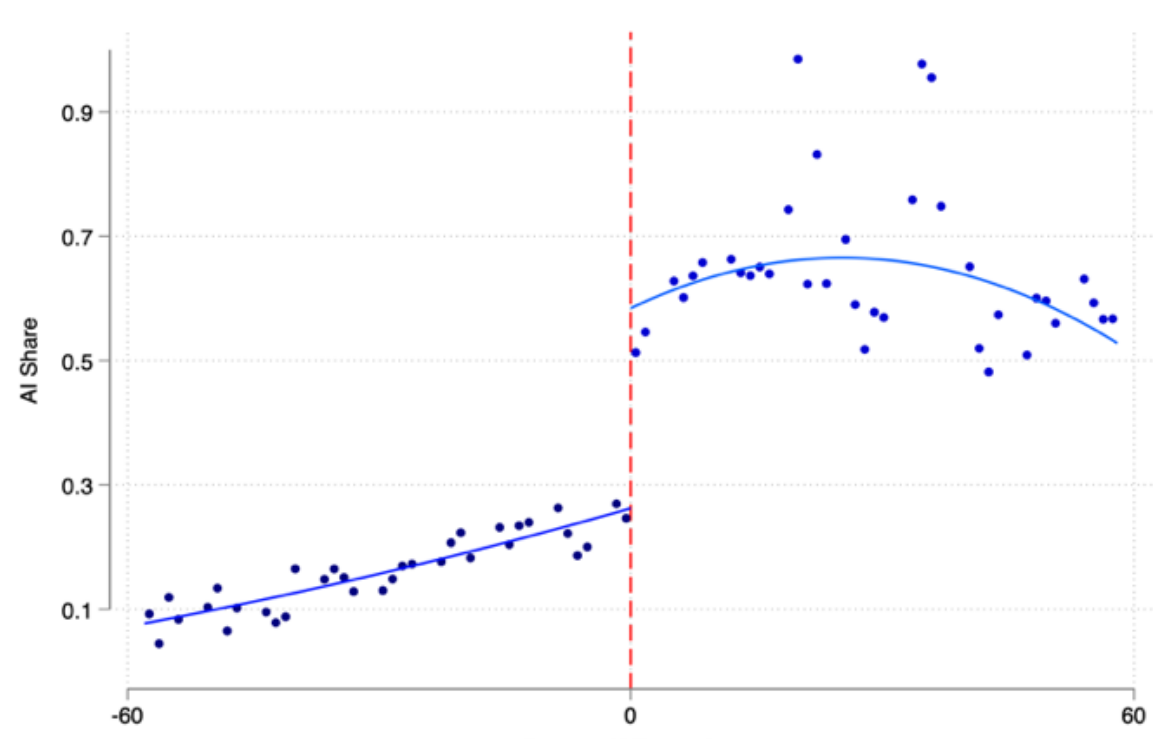

**Figure 3. First Stage: AI Share of Cold Leads Around the Expansion**
***Notes.*** Daily AI share of cold leads, ±60 days around August 28, 2024. Solid curves: local polynomial fits on each side of the cutoff.

**Table 2. Pool-Composition Shift: Probability That a Human-Handled Call is a Hot Lead**

| | (1) | (2) | (3) | (4) | (5) |
|---|---|---|---|---|---|
| | 1–3 | 4–6 | 7–9 | 10–12 | 81–120 |
| After AI Expansion | 0.234*** | 0.157*** | 0.117*** | 0.109* | −0.008 |
| | (0.033) | (0.033) | (0.043) | (0.060) | (0.055) |
| Bandwidth | 63 | 63 | 63 | 63 | 63 |
| Observations | 23,085 | 21,383 | 20,504 | 19,900 | 51,538 |
| Agent FE | Yes | Yes | Yes | Yes | Yes |
| Date FE | Yes | Yes | Yes | Yes | Yes |
| Hour-of-day FE | Yes | Yes | Yes | Yes | Yes |

***Notes.*** RDiT on P(hot lead | human-handled), by position bin. Agent, calendar-date, and hour-of-day FE; ±10-day donut. Agent-day-clustered SEs in parentheses. *p<0.10, **p<0.05, ***p<0.01.

### 4.2 Position-Graded Conversion Response

The position-graded structure of the first stage documented in Table 2 provides the identifying leverage. If AI's improvement in the human-facing pool were the only channel operating at the cutoff, human conversion should rise, most sharply at the earliest positions where the pool improvement is largest. The reservation-value mechanism predicts the opposite. Because a richer pool of subsequent opportunities makes the current lead relatively less attractive, human conversion should fall precisely where the first stage improves most. Figure 4 and the associated position-specific RDiT estimates provide the principal test of this prediction.

#### 4.2.1 Descriptive Evidence

The reservation-value mechanism predicts that the within-day conversion profile for human-handled hot leads should flatten following the AI expansion, with the largest position-specific movement occurring at the earliest call positions (where the first-stage enrichment is largest) and fading as position advances. We test this prediction using the call-level conversion indicator for hot leads. Figure 4 compares the within-day hot-lead conversion profiles before and after the AI expansion. Before the expansion, conversion declines steeply across the daily shift, from approximately 10.4% at positions 1–3 to approximately 1.1% at positions 81 and beyond. After the expansion, the conversion profile rotates and becomes substantially flatter. The pre- and post-expansion conversion profiles differ most sharply at the earliest positions, narrow across the intermediate positions, and the post-expansion conversion rate becomes modestly higher in the late tail. These patterns of raw conversion profiles therefore suggest a redistribution of conversion performance across the daily shift, not a uniform upward or downward shift.

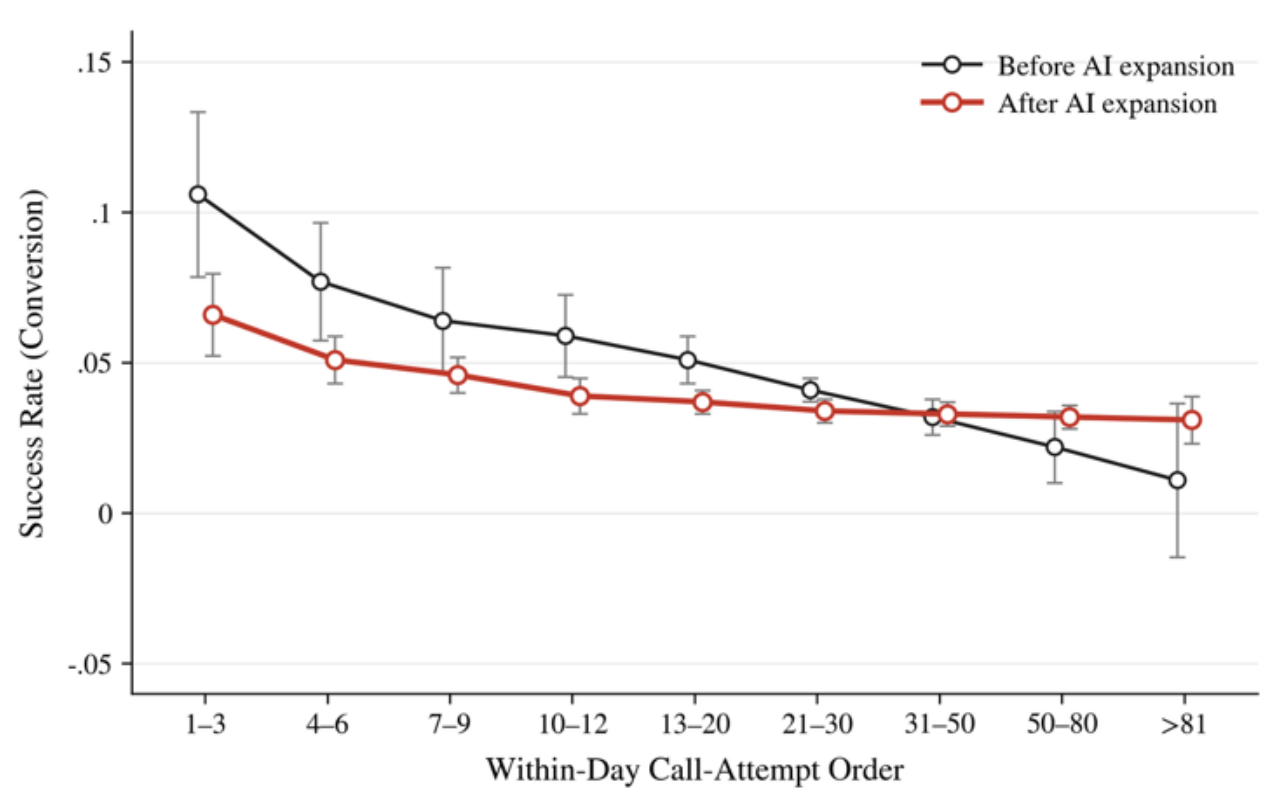

**Figure 4. Within-Day Hot-Lead Conversion Profiles Before and After the AI Expansion**

***Notes.*** Adjusted mean conversion rates for human-handled hot leads by position bin, pre and post August 28, 2024. Regression with agent, date, and hour-of-day FE. Bars: 95% CIs from agent-clustered SEs.

### 4.2.2 RDiT Estimates

Table 3 formally tests the conversion profile change using the RDiT design. All specifications examine human-handled hot leads. Panel A reports summary tests of the change in the within-day conversion profile, whereas Panel B reports the discontinuity separately for each call-attempt-position bin. The average call-level discontinuity reported in Panel A, column 1, is 0.1 percentage points and is statistically indistinguishable from zero. Thus, pooling calls across positions reveals no detectable change in average hot-lead conversion at the expansion. The profile-slope estimate in column 2, however, is positive and statistically significant at the 1% level. The estimated slope change is 0.5 percentage points per position bin, relative to a pre-expansion slope of −1.1 percentage points per bin. The steep negative conversion gradient observed before the expansion therefore becomes substantially less pronounced afterward. Column 3 reports a complementary date-level early-minus-late conversion contrast. The contrast declines by 1.4 percentage points and is statistically significant at the 10% level. This result indicates that the conversion advantage of early calls relative to late calls narrows following the AI expansion. Taken together, the profile-slope estimate and the early–late contrast provide formal evidence for the flattening predicted by the reservation-value mechanism. Panel B identifies the positions that generate this flattening. The position-specific RDiT estimate for hot-lead conversion is −4.5 percentage points at positions 1–3 and −2.5 percentage points at positions 4–6; both estimates are statistically significant at the 1% level. The estimates at positions 7–9 and 10–12 are close to zero and statistically insignificant. By contrast, hot-lead conversion at positions 81–120 increases by 3.4 percentage points and is statistically significant at the 1% level. The position-specific estimates therefore reveal a pronounced redistribution of conversion across the workday. The position-specific conversion estimates are negative at the earliest positions, statistically indistinguishable from zero at intermediate positions, and positive in the extreme late tail. These directionally opposed movements explain why the average call-level effect is close to zero even as the within-day conversion profile changes substantially.

**Table 3. Effects of the AI Expansion on Human Hot-Lead Conversion**

***Panel A. Summary Tests of Within-Day Profile Change***

| | (1)<br>Average conversion | (2)<br>Profile slope | (3)<br>Early–late contrast |
|---|---|---|---|
| After AI expansion | 0.001<br>(0.004) | 0.005***<br>(0.001) | −0.014*<br>(0.008) |
| Pre-period mean | 0.043 | −0.011 | 0.027 |
| Observations | 306,413 | 306,413 | 116 |
| Controls | Agent, hour, day-of-week FE; local linear trend | | |

***Panel B. Position-Specific Effects on Hot-Lead Conversion***

| | **(1)**<br>1–3 | **(2)**<br>4–6 | **(3)**<br>7–9 | **(4)**<br>10–12 | **(5)**<br>81–120 |
|---|---|---|---|---|---|
| After AI expansion | -0.045***<br>(0.013) | -0.025***<br>(0.009) | 0.001<br>(0.010) | 0.007<br>(0.010) | 0.034***<br>(0.010) |
| Bandwidth | 65 | 65 | 65 | 65 | 65 |
| Observations | 20,978 | 17,565 | 15,178 | 13,638 | 24,210 |
| Agent FE | Yes | Yes | Yes | Yes | Yes |
| Date FE | Yes | Yes | Yes | Yes | Yes |
| Hour-of-day FE | Yes | Yes | Yes | Yes | Yes |

***Notes.*** Panel A: summary RDiT tests (col 1 = average, col 2 = slope, col 3 = early–late contrast). Panel B: position-specific coefficients (1–3, 4–6, 7–9, 10–12, 81–120). Agent, calendar-date, and hour-of-day FE; MSE-optimal bandwidth (Calonico et al. 2014); ±10-day donut. Agent-day-clustered SEs; Panel A col 3 uses HC-robust date-level SE. $*p<0.10$, $**p<0.05$, $***p<0.01$.

### 4.2.3 Position-Specific RDiT Evidence

Figure 5 visualizes the position-specific discontinuities reported in Table 3, Panel B. The positions 1–3 and 4–6 panels exhibit clear downward breaks at the cutoff. The positions 7–9 panel shows little detectable change, whereas the positions 81–120 panel exhibits an upward break. The graphical evidence therefore closely matches the point estimates and makes the flattening of the within-day conversion profile visible in the local RDiT comparisons.

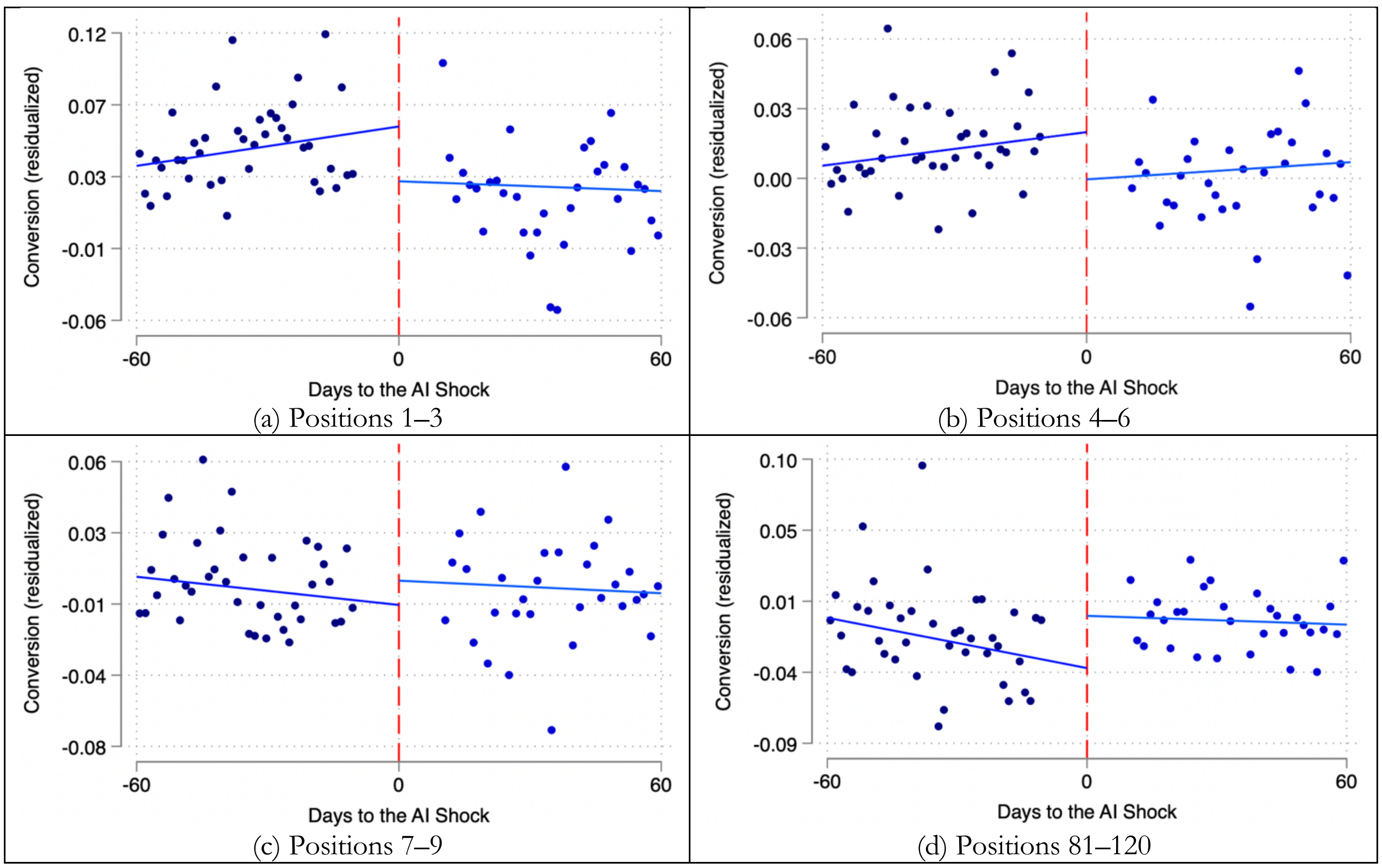

**Figure 5. RDiT Plots of Hot-Lead Conversion by Within-Day Call Position**

***Notes.*** Position-specific RDiT for Table 3, Panel B. Residualized hot-lead conversion by bin against days from cutoff. Solid lines: local linear fits on each side; ±10-day donut. Panels (a)–(b): negative discontinuities at 1–3, 4–6. Panel (c): 7–9 negligible. Panel (d): positive at 81–120. Bin 10–12 omitted.

### 4.2.4 Magnitude and Interpretation

The early-position conversion response is large relative to the 4-percentage-point increase in the overall hot-lead share of the human-facing pool. The aggregate statistic, however, masks a much stronger position-specific first stage. As documented in §4.1, the relative probability that a human-handled call is a hot lead increases by 23.4 percentage points at positions 1–3, 15.7 points at positions 4–6, and approximately 11 points at positions 7–12. Both the compositional shift and the early-position conversion response are therefore concentrated at the beginning of the daily shift. This alignment is consistent with the reservation-value mechanism. At early positions, agents anticipate many subsequent assignments, so an improvement in the expected quality of future opportunities increases the opportunity cost of continuing to work the current lead. At later positions, fewer same-day opportunities remain, and this forward-looking behavioral response should weaken. The fact that the conversion change follows this predicted position gradient is difficult to reconcile with a generic shock that affected all calls uniformly. The positive conversion discontinuity at positions 81–120 should be interpreted differently. This range is a late-position benchmark, not an untreated control group. Table 2 shows no detectable increase in hot-lead exposure at these positions relative to the date-specific overall composition, but the hot leads reaching the late tail may still change in freshness, prior-attempt history, vintage, or effective queue position. The late-position increase is therefore in line with within-hot-lead compositional reshuffling, although the conversion estimate alone does not identify the specific dimension responsible.

As a complementary analysis, Appendix Table D.5.1 examines overall human conversion without conditioning on lead type. Because this outcome combines changes in the human-facing lead mix with any associated behavioral responses, it should not be interpreted as isolating a purely mechanical composition effect. The results support the position-graded conversion prediction. The AI expansion does not produce a uniform shift in hot-lead conversion. Instead, it flattens the within-day profile through substantial declines at the earliest positions, little change at intermediate positions, and an increase in the late tail. The economically meaningful result is therefore a redistribution of conversion performance across the daily shift, not a common change in average performance.

### 4.3 Behavioral Margins

§4.2 documented a position-graded flattening of the within-day conversion profile. This section shows that the two behavioral margins predicted by the reservation-value mechanism, extensive-margin reallocation and intensive-margin effort, shift at the AI expansion, and that both margins inherit the same position gradient as the conversion result. Reallocation rises most sharply at the earliest positions and attenuates across the workday (§4.3.1, Table 4); per-lead effort compresses most at the earliest positions and less at later ones (§4.3.1, Figure 6). We report a statistical decomposition (Gelbach 2016) attributing observed conversion movements to the two behavioral margins (§4.3.2, Appendix Table E7). The residual direct effect captures the full within-call yield change net of the observable margins, and the decomposition is reported as a complementary quantification, not as the primary identification. The primary evidence consistent with a reservation-value response is the joint position gradient shared by the conversion, reallocation, and effort results, which position-invariant confounders would not naturally produce.

#### 4.3.1 Reallocation and Per-Lead Effort

Before examining the two behavioral margins jointly, we first check whether the AI expansion affected each margin as predicted by the reservation-value mechanism. The first is an extensive-margin reallocation response. Following an unsuccessful call, agents decide whether to return the current lead to the shared pool for another agent to pursue. The second is an intensive-margin effort response. Agents decide how much time and effort to devote to the current customer interaction. Both margins are endogenous responses to operational conditions, not fixed parameters, as established in prior research that treats agent effort as an unobservable strategic response to operational incentives and demand (Song et al. 2024).

Figure 6, Panel (a), plots the within-day reallocation profile before and after the AI expansion. The reallocation rate is monotonically decreasing in call-attempt order in both periods, with the largest before–after gap at the earliest positions (approximately 38% at positions 1–3 in the post-period against 25% in the pre-period). The two profiles converge by mid-workday. Although these adjusted profiles are descriptive, not causal, their within-day difference displays the front-loaded pattern articulated by the fading reallocation prediction. Figure 6, Panel (b), plots the within-day call-duration profile before and after the AI expansion. In the pre-period, call duration is steeply decreasing in call-attempt order, starting at roughly one minute at positions 1–3 and falling sharply across positions to under thirty seconds at positions 80 and beyond. In the post-period, the profile flattens. Early-position duration falls by roughly half whereas late-position duration is essentially unchanged, so the two profiles converge, exactly the within-day flattening that the fading effort prediction anticipates.

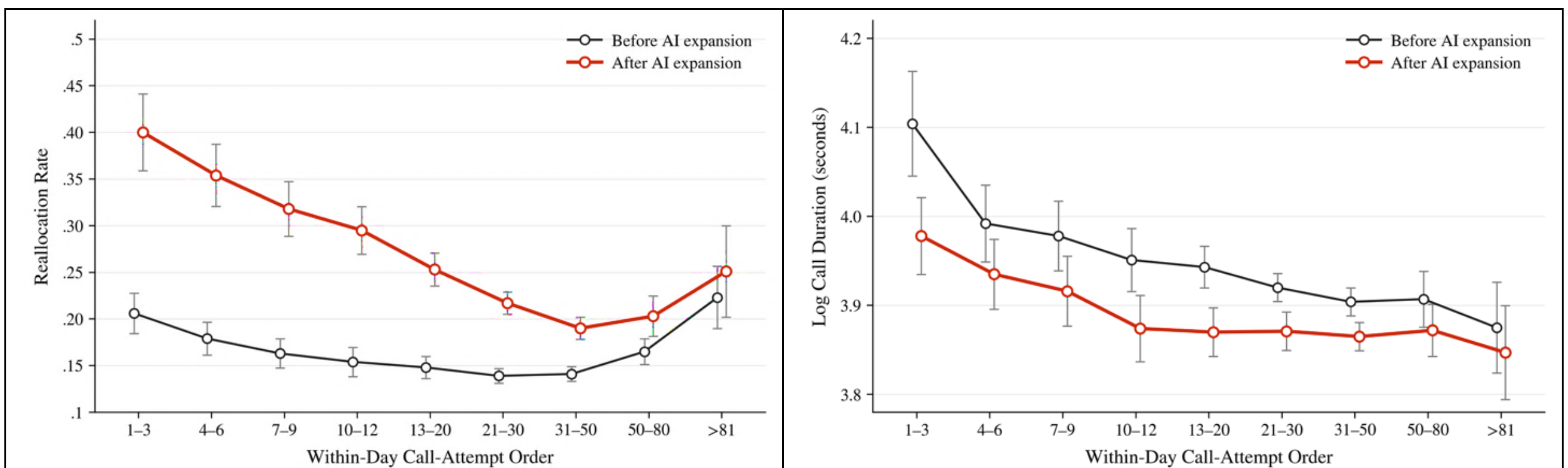

(a) Reallocation Rate (b) Log Call Duration

**Figure 6. Within-Day Profiles, Pre and Post AI Expansion**

***Notes.*** Residualized means by position bin, agent-day, pre vs post (±60-day windows). Agent, hour-of-day, and calendar-date FE; adjusted predictions at bin midpoints; 95% CIs from agent-clustered SEs. Panel (a): reallocation. Panel (b): log call duration.

Table 4, column (1) reports RDiT estimates of the reallocation-rate discontinuity at the August 28, 2024 cutoff. The headline estimate is at positions 1–3, where the reallocation rate jumps by 8.1 percentage points. Against a pre-period mean of about 25%, this is a roughly 32% increase in the rate at which agents return their earliest hot leads to the shared pool. The estimate at positions 4–6 is similar in magnitude (7.2 percentage points) and statistically significant at the 10% level. The estimates at positions 7–12 are smaller and statistically indistinguishable from zero. At positions 81–120, the coefficient is sizable and negative but imprecisely estimated (−21.4 percentage points, s.e. 14.4 percentage points), preventing a definitive

conclusion about the late-position response. The pattern of estimates is consistent with the fading reallocation prediction and with the asymmetric structure derived in §2.3. The statistically detectable increase is concentrated at the earliest positions, where the reservation-value shift is predicted to be largest. Estimates at later positions are less precise and do not establish an increase in reallocation. The economic interpretation we emphasize centers on the early-position estimate. At positions 1–3, on what should be the highest-stakes calls of the day, agents are now substantially more likely to return their leads to the queue. Consistent with offsetting movements across positions, the day-level aggregate estimate for hot-lead reallocation is negative but statistically indistinguishable from zero (Appendix Table C1: −0.119 with standard error 0.076 in the day-level RDiT on hot calls), reflecting the offset between the early-position rise documented above and the late-position decline at positions 81–120.

Table 4, column (3) reports RDiT estimates of the log-call-duration discontinuity at the cutoff. The headline estimate is at positions 1–3, where log call duration falls by 0.145, roughly a 13% decline in call length on the earliest hot-lead calls of the day. The estimate at positions 7–9 is also negative, whereas the estimates at intermediate and late positions are close to zero. The within-day duration profile thus compresses sharply. Early-position duration declines, late-position duration is unchanged, and the cross-position dispersion falls. This pattern is consistent with the fading effort prediction and with the position-graded structure derived in §2.3. The compression of the cross-position duration profile shows that the early-position advantage in effort devoted to hot leads narrows after the expansion. This pattern is consistent with, but does not directly establish, a broader compression of effort across lead types. When the human-facing pool becomes more concentrated in hot leads and the reservation value rises, agents differentiate their effort less across the leads they encounter on the early, high-intent calls of the day. The compression is concentrated at the earliest positions, where the reservation-value shift is largest, and is essentially absent at the latest positions, where the option value of continued search has collapsed.

**Table 4. RDiT Estimates at the AI Expansion Cutoff: Hot Calls**

| Within-day call position | (1) Reallocation rate | (2) Reallocation \| no conversion | (3) Log call duration |
|---|---|---|---|
| 1–3 | 0.081* | 0.075* | -0.145*** |
| | (0.044) | (0.043) | (0.054) |
| 4–6 | 0.072* | 0.072* | -0.009 |
| | (0.040) | (0.041) | (0.053) |
| 7–9 | 0.037 | 0.043 | -0.128** |
| | (0.044) | (0.041) | (0.052) |
| 10–12 | 0.068 | 0.061 | -0.035 |
| | (0.045) | (0.046) | (0.063) |
| 81–120 | -0.214 | -0.250* | 0.003 |
| | (0.144) | (0.139) | (0.050) |
| Agent FE | Yes | Yes | Yes |
| Calendar-date FE | No | No | Yes |
| Hour-of-day FE | Yes | Yes | Yes |

***Notes.*** Position-specific RDiT for hot calls. Col (1): reallocation (all hot calls). Col (2): reallocation | no conversion. Col (3): log call duration (connected calls). Agent + hour-of-day FE (cols 1–2); + calendar-date FE (col 3). MSE-optimal bandwidth (Calonico et al. 2014); ±10-day donut. Agent-day-clustered SEs. *$p$<0.10, **$p$<0.05, ***$p$<0.01.

To ensure the reallocation increase is not mechanical, driven by the concurrent conversion decline that enlarges the eligible base, we re-estimate the effect conditional on no conversion. The positions 1–3 discontinuity is essentially unchanged, at 7.5 percentage points relative to the unconditional 8.1 percentage points, confirming that the shift reflects a genuine increase in agents' willingness to release leads, not a change in the eligible base. The reservation-value mechanism predicts both the within-day flattening of the hot-lead effort profile reported here and a cross-type compression of effort between hot and cold leads, but the two predictions are not independent. Under our framework, the same reservation-value shift drives both. We focus on the within-day formulation because it ties directly to the position-bin RDiT design that organizes our other tests, and because a cross-type test estimated post-expansion would be confounded by the AI-induced change in which calls are classified as hot or cold. The within-day flattening on hot leads is therefore the more cleanly identified test of the effort-compression prediction in our setting.

**4.3.2 What Leads Are Reallocated? Downstream Outcomes of Reallocated Leads**

Reallocation rises sharply at the early within-day positions. A natural interpretation is that agents filter more selectively as the pool improves, reallocating leads perceived as less promising and retaining those judged more likely to convert. If that were the whole story, reallocated leads should display lower realized downstream value than retained leads. We assess this by comparing the realized downstream outcomes of the two groups.

Figure 7 compares the realized downstream outcomes of leads that agents reallocated with those of leads retained by the same population of agents during the post-expansion period. In the pooled comparison, reallocated leads convert at an eventual rate of approximately 5.7%, compared with 3.0% among retained leads, a ratio of approximately 1.9 to one ($p < 0.001$). Conditional on conversion, the average payment is somewhat lower for reallocated leads, approximately ¥80.97 versus ¥88.63 for retained leads ($p = 0.001$), though this margin is estimated from the small subset of calls that convert and is correspondingly imprecise. Combining the two margins, the expected realized value per lead, equal to the conversion rate times the average payment conditional on conversion, is higher for reallocated leads, approximately ¥4.6 versus ¥2.7, because the conversion advantage more than offsets the lower conditional payment. Reallocated leads are therefore not leads with low realized downstream value; if anything, the pooled association runs in the opposite direction. This pooled association is, however, largely a between-agent phenomenon, not a within-agent selection pattern. When we absorb agent fixed effects and compare reallocated and retained leads handled by the same agent, the differences collapse: the conversion gap falls to 0.6 percentage points ($p = 0.73$) and the conditional-payment gap to ¥1.4 ($p = 0.79$), neither distinguishable from zero. The pooled gap thus reflects composition. Agents who reallocate more also tend to convert at higher rates, and it is this correlation across agents, not sorting of high-intent leads within an agent's own caseload, that generates the aggregate pattern in Figure 7. We therefore read the figure as descriptive, not as evidence of deliberate within-agent selection.

Two interpretations survive this qualification. The first is minimal and is supported directly by the data: reallocated leads are not disproportionately low-value, so reallocation is not well described as agents offloading leads they have identified as worthless. The second is a behavioral hypothesis consistent with the aggregate pattern but unconfirmed by the present comparison at the within-agent level. Within the broad hot and cold categories, leads differ in the effort required to close them: some convert with a brief standard interaction, others only with sustained effort, careful objection-handling, and long calls. When the reservation value rises, the additional effort required to close a difficult but potentially valuable lead can exceed the value of giving it up and drawing a fresh one, so agents may retain easy-to-close leads and reallocate effort-intensive ones that nonetheless retain downstream value. Figure 7 is consistent with this account, but the within-agent nulls above mean it remains a conjecture the comparison does not establish, whether because any such selection is genuinely weak or because the test lacks power.

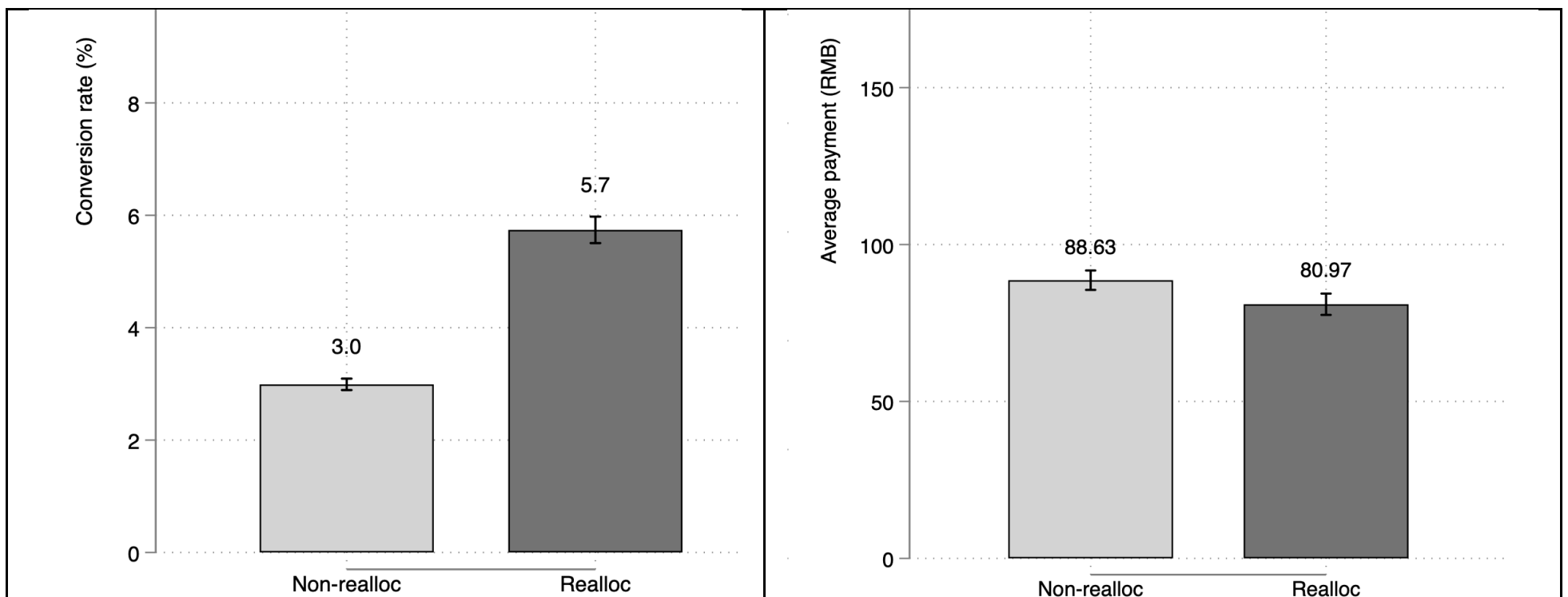

(a) Conversion Rate (b) Average Payment

**Figure 7. Realized Outcomes of Reallocated vs Retained Leads (Post-Expansion)**

***Notes.*** Pooled reallocated vs retained leads, post-expansion. Bars: group means; whiskers: 95% CIs; $p$-values from Welch t-tests. Reallocated: conversion 5.7% vs 3.0% ($p$<.001), conditional payment ¥80.97 vs ¥88.63 ($p$=.001). With agent FE, differences collapse (0.6 pp, $p$=.73; ¥1.4, $p$=.79).

Two further features of the comparison reinforce this caution. First, the higher subsequent conversion on reallocated leads partly reflects that reallocated leads receive additional attempts by other agents in the post-reallocation window, whereas retained leads receive no further attempts in the focal observation by definition of being retained. The comparison therefore measures eventual lead-level realized value, not a controlled latent-quality test that holds attempt count fixed. Second, the observation windows for the two groups differ on average, since reallocated leads are observed over the period after their reallocation, which extends further beyond the cutoff for leads reallocated earlier in the post-period. A more controlled analysis would fix both the observation window and the attempt count; we do not attempt one here, and we accordingly treat Figure 7 as suggestive, not as a direct estimate of latent-quality differences.

The option-preservation logic that would rationalize reallocating potentially valuable leads is therefore a natural feature of this institutional environment. The six-month no-recontact rule makes reallocation a comparatively less destructive way for an agent to disengage from a difficult lead: the agent can move on to another assignment without foreclosing the firm's ability to pursue the current lead later. The rule therefore shapes how the reservation-value mechanism is expressed in agents' reallocation decisions. We treat the reservation value generated by expected future assignments as the primary behavioral channel, and the no-recontact rule as an institutional feature that increases the relative attractiveness of reallocation as the way to move on. The evidence in this section is consistent with that logic without establishing it.

### 4.4 Alternative Explanations and Robustness

The position-graded pattern documented in §§4.1–4.3 is consistent with the reservation-value mechanism but does not by itself identify it. §4.4.1 assesses whether leading alternative explanations reproduce this gradient, and §4.4.2 assesses whether the position-graded estimates are robust to standard modeling choices.

#### 4.4.1 Alternative Explanations and the Position Gradient

The identification argument in §3.3 requires that no confounder shapes a matching position-graded response. Appendix D provides evidence against the leading confounders (detailed below) and documents within-hot-lead composition shifts at late positions (Appendix D.6). Appendix Table E5 shows that stratifying by clock hour does not detect the response, because agent start times are widely dispersed (Appendix Figure E1) and a given clock hour therefore pools very different within-day positions. The evidence in Appendix D reduces, but does not eliminate, the plausibility of the principal alternatives, none of which clearly reproduces the full position-graded response. Several alternative explanations could generate changes in the same direction as those predicted by the reservation-value mechanism. Three are addressed by the structure of the main results themselves. First, the concentration of the response at the earliest within-day positions is less consistent with explanations based solely on agent-level beliefs, reference points, or norms. Such mechanisms would not naturally predict a response that weakens sharply as the workday progresses, whereas the reservation-value mechanism does because the number of remaining opportunities declines with call position. Second, the absence of a corresponding increase in late-position effort (§4.3.1) provides little support for a pure effort-rebudgeting explanation, under which effort saved on early calls should reappear later in the day. Third, the much weaker response at the June 5, 2024 trial cutoff (Appendix Table E3) is less consistent with an explanation based solely on agents' awareness of AI. Agents were aware of the system at both cutoffs, but only the August 28 expansion produced a sharp change in the composition of the human-facing opportunity pool. Taken together, these patterns are more consistent with agents responding to the changing state of the shared opportunity pool, as emphasized by the sequential-search framework.

To assess whether the decline at early positions reflects a shift in observable lead composition or a change in agent behavior, we use the number of prior call attempts recorded for each lead. Conditional on agent, hour-of-day, and calendar-date fixed effects, this variable is a strong predictor of conversion (Appendix Table D.3.3, Panel A): the linear coefficient is 0.032 ($t = 21.6$), and leads with five or more prior attempts differ from first attempts by 23.4 percentage points ($t = 15.1$). A shift in the composition of leads along this dimension would therefore be detectable in our data. Panel B reports the discontinuity in prior attempts at the cutoff, and the pattern is asymmetric in the way our interpretation predicts. At the late tail, where we attribute the conversion gain to a shift in pool composition, prior attempts move significantly (0.138, SE 0.069), and the implied indirect effect accounts for 13 percent of the reduced-form gain in that bin. At the early positions where conversion declines, the shift is statistically indistinguishable from zero: −0.084 (SE 0.074) at positions 1–3 and −0.043 (SE 0.067) at positions 4–6, against a standard deviation of 2.18 attempts. The implied indirect effects account for 6.2 and 5.1 percent of the corresponding reduced-form estimates, and the upper bound implied by the 95 percent confidence interval at positions 1–3 is approximately 17 percent. The exercise thus detects a compositional shift where our interpretation predicts one and fails to detect one where it does not, and at least four-fifths of the early-position decline remains unaccounted for by prior-contact history.

Beyond observable lead composition, Appendix D reports tests of three additional explanations. First, under slower pace, agents respond to the improved pool by increasing the time between calls, not by changing their behavior within calls. The inter-call gap shows no detectable change at the earliest positions, where the reallocation, effort, and performance responses are largest; the only detectable movements are negative coefficients at positions 7–9 and 10–12, indicating a shorter gap, in the opposite direction from the slower-pace prediction (Appendix Table D.1.1). Second, a learning account predicts sharper adjustment among agents with less accumulated experience, yet the response is comparable in magnitude across experience levels (Appendix D.2; tenure moderator in Appendix G). The event-study in Figure 8 shows the early-late conversion gap reaching its new level within the first thirty days after the cutoff and holding steady over the remaining post-period, which leaves little room for belief updating that accumulates over several months. The coarser window-by-window estimates in Appendix Table D.2.1 are too imprecise to characterize the adjustment path in finer detail.

Third, under a customer-side explanation, the observed changes arise because customers respond differently following the AI expansion. We find no detectable discontinuities in early hang-up or customer-composition measures at the positions driving the main effects (Appendix Tables D.3.1 and D.3.2). Finally, Appendix Table D.4.1 shows that the within-day task stream becomes more homogeneous after the expansion, with longer contiguous runs of hot leads. This finding accords with the underlying change in the opportunity pool that activates the reservation-value mechanism.

**4.4.2 Robustness Checks**

We assess whether the position-graded response documented in §4.2 and §4.3 is robust to the principal modeling choices. Table 5 summarizes four principal robustness exercises (outcome coverage varies across exercises; the event-study check applies only to conversion). Full estimates appear in Appendix E. The position gradient survives an alternative ±60-day bandwidth (Appendix Table E1), alternative 5-call position bins (Appendix Table E2), an event-study specification of the early-late conversion gap (Figure 8), the alternative June 5, 2024 trial-launch cutoff (Appendix Table E3), and a connected-call sample-accounting check (Appendix Table E4). Stratifying by clock hour does not detect the response (Appendix Table E5), and the position-specific estimates are reported with lead freshness controlled for (Appendix Table E6). The most consistent result across specifications is a negative conversion response concentrated at the earliest within-day positions; reallocation and call duration provide directionally supportive but less monotonic evidence. The gradient in conversion is not an artifact of any single specification choice.

**Table 5. Robustness Check Summary**

| Concerns | Robustness Check | Modification from Baseline | Conversion | Reallocation | Duration | Evidence |
|---|---|---|---|---|---|---|
| Event Study | Event-study dynamics | Trace dynamics of the early-late conversion gap around the cutoff | ✓† | NA | NA | Figure 8 |
| Bandwidth | Fixed 60-day bandwidth | Replace MSE-optimal bandwidth with ±60-day fixed window | ✓ | ✓ | ✓ | Table E1 |
| Binning | Alternative bin definitions | Re-define call-order bins at widths of 5 | ✓ | ✓ | ✓ | Table E2 |
| Alternative Cutoff | June 5, 2024 trial cutoff | Re-estimate at trial-period start (weaker first stage) | ✓† | ✗ | ✓† | Table E3 |

***Notes.*** Each row: whether each finding is preserved under the modification. ✓ = directionally consistent and significant at 10% in early bins. ✓† = directionally consistent, not significant. ✗ = not directionally consistent.

The three findings are robust to the alternative bandwidth and bin definitions. The early-position increase in reallocation, reduction in call duration, and negative position-specific conversion estimate remain statistically detectable when we replace the MSE-optimal bandwidth with a fixed ±60-day window (Appendix Table E1) and redefine the call-order bins using widths of five (Appendix Table E2). At the June 5, 2024 trial cutoff, the conversion and duration estimates remain directionally consistent but are smaller in magnitude, whereas the reallocation estimates turn significantly negative at positions 4–6 and 7–9 (Appendix Table E3). The attenuated estimates at the June 5 event match its weaker first stage. The directional similarity across the two rollout stages reduces, although does not eliminate, concern that the August results reflect an unrelated break specific to that date. We also trace the dynamics around the cutoff with an event-study specification of the early-late conversion gap, defined as the difference in conversion between calls at positions 1–6 and 81–120 within the same agent-day, across intervals relative to the rollout. The gap is statistically indistinguishable from zero in both pre-cutoff intervals, satisfying parallel trends, and then drops by about 3 percentage points within the first thirty days after the cutoff and remains there (Figure 8). The gap completes its adjustment within the first post-cutoff interval, a horizon difficult to reconcile with learning or fatigue processes that build over several months.

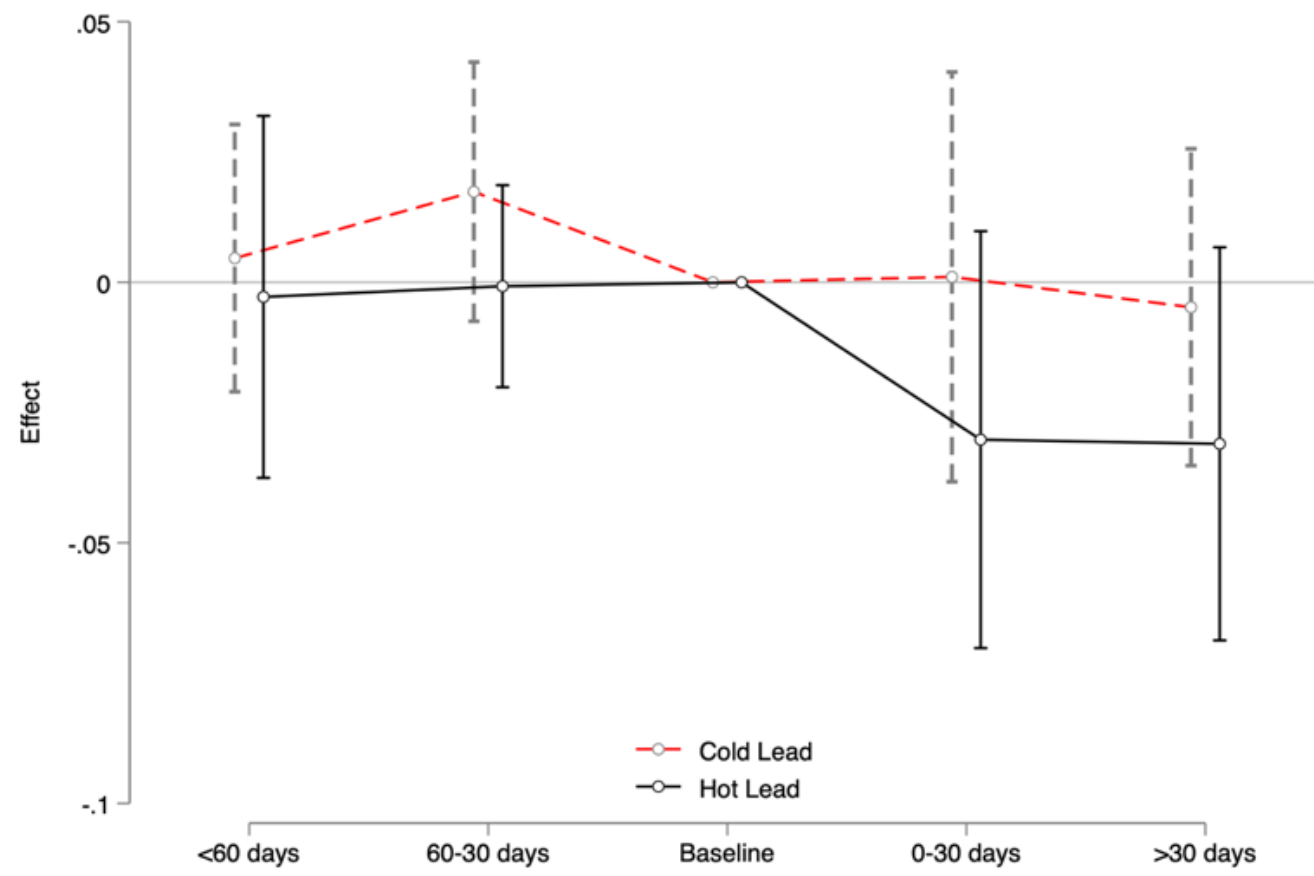

**Figure 8. Event-Study Estimates of the Early-Late Conversion-Rate Gap by Lead Type**
***Notes.*** Event-study estimates of the within-day early–late conversion gap around August 28, 2024, separately for hot and cold. Event-time bins: >60 before, 30–60 before, baseline, 0–30 after, >30 after; baseline normalized to zero. Ten-day donut. 95% CIs. Agent, day-of-week, and month FE.

### 4.5 Heterogeneity

As an exploratory diagnostic, we examine whether the position-graded conversion response varies across agents. A causal forest estimated on human-handled hot leads at positions 1–9 within the primary analytic window, following Wager and Athey (2018) and Athey et al. (2019), suggests that the conversion response is most robustly moderated by pre-expansion reallocation behavior (interquartile-shift effect: 1.4 pp less negative, $p<0.01$), with other moderators (daily call volume, call duration, tenure, gender) showing point-estimate differences that are not statistically distinguishable from zero. Because treatment varies at the calendar-date level and moderators are estimated proxies, we treat these findings as suggestive patterns, not independently identified subgroup causal effects. Full estimates from the causal-forest heterogeneity analysis and the best-linear-projection visualization appear in Appendix G.

## 5. Implications for Firm-Level Yield

In §4 we established that human workers respond to AI deployment along two behavioral margins, reallocation and within-call effort, with both the pool improvement and the response graded across the within-day sequence. This section examines what these findings imply for firm-level yield. We first decompose the change in human yield on hot leads into an ordering channel and a conduct channel (Table 6). The ordering channel isolates the within-day rearrangement of hot leads holding the hot-lead pool size fixed; it is one component of the broader mechanical enrichment described in §2.3, which also includes the hot/cold composition shift documented in Figure 3. The two are related but not identical accounting objects. We then note how the position-graded structure of the behavioral response constrains the firm's ability to translate a richer opportunity pool into higher realized output. The framework is an illustrative calibration, not a structural identification of causal channels; the full setup, calibration, and rhythm counterfactual appear in Appendix F. We decompose the change in average human yield on hot leads between the pre- and post-expansion regimes by substituting the position-order profiles one at a time. The ordering channel swaps the hot-lead profile from its no-AI to its AI shape while holding conduct fixed; it is the mechanical effect of the shift in the order distribution of hot leads. The conduct channel swaps the conversion profile while holding the new ordering fixed; it is the behavioral effect on within-call conversion, which combines the reallocation response of §4.3 and the within-call effort response. Table 6 reports the decomposition.

**Table 6. Counterfactual Analysis: AI Impact and Productivity Channels**

| Statistic | (1)<br>Pre-expansion | (2)<br>Post-expansion | (3)<br>Δ (Post − Pre) | (4)<br>Pct. Change |
|---|---|---|---|---|
| ***Panel A: Realized Conversion Composition, per 50 Human Calls*** | | | | |
| Human · hot leads | 1.336 | 1.198 | −0.138 | |
| Human · cold leads | 0.261 | 0.264 | +0.002 | |
| AI · cold leads | 0.049 | 0.188 | +0.139 | |
| AI · hot leads | 0.083 | 0.189 | +0.106 | |
| Total (arithmetic check only) | 1.730 | 1.838 | | |
| Memo: human calls in window | 243,473 | 166,149 | −77,324 | −31.8% |
| Memo: AI calls in window | 28,579 | 128,889 | +100,310 | +351.0% |
| ***Panel B: Model-Based Yield Decomposition (Hot Leads)*** | | | | |
| Avg. Yield on Hot Leads | 0.0452 | 0.0375 | −0.0077 | −17.00% |
| (1) Ordering Effect | | | +0.0003 | +0.58% |
| (2) Conduct Effect (reallocation + effort) | | | −0.0080 | −17.58% |
| Total Net Effect | | | −0.0077 | −17.00% |

***Notes.*** Panel A: realized composition of conversions per 50 human calls (accounting composition, not effect estimate). Because human volume fell 31.8% whereas AI rose 351%, per-call and firm-level totals diverge. Identified day-level estimate in Appendix Table C1. Panel B: accounting model of Appendix F.

The routing change redistributed where conversions come from; it did not add to them. Per 50 human calls, humans lose 0.138 hot-lead conversions whereas AI adds 0.139 on cold leads. At the firm's daily volume the two are not equal. Humans lose 50.5 hot-lead conversions per day against 7.7 recovered by AI on cold leads, a ratio of about seven to one. The deployment nonetheless looks successful on one standard measure. System conversions per 50 human calls rose 6.3%, because human calls fell 31.8% whereas total conversions fell 27.5%. That gain comes from reassigning volume, not from better performance. On the higher-value work humans retained, conversion fell from 4.4% to 3.7%, a decline of 15.9%, and across all human calls it fell from 3.2% to 2.9%. The accounting model-based hot-lead yield decomposition in Table 6 Panel B estimates the same margins over positions 1–50 using calibrated order weights, and reports a decline of 17%. The two differ because the model excludes the latest positions where conversion increased. The only quantity that rises is output per unit of human labor, and it rises because its denominator falls faster than its numerator.

The ordering channel is small and positive: the mechanical improvement in the order composition of hot leads raises yield by 0.58%. The conduct channel is large and negative, at −17.6%, and operates in the opposite direction. Because Appendix Table E7's statistical decomposition finds that the observed reallocation and duration margins jointly account for only a modest share of the reduced-form conversion change, we interpret the conduct channel as capturing the full residual conduct-related yield change, not as identifying reallocation and duration as the specific mediators. AI improves the composition of opportunities facing humans, whereas workers adjust within the calls they continue to attempt; the behavioral response is about thirty times the magnitude of the mechanical gain under our calibration. The decomposition reconciles the puzzle that motivates the paper. A manager would expect AI to raise productivity through a better-composed pool; the ordering channel confirms the composition does improve, but the conduct channel operates in the opposite direction with far greater magnitude, and §4.4.1 addresses the principal alternative within-match channels of slower pace, learning, and customer-side responses.

The decomposition framework can be extended to compute counterfactual firm yield as a joint function of the average hot-lead share and the within-day rhythm of hot-lead release. This exercise treats the rhythm as a strategic lever holding the pool fixed. Across the range of rhythms the data support, firm yield is decreasing in the rhythm at every hot-lead share we examine, and easing the rhythm while holding the pool fixed raises conversions per call. Within the calibrated model, returning the rhythm to its pre-expansion value increases predicted firm conversions per call by about 13.5 percent, with a 95 percent parametric-bootstrap interval of 10.5 to 16.6 percent. Appendix F reports the derivation, the calibrated surface, and the bootstrap.

## 6. Discussion

We examine how human workers respond when AI improves the composition of the opportunity pool left for them. In our setting, an AI voice agent handled a separate portion of a shared customer acquisition pipeline. By absorbing low-conversion cold leads, it enriched the opportunity pool left for human agents. Yet the customer acquisition rate declined most where this opportunity enrichment was greatest and recovered as the opportunity improvement faded across the workday. Reallocation and call duration, as alternative outcomes, provide directionally consistent, though less monotonic, evidence. Taken together, the findings are most consistent with a reservation-value behavioral response: when service workers expect better opportunities ahead, they become less willing to persist with the opportunity currently in hand. Our calibrated accounting decomposition analysis shows that the small mechanical gain from the enriched human-facing pool is more than offset by a larger negative behavioral response, leaving the aggregate day-level performance effect close to zero. The findings point to a broader principle of AI-enabled work design: AI can change human behavior not only by assisting, replacing, or advising workers, but also by resetting the expectations and outside options against which workers weigh the task currently in hand.

Our findings connect behavior-aware service operations with AI-enabled work allocation. Behavior-aware research shows that workers adjust their pace, effort, persistence, and task choices in response to operating conditions and do not act as fixed-rate production units (Donohue et al. 2020; Roels and Staats 2021). Extant studies emphasize workload and congestion (KC and Terwiesch 2009; Tan and Netessine 2014; Cho et al. 2019), scheduling and task ordering (Ibanez and Toffel 2020; Kesavan et al. 2022), and the value, priority, and discretion associated with the current task (Ibanez et al. 2018; KC et al. 2020; Kostami 2024; Beil et al. 2026). We extend this literature by identifying opportunity distribution, the quality composition and temporal ordering of tasks over a shift, as another operating condition that shapes worker behavior. Workers respond not only to how much work is present or to the value of the current task, but also to the opportunities they expect to receive next. This extension also advances research on gatekeeping, referral, and discretionary shared-pool systems. Prior work examines whether workers retain or refer a task and whether they participate in a shared pool when outside options change (Shumsky and Pinker 2003; Freeman et al. 2017; Saghafian et al. 2018; Freeman et al. 2021; Hu et al. 2022; Liu et al. 2023). Our findings show that the future opportunity pool can affect both whether workers continue with the current task and how much effort they invest before moving on. More importantly, the outside option in our setting is not fixed. It is endogenously reshaped by the firm's AI-routing policy. Staffing and routing decisions therefore affect behavior not only by changing workload or task assignment, but also by changing the alternatives against which workers evaluate the task currently in hand.

Our second contribution is to the AI-enabled work allocation body of work. Extant studies commonly examine AI as a substitute that performs the focal task in place of a human worker or as an augmenting tool that assists the worker during the focal interaction (Acemoglu and Restrepo 2019; Brynjolfsson et al. 2025; Dell'Acqua et al. 2026b; Fügener et al. 2026). Other work studies how workers respond to algorithmic recommendations, rules, and market conditions (Wang Q et al. 2023; Benjaafar et al. 2025; Snyder et al. 2026). We identify a different pathway. AI need not participate in the focal human task or directly communicate with the worker. By selectively processing one portion of a shared opportunity stream, it changes the work environment facing humans and thereby changes their behavior. This finding challenges the view that parallel AI and human work can be evaluated as independent production streams. In a selective parallel-allocation system, AI and humans may handle separate tasks, but their performance remains behaviorally connected because they draw from the same underlying pool. Routing lower-value opportunities to AI mechanically improves the work left for humans, but it can also raise the attractiveness of moving on from the current task. Human–AI specialization can therefore have a behavioral limit: assigning higher-value opportunities to humans does not necessarily improve realized human output when the resulting outside options weaken persistence and effort. This extends emerging work on the human–algorithm connection in operations (Caro et al. 2026; Cohen et al. 2026) by showing that the value of specialization depends not only on comparative task performance, but also on how the resulting allocation changes human behavior.

More broadly, we build on prior evidence that features of the service environment shape workers' responses to algorithmic advice (Snyder et al. 2026) by showing that AI deployment can itself reshape that environment. This mechanism is also distinct from another documented backfire path, which operates through competitive skill-dissemination dynamics that demotivate top performers when AI encodes their expertise (Yuan et al. 2026); our mechanism instead operates through service workers' reservation value for continuing when the AI-induced opportunity pool changes. Related recent work shows that generative-AI voice agents can outperform human workers on narrow emotional-labor tasks such as debt collection (Fang et al. 2026), and function as a teammate that reshapes coworker interaction and expertise flow (Dell'Acqua et al. 2026a).

Our third contribution is to translate a sequential-search prediction into a within-day empirical test. Sequential search theory predicts that better future opportunities raise the threshold for continuing with the current opportunity (Stigler 1961; McCall 1970; Lippman and McCall 1976). We extend this logic to a service setting in which workers can both reallocate a task and adjust their effort before moving on, yielding a clear prediction: because many future assignments remain early in the daily shift, the behavioral response to an improved opportunity pool should be strongest early and weaken as the sequence progresses. Our empirical strategy matches this prediction to the variation created by the AI expansion. The improvement in the human-facing pool is strongest early in the sequence and fades later in the day. Conversion, reallocation, and call duration exhibit corresponding position-varying responses, with the clearest behavioral changes occurring where the pool improvement is largest. A conventional before-after design to detect level shifts could establish that outcomes changed around the AI expansion, but it would not by itself distinguish a reservation-value response from other contemporaneous changes. The alignment between the position-varying first stage and the position-varying outcome responses, together with tests of customer composition, workforce composition, learning, and work pace, provides evidence most consistent with the reservation-value mechanism. Our methodological approach isolates this joint pattern, revealing how a theory-specific comparative analysis can provide stronger mechanism evidence than typical average treatment effect alone.

The findings have at least two implications for practice. First, parallel AI deployment is not behaviorally neutral. Aggregate measures such as AI task volume, automation rates, or the average quality of work left for humans likely miss critical changes in human behavior. Firms should evaluate AI-enabled routing as both a technological intervention and a behavioral intervention. Day-level averages may be misleading when positive and negative responses occur at different points in the workday. Second, effective routing requires managing not only which opportunities humans receive, but also when they receive them. Our accounting exercise suggests that concentrating high-intent opportunities early in the shift can weaken persistence precisely when the outside option of moving on is most attractive. Firms could therefore consider task routing rhythms that reduce excessive front-loading, such as staggering the release of high-intent opportunities or batching some lower-value work. Such interventions should be designed jointly with queue rules, reallocation rights, performance measures, and incentives. The lesson is not that firms should give workers worse opportunities or reverse AI deployment. It is that improvements in opportunity quality may require complementary workflow changes to preserve engagement with the task currently in hand.

Our study opens three promising directions for future research. First, future studies can examine the boundary conditions of the reservation-value response across service systems. Its strength may depend on the depth of the opportunity pool, AI capability, incentive structures, worker discretion, and the ease of task reallocation. The mechanism should be most relevant when service workers draw from a shared task pool, retain discretion over effort or task release, and expect additional opportunities after moving on. Testing these conditions in diverse workflows such as claims processing, fraud review, content moderation, customer triage, and shared customer acquisition pipelines would clarify the generalizability of our findings. Second, our position-varying identification strategy exploits a first-stage change that differs across the workday. Complementary designs could test the same theoretical logic when the first-stage change is uniform, expanding the empirical toolkit for evaluating sequential-search predictions in service operations research. Third, within-day position captures progression through the work sequence but not workers' perceived remaining horizon. Future research could measure queue conditions and worker expectations in real time or study settings with explicit shift endpoints to test the option-value logic more directly. Together, these extensions could support a broader theory of how AI-enabled routing reshapes human behavior in service systems.

## 7. Conclusion

The promise of AI deployment in service operations is that automating low-value work allows human workers to concentrate on higher-value tasks. In our customer acquisition setting, AI absorbed the weakest leads and enriched the opportunity pool facing human agents, yet agents responded by releasing leads more readily and investing less effort, with no net performance gain. This pattern suggests that AI-enabled allocation can affect human performance indirectly by changing workers' expectations about future opportunities and the outside options against which current tasks are evaluated. More broadly, AI does not merely change who performs a task; it can also change how humans value, prioritize, and engage with the tasks that remain. As a result, the performance consequences of AI deployment depend not only on the allocation of work, but also on the behavioral responses that allocation induces.

**When Better Opportunities Backfire: Service Worker Responses to Parallel AI Deployment in Customer Acquisition**

**ONLINE APPENDIX**

## Appendix A: Related Work

**Table A1. Operating Conditions and Worker Responses in Behavior-Aware Service Operations**

| **Worker-facing opportunity** | **Design lever / work structure** | **Study** | **Outcome** | **Key insight / relation to this paper** |
|---|---|---|---|---|
| AI endogenously reshapes worker-facing opportunity distribution | Parallel AI agent / shared lead pool | **This paper** | Worker behavior and conversion | AI improves future opportunities; reallocation and effort compression are consistent with higher reservation value |
| Worker-facing opportunity distribution taken as given | Staffing / workload | Cho et al. (2019) | Worker effort | Staffing should anticipate behavioral effort response |
| | | Tan and Netessine (2014) | Worker productivity | Workload changes how hard workers work |
| | | KC et al. (2020) | Worker task choice | High workload shifts workers toward easier tasks |
| | | Saghafian et al. (2018) | Service-system performance | Workload management matters in knowledge-based services |
| | Peer / coworker composition | Tan and Netessine (2019) | Worker performance | Coworker mix shapes individual performance |
| | Task choice / discretion | Ibanez et al. (2018) | Worker task ordering | Workers deviate from prescribed queue order toward faster/similar tasks |
| | | Freeman et al. (2017) | Worker gatekeeping | Workload shifts service configuration and referral decisions |
| | | Hassin and Nathaniel (2021) | Worker task choice | Workers strategically choose which tasks to accept |
| | | Kostami (2024) | Worker effort / throughput | Discretion changes effort allocation and throughput |
| | Optimal stopping | Manzoor et al. (2025) | Worker persistence / call termination | Formalizes salesperson stopping as optimal-stopping problem; complements reservation-value framing |
| | Routing / queueing | Zhan and Ward (2019) | Service-system performance | Joint staffing, routing, and payment manage the speed–quality tradeoff |
| | | Zhong et al. (2025) | Server behavior / queue outcome | Server incentives reshape queue outcomes |

**Table A2. Selected Work on Human–AI Work Allocation Relevant to Service Operations**

| **AI / work structure** | **Paper** | **Focal response** | Opportunity pool | **Key insight / relation to this paper** |
|---|---|---|---|---|
| Parallel deployment (same task, different opportunity quality) | **This paper** | Worker performance and behavior | Endogenous to AI deployment | AI improves the opportunity pool yet lowers early human conversion; reallocation and shorter calls indicate reduced persistence |
| Sequential AI→human handoff | Jia et al. (2024) | Employee creativity / performance | Upstream inputs reshaped; future pool not central | AI operates upstream before human selling; examines worker-skill heterogeneity, not future opportunity distribution |
| Augmentation | Brynjolfsson et al. (2025) | Worker productivity | Taken as given | AI improves productivity, for novices |
| | Zhang and Narayandas (2026) | Worker performance | Taken as given | AI gains concentrate among less-experienced agents |
| | Tong et al. (2021) | Worker performance | Taken as given | AI feedback helps performance; disclosure can weaken the effect |
| | Allen and Choudhury (2022) | Worker performance | Taken as given | Domain experience creates countervailing ability and aversion effects |
| | Cao et al. (2026) | Worker algorithm-use behavior | Taken as given | Workers selectively underuse AI on low-quality leads; forced use broadens adoption |
| | Dell'Acqua et al. (2026) | Worker performance | Taken as given | AI helps within the capability frontier but hurts beyond it |
| Dispatch | Kanazawa et al. (2026) | Worker performance | Assignment changes; pool not central | AI dispatch complements skill and raises output |
| | Liu et al. (2026) | Worker behavior | Assignment changes; pool not central | Drivers exhibit algorithm aversion |
| Automation/ Substitution | Beer et al. (2026) | Worker behavior | Role/process changes; pool not central | Automation in one stage spills over to human behavior elsewhere |
| | Fügener et al. (2026) | Human–AI team performance | Role/process changes; pool not central | Identifies when AI should automate versus augment |
| | Wang L et al. (2023); Luo et al. (2019) | Consumer response | Human pool not central | AI deployment changes demand and purchase responses |
| | Garvey et al. (2023); Castelo et al. (2023); Xu et al. (2026); Schanke et al. (2021) | | Human pool not central | Service evaluations depend on AI role, disclosure, and anthropomorphism |

Appendix B: Additional Evidence on the Opportunity-Distribution Shift

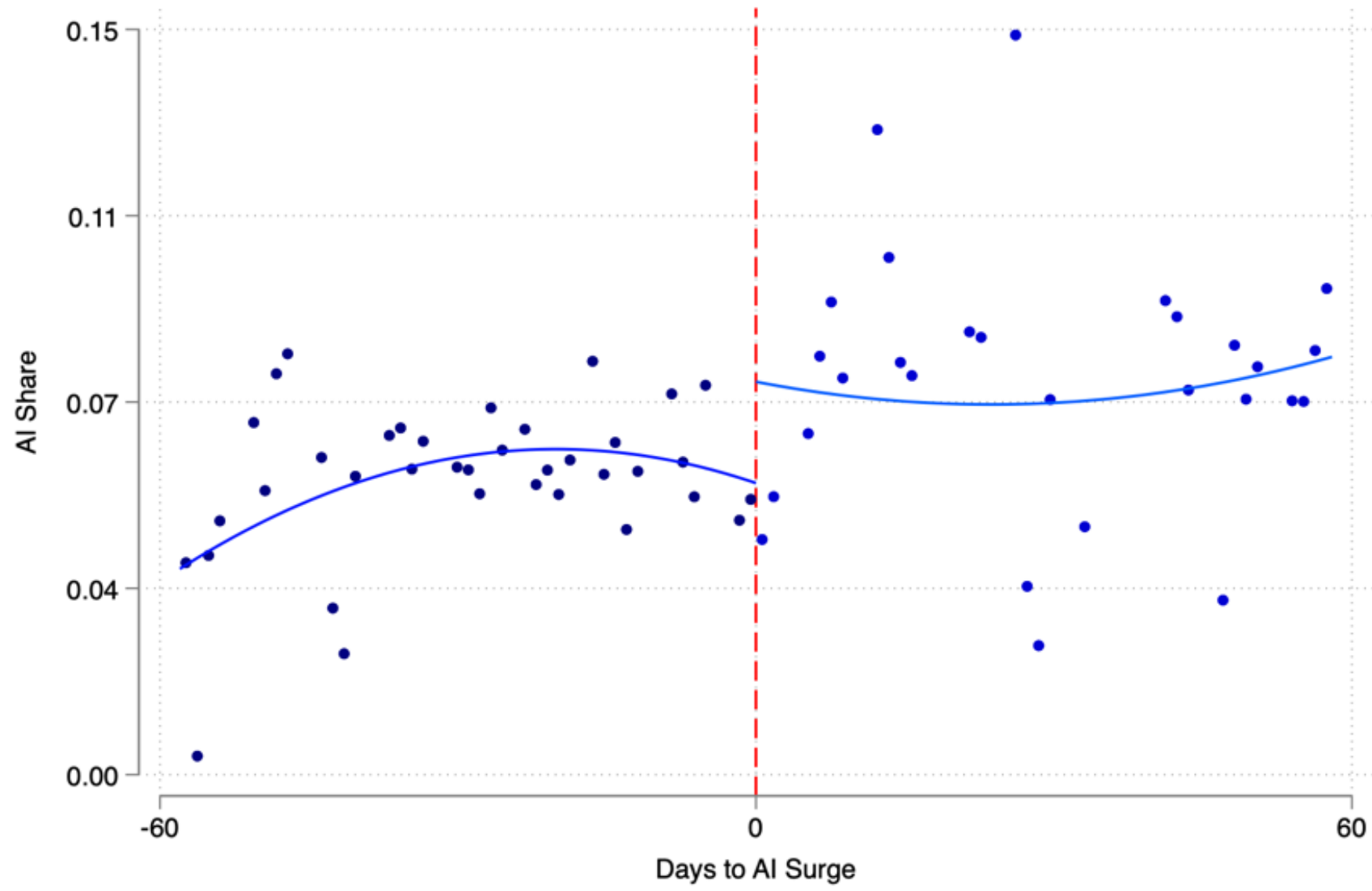

**Figure B1. Placebo First Stage: AI Share of Hot Leads**
***Notes.*** Daily AI share of hot leads, ±60 days around August 28, 2024. Local polynomial fits on each side. AI handling of hot leads is essentially flat at the cutoff (contrast with cold leads).

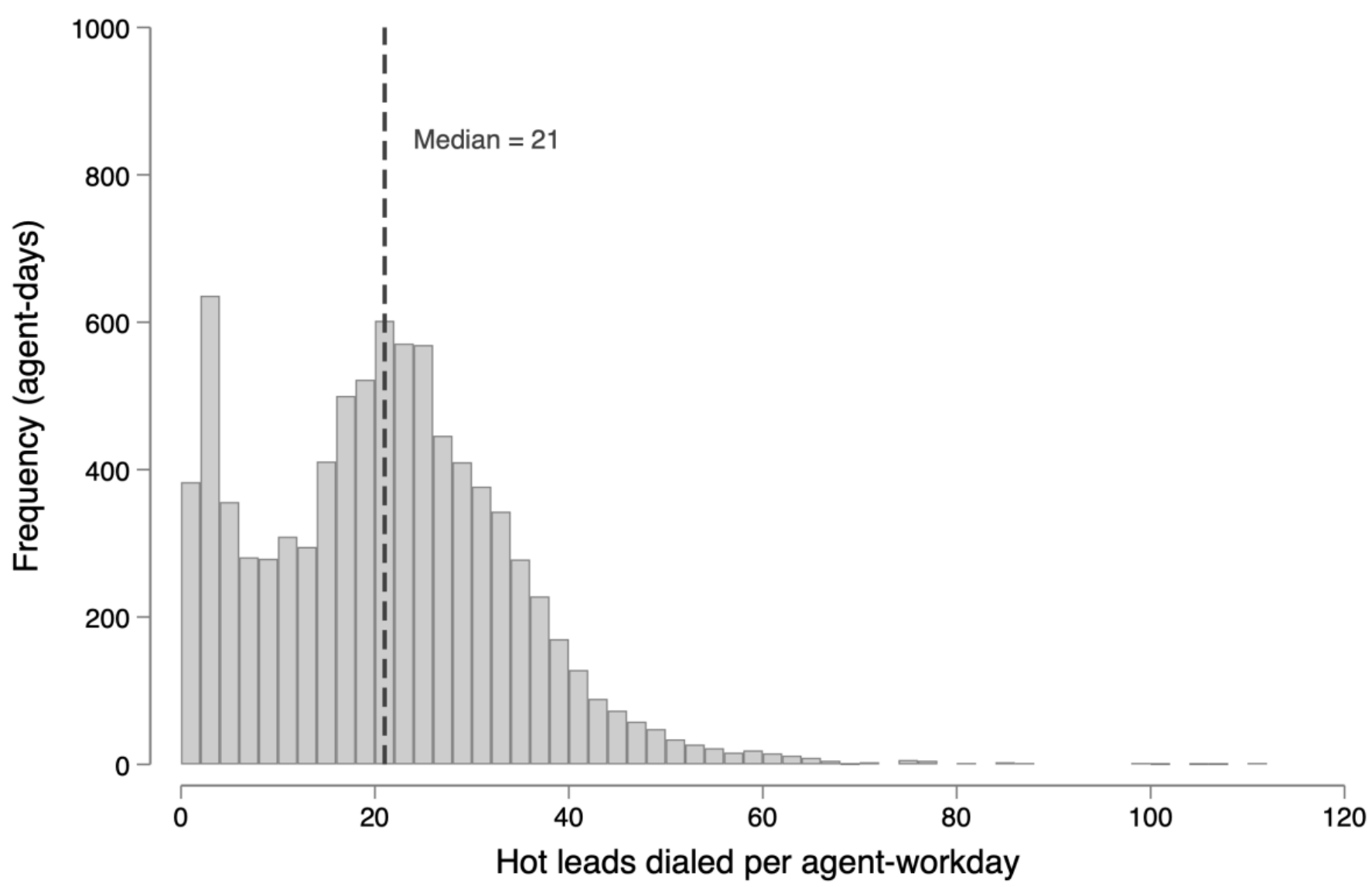

**Figure B2. Distribution of Hot-Lead Volume per Agent-Workday Before AI**
***Notes.*** Distribution of hot leads per agent-workday, pre-expansion, ±60-day window. Bin width = 2. Dashed line: median = 21. Right-skewed; 2 of 8,552 agent-days recorded >120 (127 and 137) and are omitted for legibility.

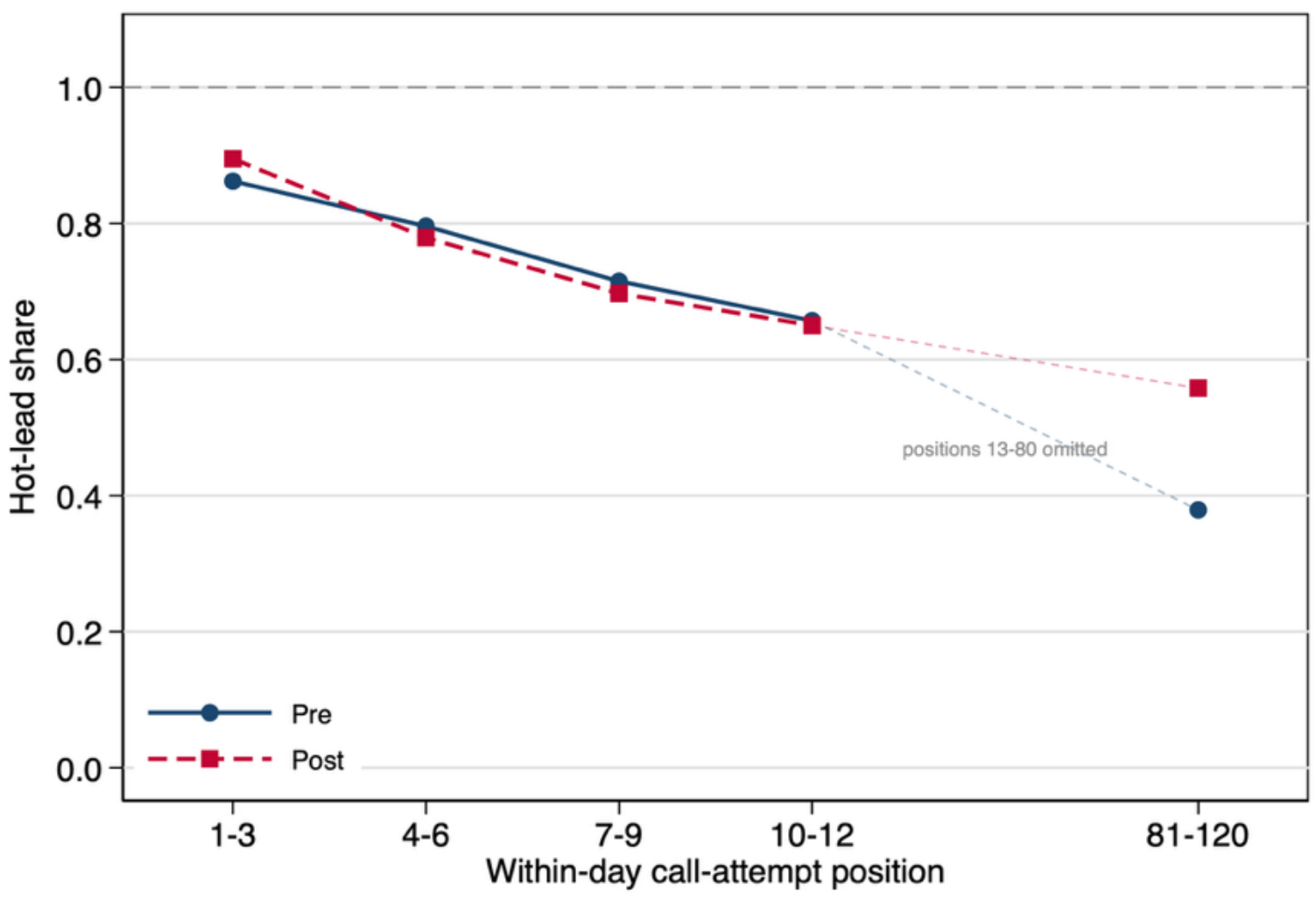

**Panel A.** Raw hot-lead shares, pre and post

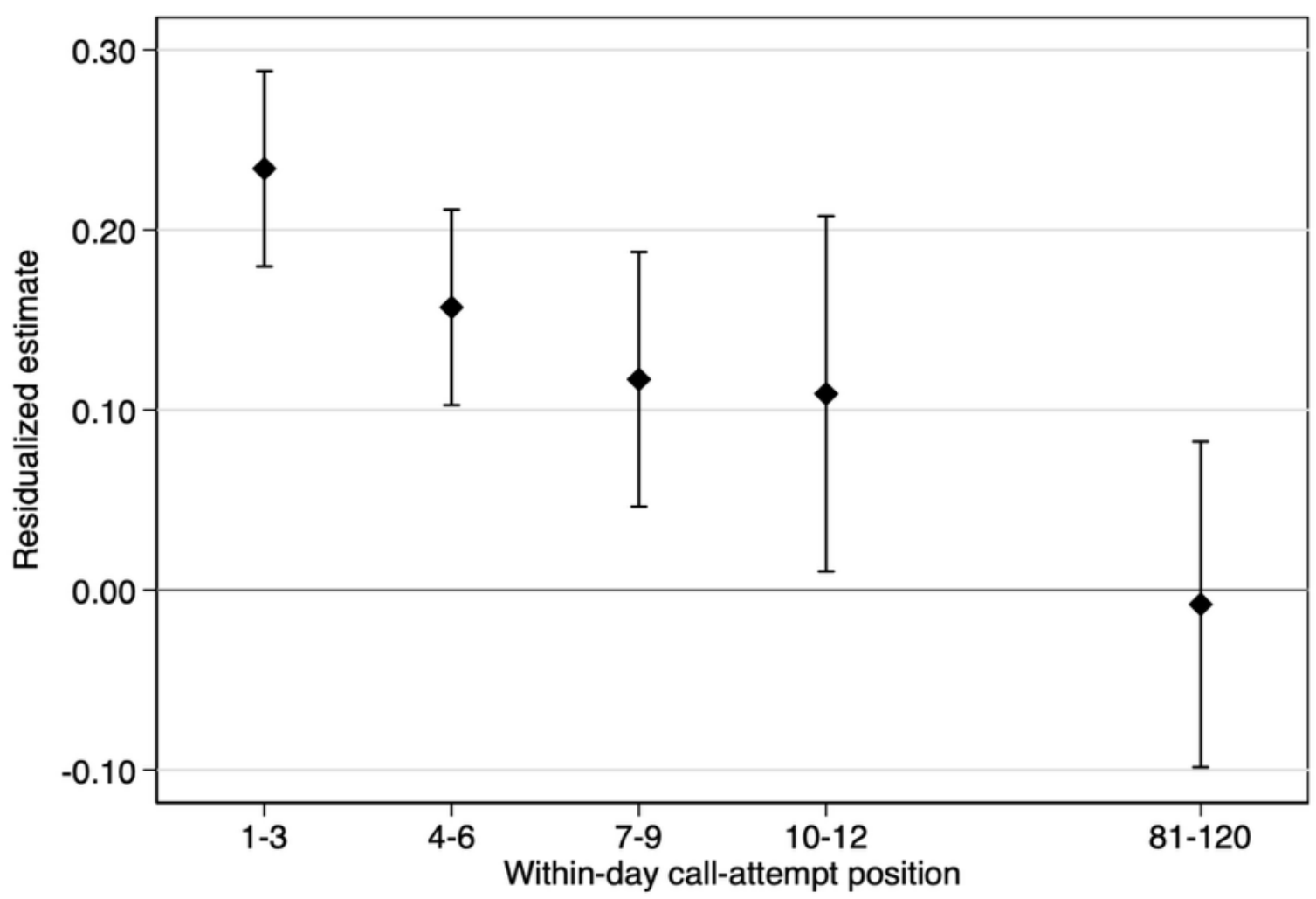

**Panel B.** Residualized estimates, relative to the same-day average

**Figure B3. Hot-Lead Exposure by Within-Day Call-Attempt Position: Levels versus Residualized Estimates**

***Notes.*** Panel A plots raw pre- and post-expansion shares of hot leads by within-day call-attempt position among human-handled calls, over a ±60-day window around the August 28, 2024 cutoff with a ±10-day donut, reproducing Table 1, Panel D. Sampling confidence intervals on these means are below 0.006 and are smaller than the plot markers. The dashed horizontal line at one marks the upper bound on the share. Panel B plots the estimates from Table 2, in which the dependent variable is the hot-lead indicator residualized on agent, hour-of-day, and calendar-date fixed effects. Because the residual is orthogonal to calendar date by construction, Panel B measures each bin's hot-lead composition relative to the same-day average, not a change in level, and the estimates sum to approximately zero across all position bins. Estimates are the discontinuity at the cutoff from local linear fits on either side; whiskers are 90 percent confidence intervals, corresponding to the $p < 0.10$ threshold used in Table 2. Standard errors are clustered by agent-day.

## Appendix C: Average Effects Results

Table C1 reports day-level RDiT estimates of the average effect of the AI expansion on the four outcomes. Aggregate effects on conversion and on log call duration are small and statistically indistinguishable from zero, consistent with the asymmetric within-day pattern documented in Section 4.2. The behavioral channel on retained hot leads and any compositional support on the enriched cold-lead margin sum to a small day-level aggregate. The day-level aggregate effect on reallocation is small and statistically indistinguishable from zero at the day level on hot calls (Table C1 column 2 Panel A: −0.119, s.e. 0.076), reflecting the offset between the early-position rise documented in §4.3.1 and the late-position attenuation at positions 81–120; the cold-call aggregate is positive but also insignificant (+0.050, s.e. 0.045).

**Table C1. Day-Level Regression Discontinuity in Time Estimates: Average Effects**

| | (1) Conversions | (2) Re-alloc | (3) Duration | (4) Gap |
|---|---|---|---|---|
| ***Panel A: Hot Calls*** | | | | |
| After AI Expansion | 0.011<br>(0.008) | -0.119<br>(0.076) | -0.047<br>(0.076) | 0.117<br>(0.524) |
| Observations | 62 | 66 | 102 | 113 |
| ***Panel B: Cold Calls*** | | | | |
| After AI Expansion | -0.001<br>(0.004) | 0.050<br>(0.045) | 0.016<br>(0.049) | -0.207<br>(0.404) |
| Observations | 78 | 92 | 90 | 102 |
| First-stage Agent FE | Yes | Yes | Yes | Yes |
| First-stage Hour FE | Yes | Yes | Yes | Yes |
| DOW FE | Yes | Yes | Yes | Yes |

***Notes.*** Day-level RDiT of average discontinuity, human-handled hot leads. Call-level outcomes residualized on agent and hour-of-day FE, averaged to daily; day-level regression with day-of-week FE. MSE-optimal bandwidth; ±10-day donut. HC-robust SEs. **p*<0.10, ***p*<0.05, ****p*<0.01.

**Table C2. Day-Level Regression Discontinuity in Time Estimates: Workload**

| | (1) Total Calls per Agent | (2) Hot Calls per Agent | (3) Cold Calls per Agent |
|---|---|---|---|
| After AI Expansion | 0.750<br>(2.987) | 4.086**<br>(1.872) | 1.474<br>(1.465) |
| Bandwidth | 48 | 41 | 59 |
| Observations | 78 | 64 | 100 |
| DOW FE | Yes | Yes | Yes |

***Notes.*** Day-level RDiT on per-agent workload. Cols (1)–(3): total, hot, cold calls per agent-day. Day-of-week FE; ±10-day donut; MSE-optimal bandwidth per outcome (so col 1 ≠ col 2 + col 3). HC-robust SEs. Total calls unchanged; hot ↑ ~4/day; cold unchanged. **p*<0.10, ***p*<0.05, ****p*<0.01.

**Table C3. RDiT: Agent Characteristics**

| | (1) Tenure | (2) Start Hour | (3) New-hire Share | (4) Veteran Share |
|---|---|---|---|---|
| After AI Expansion | −0.601<br>(0.446) | −0.086<br>(0.135) | −0.008<br>(0.026) | −0.011<br>(0.007) |
| Bandwidth | 60 | 60 | 60 | 60 |
| DOW FE | Yes | Yes | Yes | Yes |

***Notes.*** N=13,199 agent-days. RDiT on predetermined agent characteristics: (1) Tenure (mo), (2) Start Hour, (3) New-hire share (tenure ≤1 mo), (4) Veteran share (≥24 mo). ±60-day bandwidth; ±10-day donut; day-of-week FE; SEs clustered by calendar date (Lee and Card 2008). None significant. **p*<0.10, ***p*<0.05, ****p*<0.01.

## Appendix D: Detailed Results about Alternative Explanations

Section 4 of the main text presents the central mechanism evidence. Among hot leads agents return to the shared pool, the unconditional conversion rate and average payment are higher, not lower, than among leads they retain. This pattern is less consistent with rational filtering on observed quality. This appendix presents the formal evidence addressing four additional alternative explanations of the within-day behavioral pattern documented in Section 4: a slower-pace explanation, a learning explanation, a customer-side-response explanation, and a task-sequence-consolidation explanation. The first three are alternatives to the reservation-value channel; the fourth is complementary to it.

## D.1 Slower-Pace Explanation

A second potential explanation for the decline in human conversion is intertemporal. Agents respond to the improved pool by slowing their dialing pace, taking longer breaks between calls, and consequently completing fewer calls per unit of working time. Under this explanation, the within-call effort and persistence margins might appear unchanged, while the operative margin is the spacing between calls.

We test this explanation directly using the inter-call gap, defined as the time elapsed between the completion of one call and the initiation of the next call by the same agent. Table D.1.1 reports RDiT estimates of the discontinuity in the inter-call gap at the August 28, 2024 expansion, separately by call-attempt-order bin. The estimated discontinuities are small and statistically indistinguishable from zero at the earliest positions; the only detectable movements are negative coefficients at positions 7–9 and 10–12, indicating a *shorter* gap, in the opposite direction of the slower-pace prediction. At the early positions where the reallocation, effort-compression, and performance effects are largest, the inter-call gap shows no significant change.

**Table D.1.1. RDiT Estimates: Inter-Call Gap (Hot Calls, in Minutes)**

| | (1)<br>1–3 | (2)<br>4–6 | (3)<br>7–9 | (4)<br>10–12 | (5)<br>81–120 |
|---|---|---|---|---|---|
| After AI Expansion | 0.588<br>(1.167) | 0.115<br>(1.111) | −2.774*<br>(1.453) | −2.922**<br>(1.433) | 0.118<br>(0.727) |
| Bandwidth | 65 | 65 | 65 | 65 | 65 |
| Observations | 15,650 | 17,382 | 15,086 | 13,570 | 24,170 |
| Agent FE | Yes | Yes | Yes | Yes | Yes |
| Date FE | Yes | Yes | Yes | Yes | Yes |
| Hour FE | Yes | Yes | Yes | Yes | Yes |

***Notes.*** RDiT on inter-call gap (minutes between consecutive calls within agent-day), by position bin. Agent and time FE; ±10-day donut. Agent-day-clustered SEs. **p*<0.10, ***p*<0.05, ****p*<0.01.

Agents do not slow their pace after AI deployment. The behavioral response is concentrated within calls, in the form of shorter and less differentiated effort and a higher probability of returning the lead to the pool, and not between calls. This evidence weighs against the slower-pace explanation and locates the observed adjustment within calls.

## D.2 Learning Explanation

A third potential explanation is that the patterns reflect agent learning. As agents observe a richer pool, they update their beliefs about what works and adjust their effort allocation accordingly, with the adjustment becoming most apparent at the start of each workday when agents enter the system fresh. Under this explanation, the front-loaded pattern documented in Section 4 would reflect a front-loaded learning response that diminishes over the course of each workday, not an immediate behavioral reoptimization to a changed pool composition.

Two pieces of evidence weigh against the learning explanation. First, learning would predict a gradual evolution of behavior in the weeks following the rollout, as agents accumulate experience with the new pool composition. The RDiT design, by contrast, identifies a sharp discontinuous change at the rollout date, and the event-study results show that the early-late conversion gap shifts within the first thirty days after the expansion and then stabilizes. The timing is more consistent with an immediate reoptimization in response to a change in pool composition than with a gradual accumulation of experience. Second, as documented in the heterogeneity analysis, the response is comparable in magnitude across agents with different levels of accumulated experience. A learning channel would predict that less-experienced agents adjust more sharply, since they have more belief-updating to do; a reservation-value channel predicts that all reoptimizing agents respond regardless of experience. The pattern is more consistent with the latter.

**Table D.2.1. Learning Time Path: Conversion RDiT at Positions 1–3 by Post-Rollout Window**

| Post-rollout window | Estimate | SE | N |
|---|---|---|---|
| Weeks 1–2 (Aug 29 – Sep 11) | −0.009 | (0.019) | 14,515 |
| Weeks 3–4 (Sep 12 – Sep 25) | −0.031 | (0.032) | 14,556 |
| Weeks 5–8 (Sep 26 – Oct 24) | −0.061** | (0.024) | 17,064 |
| Week 9 (Oct 25 – Oct 31) | −0.029 | (0.192) | 13,589 |
| Joint test of equality across windows | | | p = 0.24 |
| F(3, 908) = 1.40 | | | |

***Notes.*** RDiT on conversion at positions 1–3 for four post-rollout windows, each estimated against the common pre-period (60 days before the cutoff). Agent and hour-of-day fixed effects; calendar-date effects are omitted because they are collinear with the running variable within a bin. Agent-day-clustered standard errors. The sample ends 64 days after the cutoff, so the final window contains seven days, not the five weeks its calendar definition would span. Within that window the running variable varies by only 2.1 days while the local linear fit is extrapolated 61 days back to the cutoff; the implied variance inflation for the intercept is roughly two orders of magnitude larger than in the first window, which accounts for its standard error. $*p<0.10$, $**p<0.05$, $***p<0.01$.

The behavioral response we document is therefore better characterized as endogenous reoptimization to a shifted pool than as deliberate learning about a new environment.

To test the learning interpretation directly, we estimate the local linear RDiT model separately on early and late post-rollout windows. If learning drove the behavioral response, the early-position effects should be small in the first weeks after the cutoff and grow as agents accumulate experience with the enriched pool. Appendix Table D.2.1 reports the estimated discontinuity at positions 1–3 across four post-period windows (weeks 1–2, 3–4, 5–8, and week 9). The point estimates are −0.9, −3.1, −6.1, and −2.9 percentage points respectively (the weeks 5–8 estimate is statistically significant at the 5% level on its own), with a joint test of equality across windows yielding p = 0.24. The window estimates are thus statistically indistinguishable from one another, and the response is already precisely estimated by weeks 5–8; together with the event-study evidence that the early-late conversion gap shifts within the first thirty days and then stabilizes (Figure 8 in the main text), the timing is more consistent with immediate reoptimization than with gradual learning.

### D.3 Customer-Side Responses

A fourth potential explanation is that the documented changes reflect customer-side responses to AI deployment, not agent-side responses. Customers who became aware that the firm uses an AI agent for some outbound calls might react differentially when contacted by a human agent. This could in principle generate changes in conversion rates without any change in agent behavior.

The effects we document could in principle reflect changes on the customer side, not the agent side. If the expansion altered the composition or behavior of the customers reaching human agents, for example by routing more difficult customers to early within-day positions, the change in conversion might stem from customer responses, not from any change in agent effort. We probe this possibility by examining two customer-side margins directly, both estimated by the same position-bin RDiT used throughout. The first margin is the short-call rate, defined as the probability that a connected call ends within sixty seconds. A discontinuous increase in early termination, concentrated at the positions where the conversion effect is largest, would be consistent with customers becoming quicker to disengage. We note, however, that this measure cannot distinguish a customer hang-up from an agent ending the call, so an increase is not by itself evidence of a customer-side change. The second margin asks whether the conversion effect differs by customer demographics. Because customer age is unavailable for the bulk of the sample, we use customer gender, interacting the full RD polynomial with a female-customer indicator and reading the coefficient on the post-cutoff interaction as the differential effect for female relative to male customers. A customer-side account in which the expansion changed who agents face, or how customers respond, would tend to generate heterogeneity in the effect across these observable customer characteristics. Both specifications absorb agent and hour-of-day fixed effects.

The results, reported in Appendix Table D.3.1, point to adjustment on the agent side, not the customer side. At the positions where the conversion effect is largest, positions 1 to 3 and 4 to 6, the short-call rate shows no significant discontinuity (0.021 and −0.006, both insignificant); a small significant increase appears at positions 7 to 9 (0.051, significant at the 5% level), and because the short-call measure does not separate customer hang-ups from agent-initiated endings, an increase there is at least as consistent with agents shortening their own effort as with any change in customer behavior. The differential effect by customer gender is insignificant in every position bin, with point estimates ranging from −0.024 to 0.028, indicating that the conversion discontinuity does not vary with observable customer characteristics. Taken together, the absence of a customer-side discontinuity at the positions that drive our results and the absence of effect heterogeneity across customers are more consistent with adjustment on the agent side than with observable customer-side changes.

**Table D.3.1. Customer-Side Margins: Short-Call Rate and Effect Heterogeneity by Customer Gender**

| Position bin | Short-call rate | Effect × female customer |
|---|---|---|
| 1–3 | 0.021 (0.024) | −0.007 (0.018) |
| 4–6 | -0.006 (0.024) | 0.028 (0.020) |
| 7–9 | 0.051** (0.026) | −0.024 (0.021) |
| 10–12 | 0.011 (0.030) | −0.004 (0.017) |
| 1–120 | -0.019 (0.022) | 0.003 (0.015) |
| Bandwidth | 60 | 60 |
| Agent FE | Yes | Yes |
| Hour-of-day FE | Yes | Yes |

***Notes.*** Position-bin RDiT for two customer-side margins on human-handled hot leads. Col 1: short-call rate (P(connected call ≤60 sec)). Col 2: post-cutoff × female-customer interaction. ±60-day bandwidth; ±10-day donut; agent and hour-of-day FE. Neither significant at bins driving the main effect. Agent-day-clustered SEs. **p*<0.10, ***p*<0.05, ****p*<0.01.

A complementary customer-side check examines whether customer composition itself shifts at the cutoff within each within-day position bin. Table D.3.2 reports position-bin RDiT estimates of an indicator for female customer within each bin. Statistically insignificant coefficients across bins indicate that the customer mix is balanced across the cutoff within each position, with no evidence of a within-position selection shift on the customer side as a driver of the position-bin conversion effects.

**Table D.3.2. Within-Bin Balance: Customer Gender by Position**

| | (1)<br>1–3 | (2)<br>4–6 | (3)<br>7–9 | (4)<br>10–12 |
|---|---|---|---|---|
| After AI Expansion | 0.019 (0.021) | 0.032 (0.024) | 0.032 (0.025) | -0.010 (0.030) |
| Bandwidth | 60 | 60 | 60 | 60 |
| Agent FE | Yes | Yes | Yes | Yes |
| Hour-of-day FE | Yes | Yes | Yes | Yes |
| Observations | 17,922 | 14985 | 12,838 | 11,565 |

***Notes.*** Dependent variable: female-customer indicator, on human-handled hot leads. ±60-day bandwidth; ±10-day donut; agent and hour-of-day FE (calendar-date collinear within a bin). Insignificant = customer gender balanced across cutoff within bin. Agent-day-clustered SEs. **p*<0.10, ***p*<0.05, ****p*<0.01.

**Table D.3.3. Prior Call Attempts as a Measure of Lead Quality: Predictive Power and Discontinuity at the Cutoff**

| *Panel A: Prior attempts as a predictor of conversion* | | | |
|---|---|---|---|
| | Coefficient | SE | t |
| Prior attempts (linear) | 0.032 | (0.001) | 21.56 |
| *Relative to first attempt:* | | | |
| 1 prior attempt | 0.015 | (0.003) | 4.95 |
| 2 prior attempts | 0.033 | (0.005) | 6.82 |
| 3 prior attempts | 0.055 | (0.006) | 8.75 |
| 4 prior attempts | 0.086 | (0.008) | 10.81 |
| 5+ prior attempts | 0.234 | (0.015) | 15.10 |

| *Panel B: Discontinuity in prior attempts, by position bin* | | | | | |
|---|---|---|---|---|---|
| Position bin | Discontinuity | SE | Implied indirect effect | Reduced-form effect | Share (%) |
| 1–3 | −0.084 | (0.074) | −0.0027 | −0.043 | 6.2 |
| 4–6 | −0.043 | (0.067) | −0.0014 | −0.027 | 5.1 |
| 7–9 | 0.004 | (0.075) | 0.0001 | 0.001 | — |
| 10–12 | 0.081 | (0.076) | 0.0026 | 0.001 | — |
| 81–120 | 0.138 | (0.069) | 0.0044 | 0.034 | 13.0 |

***Notes.*** Sample: human-handled calls on hot leads. Prior attempts is the number of calls previously placed on the same lead. Panel A regresses conversion on prior attempts with agent, hour-of-day, and calendar-date fixed effects; standard errors are clustered by agent, and the omitted category in the lower block is the first attempt. Panel B reports the RDiT discontinuity in prior attempts by within-day position bin (bandwidth 63, ±10-day donut), with the mediator residualized on the same fixed effects as the outcome so that the implied shares are internally consistent; standard errors are clustered by agent-day. The implied indirect effect is the product of the discontinuity and the linear coefficient in Panel A, and the share is that quantity relative to the reduced-form estimate in the same bin. Shares are omitted where the reduced-form effect is itself indistinguishable from zero. Prior attempts have a standard deviation of 2.18 in the estimation sample. The sign of the relationship in Panel A does not affect the conclusion for early positions, where the discontinuity is itself indistinguishable from zero.

## D.4 Within-Day Task-Stream Consolidation

A complementary lens on the behavioral mechanism examines how the structure of the within-day task stream itself changes after AI deployment. Two indices summarize this structure at the agent-day level. The alternation rate is the fraction of consecutive call pairs that switch lead type (hot to cold or cold to hot) within an agent's daily shift. A high alternation rate indicates a sequence that mixes lead types frequently. The good-run share is the length of the longest contiguous block of hot leads within the agent-day, divided by the total number of calls that day. A high good-run share indicates that hot leads cluster into long contiguous stretches and are not dispersed across the workday. Table D.4.1 reports RDiT estimates of changes in these two indices at the AI expansion cutoff. The alternation rate falls by 3.5 to 5.3 percentage points, depending on specification, and the good-run share rises by 9.8 to 11.7 percentage points. Both changes are statistically significant.

**Table D.4.1. RDiT Estimates: Within-Day Task-Sequence Structure**

| | Alternation Rate | | Good-Run Share | |
|---|---|---|---|---|
| | (1) | (2) | (3) | (4) |
| After AI expansion | $-0.035^{***}$ (0.012) | $-0.053^{***}$ (0.009) | $0.098^{**}$ (0.047) | $0.117^{***}$ (0.040) |
| Bandwidth | 31 | 44 | 42 | 31 |
| Agent FE | No | Yes | No | Yes |
| Observations | 5,040 | 8,340 | 8,027 | 5,277 |

***Notes.*** RDiT on within-day task-sequence structure at agent-day level. Alternation rate: share of consecutive call pairs switching lead type. Good-run share: longest hot-lead run / total daily calls. MSE-optimal bandwidth. Agent-clustered SEs. **p*<0.10, ***p*<0.05, ****p*<0.01.

The within-day task stream becomes less alternating and more clustered. Before the expansion, agents face a sequence in which hot and cold leads frequently interleave, requiring repeated cognitive transitions between two distinct types of customer interactions. After the expansion, the sequence consolidates into longer contiguous blocks of hot leads. This consolidation reinforces the cross-type compression mechanism documented in Section 4. When the agent's recent and prospective sequence is dominated by a single lead type, the cross-type effort differentiation that would otherwise be available has fewer opportunities to manifest, and the agent's effort allocation converges further toward a uniform standard. The change in task-sequence structure thus complements the reservation-value channel by documenting that the post-AI environment offers fewer occasions for differentiated effort even before any behavioral response.

### D.5 Overall Human Conversion as a Secondary Outcome

The position-bin RDiT in the main text uses conditional hot-lead conversion as the outcome, residualized on calendar-date fixed effects. A complementary test of the mechanical pool-quality channel uses overall human conversion as the outcome, unconditional on lead type. Under this outcome, the mechanical channel operates directly through the rising hot-lead share in human queues documented in Section 4.1, and the position-bin discontinuities capture the unconditional pool-quality improvement that the hot-only outcome conditions out. Table D.5.1 reports the estimates.

**Table D.5.1. Overall Human Conversion by Within-Day Position (Unconditional on Lead Type)**

| | (1) Day-level | (2) 1–3 | (3) 4–6 | (4) 7–9 | (5) 10–12 | (6) 81–120 |
|---|---|---|---|---|---|---|
| After AI Expansion | 0.0293 | −0.0386*** | −0.0202** | 0.0023 | −0.0002 | 0.0250*** |
| | (0.0193) | (0.0132) | (0.0081) | (0.0075) | (0.0072) | (0.0075) |
| Bandwidth | 31 | 65 | 65 | 65 | 65 | 65 |
| Observations | 44 | 24,005 | 22,214 | 21,268 | 20,623 | 54,369 |

***Notes.*** RDiT on overall conversion (not conditioned on hot/cold), capturing composition + effort channels jointly. Col (1): day-level with day-of-week FE. Cols (2)–(6): call-level by position bin, residualized on agent, hour-of-day, and date FE. ±10-day donut. Cluster or HC-robust SEs. **p*<0.10, ***p*<0.05, ****p*<0.01.

### D.6 Within-Hot-Lead Composition Shifts at Late Positions

The mechanical-channel interpretation of the late-position estimates holds that the residual hot leads reaching late within-day positions are different, fresher, less worn-down, and circulating faster, after the expansion. We document the three dimensions on which this composition shifts (or fails to shift): waiting time to first contact (a proxy for lead freshness), prior-attempt count, and effective queue position. Tables D.6.1 through D.6.3 report pre- and post-expansion means by within-day position bin for each margin.

**Table D.6.1. Within-Hot-Lead Composition: Queue Waiting Time (Freshness)**

| Position bin | Pre mean | Pre SD | Post mean | Post SD | Difference |
|---|---|---|---|---|---|
| 1–3 | 63.773 | 212.817 | 48.731 | 173.350 | −15.042*** |
| 4–6 | 42.839 | 133.227 | 33.107 | 106.568 | −9.733*** |
| 7–9 | 44.487 | 132.471 | 34.116 | 112.525 | −10.371*** |
| 10–12 | 49.391 | 154.698 | 34.768 | 119.329 | −14.623*** |

| Position bin | Pre mean | Pre SD | Post mean | Post SD | Difference |
|---|---|---|---|---|---|
| 81–120 | 98.241 | 368.488 | 23.535 | 79.131 | −74.706*** |

***Notes.*** Pre and post means/SDs of queue waiting time (elapsed time from lead creation to first call), human-handled hot leads, ±60-day window, by position bin. Two-sample t-tests. Late tail (81–120) shows the largest reduction. *$p$<0.10, **$p$<0.05, ***$p$<0.01.

**Table D.6.2. Within-Hot-Lead Composition: Prior Attempt Count**

| Position bin | Pre mean | Pre SD | Post mean | Post SD | Difference |
|---|---|---|---|---|---|
| 1–3 | 2.290 | 1.746 | 2.189 | 1.513 | −0.101*** |
| 4–6 | 2.199 | 1.501 | 2.166 | 1.385 | −0.032 |
| 7–9 | 2.231 | 1.534 | 2.181 | 1.492 | −0.050** |
| 10–12 | 2.227 | 1.479 | 2.209 | 1.501 | −0.017 |
| 81–120 | 1.621 | 1.036 | 1.647 | 0.981 | 0.026** |

***Notes.*** Pre and post means/SDs of prior attempt count (times the lead was dialed previously), human-handled hot leads, ±60-day window, by position bin. Two-sample t-tests. Uniformly small differences. *$p$<0.10, **$p$<0.05, ***$p$<0.01.

**Table D.6.3. Within-Hot-Lead Composition: Effective Queue Position**

| Position bin | Pre mean | Pre SD | Post mean | Post SD | Difference |
|---|---|---|---|---|---|
| 1–3 | 1.531 | 0.671 | 1.561 | 0.689 | 0.030*** |
| 4–6 | 3.076 | 1.150 | 3.225 | 1.232 | 0.149*** |
| 7–9 | 4.591 | 1.561 | 4.788 | 1.734 | 0.197*** |
| 10–12 | 5.971 | 1.998 | 6.135 | 2.261 | 0.163*** |
| 81–120 | 35.245 | 15.643 | 33.644 | 12.578 | −1.601*** |

***Notes.*** Pre and post means/SDs of effective queue position (rank of a hot lead among all hot leads dialed by the same agent that day), human-handled hot leads, ±60-day window, by position bin. Two-sample t-tests. *$p$<0.10, **$p$<0.05, ***$p$<0.01.

## Appendix E: Detailed Results for Robustness Checks

Table E1 confirms the main findings under a fixed ±60-day bandwidth in place of the MSE-optimal bandwidth used in Section 4. The early-position discontinuities in conversion, reallocation, and log call duration retain the signs and approximate magnitudes of the baseline specification at conventional significance levels. The choice of bandwidth rule therefore does not alter the qualitative or quantitative conclusions of the main analysis.

**Table E1. Robustness: Alternative Bandwidth (±60 Days, Hot Calls)**

| | (1)<br>1–3 | (2)<br>4–6 | (3)<br>7–9 | (4)<br>10–12 | (5)<br>81–120 |
|---|---|---|---|---|---|
| ***Panel A: Conversion Rate*** | | | | | |
| After AI Expansion | −0.043*** | −0.029*** | 0.002 | 0.000 | 0.032*** |
| | (0.015) | (0.010) | (0.010) | (0.010) | (0.011) |
| Observations | 18,687 | 15,578 | 13,413 | 12,063 | 20,907 |
| ***Panel B: Reallocation Rate*** | | | | | |
| After AI Expansion | 0.057 | 0.069* | 0.028 | 0.045 | 0.097 |
| | (0.038) | (0.039) | (0.041) | (0.042) | (0.087) |
| Observations | 18,684 | 15,569 | 13,410 | 12,055 | 20,899 |
| ***Panel C: Call Duration*** | | | | | |
| After AI Expansion | −0.154*** | −0.037 | −0.123** | −0.040 | 0.028 |
| | (0.058) | (0.057) | (0.056) | (0.069) | (0.057) |
| Observations | 18,680 | 15,570 | 13,405 | 12,060 | 20,893 |
| Bandwidth | 60 | 60 | 60 | 60 | 60 |
| Agent FE | Yes | Yes | Yes | Yes | Yes |
| Date FE (Panels A, C) | Yes | Yes | Yes | Yes | Yes |
| Hour-of-day FE | Yes | Yes | Yes | Yes | Yes |

***Notes.*** RDiT for hot calls, fixed ±60-day bandwidth (in place of MSE-optimal). Outcome: conversion (Panel A), reallocation (Panel B), log duration (Panel C). Panels A/C: agent + calendar-date + hour-of-day FE; Panel B: agent + hour-of-day only. Agent-day-clustered SEs. **p*<0.10, ***p*<0.05, ****p*<0.01.

**Table E2. Robustness: Alternative Call-Order Bins (5-Call Bins, Hot Calls)**

| | (1)<br>1–5 | (2)<br>6–10 | (3)<br>11–15 | (4)<br>16–20 | (5)<br>81–120 |
|---|---|---|---|---|---|
| ***Panel A: Conversion Rate*** | | | | | |
| After AI Expansion | −0.038*** | −0.006 | 0.008 | −0.008 | 0.034*** |
| | (0.010) | (0.008) | (0.008) | (0.007) | (0.010) |
| Bandwidth | 65 | 65 | 65 | 65 | 65 |
| Observations | 33,048 | 25,291 | 22,555 | 22,511 | 24,210 |
| ***Panel B: Reallocation Rate*** | | | | | |
| After AI Expansion | 0.067 | 0.076* | 0.021 | 0.003 | −0.214 |
| | (0.042) | (0.042) | (0.042) | (0.039) | (0.144) |
| Bandwidth | 54 | 51 | 55 | 54 | 35 |
| Observations | 25,480 | 18,175 | 17,606 | 16,972 | 11,443 |
| ***Panel C: Call Duration*** | | | | | |
| After AI Expansion | −0.070 | −0.135*** | −0.005 | −0.031 | 0.003 |
| | (0.043) | (0.042) | (0.046) | (0.047) | (0.050) |
| Bandwidth | 65 | 65 | 65 | 65 | 65 |
| Observations | 33,030 | 25,281 | 22,546 | 22,500 | 24,192 |
| Agent FE | Yes | Yes | Yes | Yes | Yes |
| Date FE (Panels A, C) | Yes | Yes | Yes | Yes | Yes |
| Hour-of-day FE | Yes | Yes | Yes | Yes | Yes |

***Notes.*** RDiT for hot calls with five-call position bins (in place of three-call). Outcome and FE structure as in Table 4. ±10-day donut. Agent-day-clustered SEs. **p*<0.10, ***p*<0.05, ****p*<0.01.

**Table E3. Robustness: AI Trial-Launch Cutoff (Hot Calls)**

| | (1)<br>1–3 | (2)<br>4–6 | (3)<br>7–9 | (4)<br>10–12 | (5)<br>81–120 |
|---|---|---|---|---|---|
| ***Panel A: Conversion Rate*** | | | | | |
| After AI Launch | −0.020** | −0.009 | −0.004 | 0.019*** | 0.011 |
| | (0.008) | (0.007) | (0.007) | (0.007) | (0.007) |
| Bandwidth | 74 | 74 | 74 | 74 | 74 |
| Observations | 32,076 | 28,649 | 25,364 | 22,912 | 31,580 |
| ***Panel B: Reallocation Rate*** | | | | | |
| After AI Launch | 0.058 | −0.200*** | −0.203*** | 0.010 | 0.610* |
| | (0.038) | (0.065) | (0.073) | (0.045) | (0.298) |
| Bandwidth | 40 | 23 | 21 | 30 | 12 |
| Observations | 15,546 | 7,043 | 5,245 | 8,274 | 1,594 |
| ***Panel C: Call Duration*** | | | | | |
| After AI Launch | −0.015 | 0.033 | −0.126*** | 0.043 | 0.010 |
| | (0.038) | (0.039) | (0.039) | (0.041) | (0.040) |
| Bandwidth | 74 | 74 | 74 | 74 | 74 |
| Observations | 32,058 | 28,637 | 25,353 | 22,904 | 31,559 |
| Agent FE | Yes | Yes | Yes | Yes | Yes |
| Date FE (Panels A, C) | Yes | Yes | Yes | Yes | Yes |
| Hour-of-day FE | Yes | Yes | Yes | Yes | Yes |

***Notes.*** RDiT using the June 5, 2024 AI trial-launch date as an alternative cutoff. Outcome: conversion (Panel A), reallocation (Panel B), log duration (Panel C). Panels A/C: agent + calendar-date + hour-of-day FE; Panel B: agent + hour-of-day. ±10-day donut. Agent-day-clustered SEs. $*p<0.10$, $**p<0.05$, $***p<0.01$.

A potential concern is that the decline in call duration may be mechanically driven by non-connected calls, whose recorded duration is zero. Restricting the duration analysis to connected calls leaves the estimates unchanged, because zero-duration no-connect calls are already excluded by the log transformation: ln(0) is undefined, and the set of calls with a missing log duration coincides exactly with the set of non-connected calls (Table E4). The duration result therefore cannot be an artifact of unsuccessful connection attempts, and the estimates in Table E4 are identical to column (3) of Table 4 by construction, so the table serves as an accounting check on the sample, not an independent robustness test.

**Table E4. Robustness: Call Duration RDiT, Connected Calls Only**

| | (1)<br>1–3 | (2)<br>4–6 | (3)<br>7–9 | (4)<br>10–12 | (5)<br>81–120 |
|---|---|---|---|---|---|
| After AI Expansion | −0.145***<br>(0.054) | −0.009<br>(0.053) | −0.128**<br>(0.052) | −0.035<br>(0.063) | 0.003<br>(0.050) |
| Bandwidth | 65 | 65 | 65 | 65 | 65 |
| Agent FE | Yes | Yes | Yes | Yes | Yes |
| Date FE | Yes | Yes | Yes | Yes | Yes |
| Hour-of-day FE | Yes | Yes | Yes | Yes | Yes |
| Observations | 20,968 | 17,556 | 15,170 | 13,634 | 24,192 |

***Notes.*** Accounting check: 1,880 of 2,148,929 hot-lead calls are non-connected (recorded duration 0). We verified this share directly in the raw call-level records; a full tabulation of all 2,148,929 hot-lead call attempts returns the same 1,880 non-connected calls (0.09%). Because ln(0) is undefined, these are already excluded from the baseline log-duration specification. This table reproduces Table 4 col (3) on the connected-call sample by construction; coefficients, SEs, and Ns identical. $*p<0.10$, $**p<0.05$, $***p<0.01$.

The within-day position pattern could reflect clock time, not a worker's position in the daily task sequence. For example, customer availability or agent fatigue may change over the morning. Table E5 therefore re-estimates the conversion discontinuity separately by hour of day. None of the hour-specific estimates is statistically distinguishable from zero. Although these null estimates do not rule out all time-of-day effects, they suggest that sequence position is more informative than clock time alone as a dimension of heterogeneity in our setting.

**Table E5. Robustness: Conversion RDiT by Hour of Day**

| | (1)<br>9AM | (2)<br>10AM | (3)<br>11AM | (4)<br>12PM |
|---|---|---|---|---|
| After AI Expansion | −0.006<br>(0.006) | 0.003<br>(0.006) | 0.000<br>(0.006) | −0.007<br>(0.011) |
| Bandwidth | 65 | 65 | 65 | 65 |
| Agent FE | Yes | Yes | Yes | Yes |
| Date FE | Yes | Yes | Yes | Yes |
| Observations | 38,250 | 37,252 | 37,718 | 17,604 |

***Notes.*** RDiT on hot-lead conversion, separately by hour of day. Outcome residualized on agent and date FE; local-linear discontinuity within each hour group. ±10-day donut. None significant. Agent-day-clustered SEs. **p*<0.10, ***p*<0.05, ****p*<0.01.

Clock-hour groups do not detect the response because, as Figure E1 illustrates, agent start times are highly dispersed. Although the median agent begins shortly after 9:00 AM, a substantial tail of agent-days starts later in the morning and into the afternoon. A given clock hour therefore corresponds to very different within-day call positions across agents, the first few calls for an early starter and a much later position for an agent who began later in the day. Clock time alone cannot capture the within-day position-bin effects documented in the main analysis, which is why within-day call position is the more informative organizing dimension in this setting.

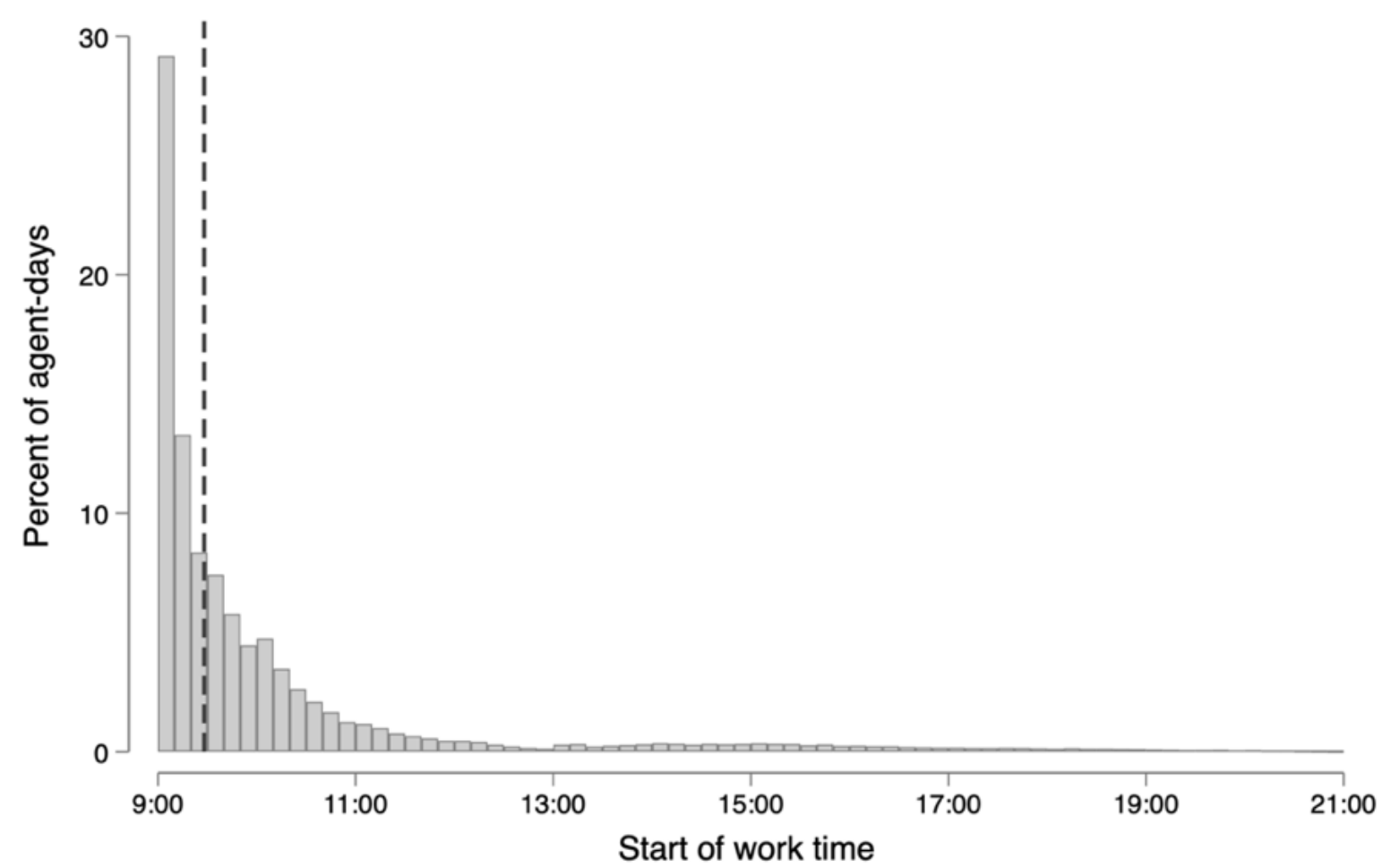

**Figure E1. Distribution of Agent Start Time (10-min bins)**

***Notes.*** Distribution of agent start times (clock time of first call), 10-min bins, ±60-day window. Median ≈ 9:30 AM (dashed line). ≈29% of agent-days start in the 9:00–9:10 bin, with a long tail through the afternoon. Motivates within-day position (not clock time) as the analysis dimension.

Another concern is that the position-specific conversion pattern may reflect changes in lead vintage after the AI expansion. We address this possibility by controlling for the elapsed time from lead creation to initial contact. Table E6 shows that the estimated conversion profile is unchanged at the reported three-decimal precision. Position-specific conversion estimates are significantly negative at positions 1–3 and 4–6, are unchanged at intermediate positions, and increase at positions 81–120. Thus, the observed measure of lead vintage does not account for the main early-to-late conversion pattern, although unobserved dimensions of within-hot-lead composition may remain. The exact match with Table 3, Panel B follows from arithmetic. The residuals from the specification with freshness differ from those without it by the freshness coefficient times residualized freshness, so each position-bin estimate moves by that coefficient times the jump in freshness at the cutoff. Both terms are tiny. Conditional on agent, hour-of-day, and calendar-date fixed effects, freshness has essentially no incremental predictive power for conversion: an interquartile increase, from 8 to 51 seconds, implies a change of 0.0065 percentage points. Freshness itself rises by roughly 14 seconds at the cutoff (SE 6.1), and the implied movement in the coefficients is about 0.00002, far below the three-decimal precision of the table.

**Table E6. Robustness: Conversion RDiT Controlling for Lead Freshness**

| | (1)<br>1–3 | (2)<br>4–6 | (3)<br>7–9 | (4)<br>10–12 | (5)<br>81–120 |
|---|---|---|---|---|---|
| After AI Expansion | −0.045*** | −0.025*** | 0.001 | 0.007 | 0.034*** |
| | (0.013) | (0.009) | (0.010) | (0.010) | (0.010) |
| Bandwidth | 65 | 65 | 65 | 65 | 65 |
| Freshness control | Yes | Yes | Yes | Yes | Yes |
| Agent FE | Yes | Yes | Yes | Yes | Yes |
| Date FE | Yes | Yes | Yes | Yes | Yes |
| Hour-of-day FE | Yes | Yes | Yes | Yes | Yes |
| Observations | 20,978 | 17,565 | 15,178 | 13,638 | 24,210 |

***Notes.*** Conversion RDiT by position bin, adding lead freshness (elapsed time from creation to first call) as a control. Human-handled hot leads; agent, hour-of-day, and date FE + freshness. ±10-day donut. Agent-day-clustered SEs. $*p<0.10$, $**p<0.05$, $***p<0.01$.

We next assess how much of the early-position decline in hot-lead conversion is statistically associated with the observed changes in reallocation and within-call effort. Our approach is a sequential-regression decomposition in the Baron and Kenny (1986) tradition, specifically the linear conditional decomposition of Gelbach (2016). It is not a nonparametric Imai et al. (2010) procedure, nor an instrumental-variables strategy. We do not instrument the mediators. The decomposition is exact and additive by construction, attributing the gap between the total and the direct effect to the included mediators in proportion to their first-stage movements and their partial associations with conversion. We pool the first three position bins, the range over which Section 4.2 documents the largest reduced-form conversion coefficients, and estimate the decomposition on a common sample of 47,655 calls.

Table E7 reports the results. Panel A presents the decomposition regressions. The total, reduced-form effect on conversion is −0.026 (column 1). The two first stages move in the directions predicted by the reservation-value mechanism. The AI expansion raises the reallocation rate by 5.1 percentage points (column 2), although this pooled first stage is imprecisely estimated, and compresses log call duration by 10.7 percent (column 3). In the long regression that includes both mediators (column 4), each remains a strong call-level predictor of conversion, with partial associations of 0.063 for reallocation and 0.062 for log duration, and the null that both mediator coefficients are zero is decisively rejected ($F = 700.4$, $p < 0.001$). The coefficient on the post-cutoff indicator in column (4) is the direct effect, −0.022, which retains most of the total effect of −0.026.

Panel B reports the indirect effects, each computed as the product of a first-stage coefficient and the corresponding mediator coefficient in the long regression, with bias-corrected 95 percent confidence intervals from an agent-day cluster bootstrap (2,000 replications). The decomposition clarifies how the two observed channels relate to the reduced-form pattern. The two observable channels enter with opposite signs and partially offset. The duration channel contributes −0.0066 (95% CI [−0.0109, −0.0023]); the reallocation channel contributes +0.0032, with a confidence interval that includes zero ([−0.0014, 0.0074]). The total indirect effect is −0.0034, with a 95% confidence interval that spans zero ([−0.0098, 0.0028]). Because the total indirect component is small and imprecisely estimated, the two observed variables jointly account for only a modest share of the reduced-form coefficient. The coefficient remaining after conditioning on them is −0.022, against a total coefficient of −0.026. Results are qualitatively unchanged with a ±5-day donut (total effect −0.017; total indirect −0.0006, with a confidence interval spanning zero). The two opposite-signed channels have a clear economic reading. The reallocation component enters with a positive sign because reallocated leads exhibit higher realized downstream conversion in the data. Because reallocated leads may receive additional attempts and longer observation windows, this positive association should not be interpreted as the causal effect of reallocation on conversion. As Section 4.3.3 documents, previously returned leads convert at roughly 1.9 times the rate of retained leads; they appear to be high-intent customers whom an earlier agent failed to close, not leads that deserved to be abandoned. When the AI expansion raises the share of such leads circulating to early positions, it enriches the early-position pool and partially offsets the within-day profile reorganization, which is why the reallocation channel works against the total. The duration channel enters with a negative sign, in line with shorter calls translating mechanically into fewer completed conversions. The partial offset between the two channels is why the total indirect component remains small.

**Table E7. Statistical Decomposition of the Reduced-Form Effect on Hot-Lead Conversion**

| | (1) Total Conversion | (2) FS: M1 Realloc. | (3) FS: M2 Log dur. | (4) Direct Conversion |
|---|---|---|---|---|
| ***Panel A: Decomposition Regressions*** | | | | |
| After AI Expansion | −0.026*** | 0.051 | −0.107*** | −0.022*** |
| | (0.008) | (0.036) | (0.035) | (0.008) |
| Reallocation (M1) | | | | 0.063*** |
| | | | | (0.007) |
| Log duration (M2) | | | | 0.062*** |
| | | | | (0.002) |
| ***Panel B: Indirect Effects (cluster bootstrap, 2,000 reps)*** | | | | |
| | Estimate | | BC 95% CI | |
| via Reallocation (M1) | 0.0032 | | [−0.0014, 0.0074] | |
| via Log duration (M2) | −0.0066** | | [−0.0109, −0.0023] | |
| Total indirect | −0.0034 | | [−0.0098, 0.0028] | |
| Joint F (M1, M2) | 700.40 | | | |
| Joint F p-value | <0.001 | | | |
| Agent FE | Yes | Yes | Yes | Yes |
| Date FE | Yes | Yes | Yes | Yes |
| Hour-of-day FE | Yes | Yes | Yes | Yes |
| Observations | 47,655 | 47,655 | 47,655 | 47,655 |

***Notes.*** Gelbach (2016) linear decomposition of the discontinuous change in hot-lead conversion at the cutoff. Sample: human-handled calls on hot leads at positions 1–9, ±60-day window, ±10-day donut. Outcome and both mediators are residualized on agent, calendar-date, and hour-of-day fixed effects on the full hot-lead sample, and the reported regressions are estimated on the residuals. Panel A reports the reduced form, the two mediator first stages, and the long regression. Panel B reports indirect effects with bias-corrected 95% confidence intervals from 2,000 agent-day cluster bootstrap replications; the bootstrap resamples the second stage only, so the intervals do not reflect first-stage estimation error. Standard errors in parentheses are clustered by agent-day. In Panel B, ** denotes a confidence interval excluding zero. **p*<0.10, ***p*<0.05, ****p*<0.01.

We interpret the small total indirect component as evidence that the measured changes in reallocation and call duration do not jointly account for the bulk of the early-position conversion response. The remaining coefficient may reflect other dimensions of agent behavior not captured by duration, such as persuasion intensity, patience in handling objections, or willingness to sustain a difficult interaction, but it may also incorporate within-hot-lead compositional changes, measurement error, or other omitted factors. The decomposition is therefore consistent with a broader reservation-value response, but it does not independently identify that mechanism.

## Appendix F: Technical Appendix: Accounting Model and Simulation Details

### F.1 Exogenous assumptions and calibrated objects

Throughout, the functional form is the assumption and the value is calibrated. Assumed functional forms and structure: The linear hot-lead profile, which defines the level and rhythm levers; the four-part expression for ,

an accounting identity, not a behavioral assumption; the linear interpolation of the conduct profiles in ; the structure of AI conversions, namely a cold-lead rate constant across order, a constant AI hot-lead conversion rate, and an AI cold-lead rate proportional to the human one; and the benchmark . None is drawn from a particular reference, and the qualitative results do not depend on their precise form.$\bar{S}^G \phi L T = 50$

Calibrated from the data: The by-order conduct profiles , , and , hence , by regime (Table 2); the order weights ; the rhythm values and , from the observed pre- and post-AI slopes; the level and its observed range; and the cold-lead rate together with the AI conversion and volume parameters , , and (Table 3). The decomposition channels, the yield surface, and the optimal-rhythm ridge are outputs of the model, not inputs.$P_s^R(j) S_s^{G,R}(j) S_s^{G,NR}(j) \bar{S}_s^G(j) w_j \phi_{\text{pre}} = 0.10 \phi_{\text{post}} = 0.22 \alpha L = 0.0150 s_{AI}^G \text{Ratio}_{AI}^B \rho_s^B$
The model is an illustrative calibration, not a structural estimation, and six of its inputs are held fixed. (i) Human conversion on cold leads is a constant L = 0.015, whereas the realized rate moves from 1.326 to 1.491 percent. (ii) AI conversion on hot leads is a constant 0.024 and AI conversion on cold leads is 0.0105, the ratio parameter times L, whereas the realized AI rates are 3.957 and 2.988 percent on hot leads and 1.303 and 0.579 percent on cold leads. (iii) The benchmark T = 50 means the decomposition iterates call positions 1 to 50 and so uses the first seven of the nine order bins. (iv) The decomposition therefore runs on a narrower domain than the counterfactual figure, which uses all nine bins. (v) The rhythm is calibrated at 0.10 before and 0.22 after, not estimated. On the decomposition's domain of positions 1 to 50 the profile slope in the data moves from 0.086 to 0.109, in the same direction as the calibration. Fitted across all nine order bins, which is the counterfactual figure's domain, the slope instead moves from 0.251 to 0.167. The difference is a level shift confined to the two bins the decomposition excludes; the within-day rhythm does not reverse. The hot-lead share at positions 51 to 80 rises from 0.439 to 0.542 and above position 80 from 0.379 to 0.558. A single linear slope through all nine bins reads that increase in the late tail as global flattening, whereas the mechanism the paper documents operates at early positions, where the profile steepens. We hold the rhythm fixed at the calibrated values, and Appendix F.5 reports results as a comparative static within the observed range so that no conclusion depends on the fitted value. (vi) The conduct profiles are observed only at the two regimes, so we let each move linearly in the rhythm between them. Values of ϕ outside the interval from 0.10 to 0.22 therefore rest on linear extrapolation, and the unconstrained conversion-maximizing rhythm lies below both observed regimes. At that rhythm the interpolation weight is −0.19, which pushes every profile past its no-AI endpoint, and all twenty-seven estimated profile values fall outside their observed ranges. Appendix F.5 accordingly reports the comparative static within the observed range and the gain from returning to the pre-expansion rhythm as the result, with the interior optimum reported only as the unconstrained value. The bootstrap band reported there reflects sampling error in the estimated conduct profiles. It does not cover the frozen rhythm calibration in (v) or the linear-interpolation assumption itself.

### F.2 Estimated order profiles

Table F1 reports the reallocation profile $P_s^R$ and the subgroup conversion profiles $S_s^{G,R}$ (reallocated) and $S_s^{G,NR}$ (retained), estimated as piecewise functions of order under each regime; the average conversion $\bar{S}_s^G$ weights the reallocated and retained subgroup rates by the reallocation probability at that order, and $P^G$ is the average hot-lead share $\alpha$ shifted by the rhythm $\phi$ times a position score that runs from positive at the earliest orders to negative at the latest.

**Table F1. Estimated Order Profiles, Before vs After AI**

| Order Bin | Reallocation $P^R$ | | Realloc. conversion $S^{G,R}$ | | Retained conversion $S^{G,NR}$ | |
|---|---|---|---|---|---|---|
| | No AI | With AI | No AI | With AI | No AI | With AI |
| 1–3 | 0.243 | 0.389 | 0.300 | 0.138 | 0.031 | 0.026 |
| 4–6 | 0.188 | 0.325 | 0.173 | 0.085 | 0.030 | 0.032 |
| 7–9 | 0.173 | 0.289 | 0.134 | 0.073 | 0.025 | 0.030 |
| 10–12 | 0.168 | 0.269 | 0.090 | 0.057 | 0.027 | 0.027 |
| 13–20 | 0.159 | 0.221 | 0.089 | 0.057 | 0.030 | 0.026 |
| 21–30 | 0.134 | 0.195 | 0.083 | 0.050 | 0.034 | 0.027 |

| Order Bin | Reallocation $P^R$ | | Realloc. conversion $S^{G,R}$ | | Retained conversion $S^{G,NR}$ | |
|---|---|---|---|---|---|---|
| | No AI | With AI | No AI | With AI | No AI | With AI |
| 31–50 | 0.126 | 0.205 | 0.074 | 0.045 | 0.035 | 0.030 |
| 51–80 | 0.153 | 0.247 | 0.069 | 0.039 | 0.033 | 0.029 |
| > 80 | 0.199 | 0.374 | 0.028 | 0.022 | 0.033 | 0.042 |

***Notes.*** Human-handled hot calls by call-attempt order. Inputs to the firm-yield calculation and Figure F1.

### F.3. Aggregation, decomposition, and parametrization

Aggregating the per-order yield over order gives total human conversions  and the average hot-lead yield  in Panel B of Table 6. Let  denote  under a state  that specifies which of  and  are at their AI-regime values. With (both no-AI),  ( at AI), and  (both at AI), the two channels are$Y_s^H = \sum_j$ $w_j\ \ y_s(j) \bar{Y}_s^{H,G} Y(\mathcal{S}) \bar{Y}^{H,G} \mathcal{S} P^G \bar{S}^G \mathcal{S}_1 \mathcal{S}_2 P^G \mathcal{S}_3$

$$\Delta Y_{\mathrm{Ord}} \equiv Y(\mathcal{S}_2) - Y(\mathcal{S}_1), \qquad \Delta Y_{\mathrm{Cond}} \equiv Y(\mathcal{S}_3) - Y(\mathcal{S}_2),$$

which sum to $\Delta Y_{\mathrm{Total}} = Y(\mathcal{S}_3) - Y(\mathcal{S}_1)$.[1]

Each of the three conduct profiles responds to the rhythm by the same linear interpolation between the two regimes. Writing $X$ for any one of $P^R$, $S^{G,R}$, and $S^{G,NR}$,

$$X(j;\phi) = X^{\mathrm{NoAI}}(j) + \frac{\phi - \phi_{\mathrm{pre}}}{\phi_{\mathrm{post}} - \phi_{\mathrm{pre}}} \left(X^{\mathrm{AI}}(j) - X^{\mathrm{NoAI}}(j)\right),$$

which moves each profile from its no-AI shape at $\phi = \phi_{\mathrm{pre}}$ to its with-AI shape at $\phi = \phi_{\mathrm{post}}$. These rhythm-dependent profiles then deliver the conversion profile $\bar{S}^G(j;\phi)$ through the four-part identity. Substituting the hot-lead profile and the rhythm-dependent conduct profiles into the yield expression gives the surface in Panel (a) of Figure F1. The dotted line traces the unconstrained maximizer $\phi^*(\alpha)$, which lies below the observed support at every hot-lead share and is shown for reference only. Panel (b) evaluates the same surface at the observed post-expansion pool and plots firm conversions per call against $\phi$, with a 95 percent bootstrap band.[2]

### F.4 Calibration

Table F2 reports the calibrated scalars.

**Table F2. Model Calibration and Parameter Settings**

| Notation | Description | Value | Source |
|---|---|---|---|
| $L$ | Conversion rate on cold leads | .0150 | Firm data; observed human cold-lead conversion rate |
| $s_{AI}^G$ | AI conversion rate on hot leads | .0240 | Firm data; observed AI hot-lead conversion rate |
| $\mathrm{Ratio}_{AI}^B$ | AI vs. human conversion ratio, cold leads | .7000 | Firm data; ratio of AI to human cold-lead rates |
| $\rho_{\mathrm{NoAI}}^B$ | AI-to-human volume ratio, cold leads (no-AI) | .1913 | Firm data; observed AI/human cold-lead volume, pre-AI |
| $\rho_{\mathrm{AI}}^B$ | AI-to-human volume ratio, cold leads (with-AI) | 1.8354 | Firm data; observed AI/human cold-lead volume, post-AI |
| $\phi_{\mathrm{pre}}$, $\phi_{\mathrm{post}}$ | Rhythm, before and after AI | .10, .22 | Calibrated to the slope of the hot-lead profile by order, before/after AI |
| $T$ | Calls per worker-day (benchmark) | 50 | Exogenous normalization; ≈ average calls per worker-day |

[1] The conduct channel splits further by swapping $P^R$ before the subgroup conversion profiles, separating a reallocation sub-channel from a residual-effort one; we report them jointly because the reallocation premium is largely selection, not a causal return, so the split is not economically identified. Panel A of Table 1 adds a step that swaps the AI-to-human volume ratios $\rho_s^G$ and $\rho_s^B$, with $s_{AI}^G$ and $s_{AI}^B = \mathrm{Ratio}_{AI}^B \cdot L$; it is negligible for the human-yield outcome.

[2] The interpolation is a first-order approximation to how conduct responds to the rhythm, not drawn from a specific source, and most reliable inside $[\phi_{\mathrm{pre}}, \phi_{\mathrm{post}}]$.

### F.5 Balancing AI and the Rhythm of the Day (Counterfactual Rhythm Analysis)

The same position gradient also defines the firm's room to maneuver. Because the behavioral pullback is concentrated where hot leads are front-loaded, re-pacing the within-day rhythm of hot-lead assignment can recover part of the lost yield without changing AI penetration itself. This subsection treats the rhythm as a choice variable.

The decomposition takes the AI-induced change in rhythm as given. We now treat the rhythm as a strategic choice and solve the firm's optimization problem jointly over the two dimensions of Figure F1, the hot-lead pool $\alpha$ set by AI penetration and the within-day rhythm $\phi$ the firm controls. For each pool $\alpha$ the firm picks the rhythm that maximizes conversions per call $\Pi(\alpha, \phi)$, and Panel (a) plots this surface, with the unconstrained maximizer $\phi^*(\alpha)$ as the dotted line.

The bright high-conversion region sits at a large pool and a moderate rhythm, so AI's expansion of the hot-lead pool is valuable only when it is paired with a calm within-day rhythm; the same pool run at a steep rhythm falls into the lower-conversion region that lies above the ridge. The dotted ridge $\phi^*(\alpha)$ slopes gently downward, so a fuller pool calls for a slightly flatter day, not a steeper one. It lies below the observed support at every hot-lead share, so we read it as a direction, not a target, and the recommendation this figure supports is the downward move inside the support.

The firm's after-AI point lies above the ridge, whereas before AI it sat close to it. The two levers differ in who controls them. The pool is set by AI, but the rhythm is the firm's own choice, so the calibrated model treats the loss as potentially recoverable without changing the pool. The lesson is to consider routing levers, such as batching cold leads or staggered hot-lead release, that would re-pace the within-day arrival of hot leads, and not to roll back AI, recognizing that the model treats the rhythm as a control variable the firm can in principle adjust.

We report this as a comparative static, not an optimum. The conduct profiles are estimated at two rhythms only, $\phi_{pre} = 0.10$ and $\phi_{post} = 0.22$, so that interval is the empirically supported range and everything outside it rests on linear extrapolation. Within the supported range the result is unambiguous. Firm yield is strictly decreasing in $\phi$ at every hot-lead share we examine, from $\alpha = 0.40$ to $\alpha = 0.90$, so a steeper day always lowers conversions per call and the model implies easing the rhythm, not steepening it.

The magnitude is economically meaningful and statistically stable. AI raised the rhythm from about 0.10 before the expansion to about 0.22 after it. Within the calibrated model, easing it back to 0.10, a rhythm the firm itself ran before the expansion, increases predicted conversions per call by 13.5 percent. A parametric bootstrap over the estimated conduct profiles, using 2,000 draws with binomial resampling at the observed cell counts, puts the 95 percent interval at 10.5 to 16.6 percent, and every draw yields a positive gain.

Extending the search below the observed range produces an interior optimum at $\phi^* \approx 0.077$ with a gain of 13.8 percent, but we do not treat that value as the recommended target. It lies below both observed regimes, and at that rhythm all twenty-seven estimated profile values are pushed outside their observed ranges. The bootstrap places the unconstrained optimum inside the supported range in fewer than one percent of draws. The two gains are in any case nearly identical, 0.0332 conversions per call at $\phi = 0.10$ against 0.0333 at $\phi = 0.077$, because yield is very flat near the top. We therefore report the return to the pre-expansion rhythm as the result and the interior optimum only as the unconstrained value.

Panel (b) plots firm yield against the rhythm at the observed post-expansion pool, with the 95 percent bootstrap band. Yield declines monotonically across the supported range, and the band excludes zero gain throughout, so the direction of the recommendation does not depend on where the unconstrained optimum lies.

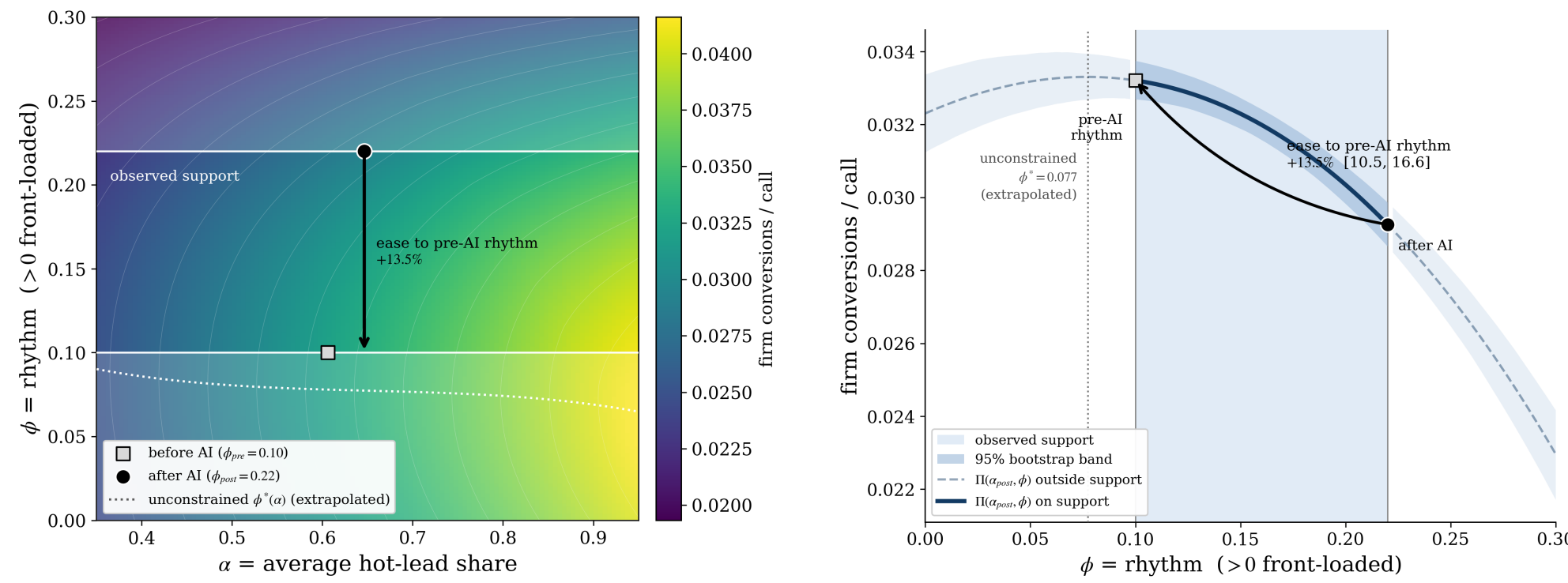

(a) (b)

**Figure F1. Firm Yield and the Within-Day Rhythm**

***Notes.*** Both panels use the accounting model of §5. α is the average hot-lead share and ϕ the within-day rhythm, with ϕ greater than zero meaning front-loaded. The conduct profiles are estimated at two rhythms only, $\phi_{pre} = 0.10$ and $\phi_{post} = 0.22$. That interval is the observed support, and values outside it are linear extrapolations. Panel (a) plots firm conversions per call. White horizontal lines at ϕ = 0.10 and ϕ = 0.22 bound the observed support, and the regions outside it carry a light veil. Markers give the before-AI and after-AI operating points, and the arrow shows easing the rhythm to the pre-expansion value at the observed post-expansion pool. The dotted line is the unconstrained ϕ*(α), which lies entirely below the observed support and is shown for reference only. Panel (b) plots firm conversions per call against ϕ at the observed post-expansion pool α = 0.646, with a 95 percent band from a parametric bootstrap of the conduct profiles using 2,000 draws and binomial resampling at the observed cell counts. The curve is drawn solid on the observed support and dashed outside it. Yield is strictly decreasing in ϕ throughout the observed support. Within the calibrated model, easing from 0.22 to 0.10 increases predicted conversions per call by 13.5 percent, with a 95 percent interval of 10.5 to 16.6 percent.

More broadly, the productivity gain a manager expects from a better-composed pool can be offset by a behavioral response of comparable magnitude operating in the opposite direction, and our calibrated model suggests the firm can compound that offset by front-loading the enlarged pool. Across the range of rhythms the data support, a steeper day always lowers conversions per call.

This appendix states the model's assumptions and calibrated inputs, reports the estimated order profiles, and gives the aggregation, decomposition, and parametrization behind the figures and tables. All quantities are normalized to a benchmark of $T = 50$ calls per worker-day.

## Appendix G: Additional Heterogeneity Results

### G.1 Heterogeneity Analysis: Detailed Results

The average within-day response documented above may mask meaningful differences across agents. We therefore conduct an exploratory heterogeneity analysis of whether the conversion response to the AI expansion varies with agents' pre-expansion characteristics. We estimate a causal forest following Wager and Athey (2018) on human-handled calls on hot leads at within-day positions 1 – 9 within the sixty-day window around the August 28, 2024 cutoff. The treatment is the post-cutoff indicator and the outcome is conversion residualized on agent and hour-of-day fixed effects. We allow the effect to vary with six agent-level moderators constructed exclusively from the pre-expansion period: a leave-out measure of hot-lead conversion ability, computed from each agent's call history that precedes the estimation window and therefore disjoint from the calls used to estimate the effect; the reallocation rate; tenure; daily call volume; gender; and average log call duration. Standard errors are clustered at the agent-day level.

The data reject a homogeneous effect. A calibration test returns a differential-forest-prediction coefficient of 1.10 that is precisely estimated, indicating that the forest recovers a genuine ranking of agents by treatment effect, and a rank-weighted average treatment effect, evaluated as the area under the TOC curve, of 0.043 confirms that this ranking has content; both are significant at the one-percent level. The average effect over this sample is itself small and statistically indistinguishable from zero (−0.004), consistent with pooling across within-day positions whose effects offset, so our interest lies in the dispersion around that average, not in its level.

Figure G1 quantifies that dispersion. Ranking agents by their out-of-bag predicted effect and grouping them into thirds, the doubly-robust average treatment effect rises monotonically, from −0.024 (standard error 0.009) in the bottom third to −0.008 (0.008) in the middle third and +0.020 (0.007) in the top third. The two extreme groups are individually significant and fall on opposite sides of zero: AI raises conversion for one set of agents and lowers it for another, a spread of roughly four percentage points that the average alone conceals. Figure G2 asks which characteristics predict these gains, through a best linear projection of the forest-predicted conditional average treatment effects onto the six moderators (Athey, Tibshirani, and Wager 2019). The pre-expansion reallocation rate is the most robust predictor: an interquartile increase is associated with an approximately 1.4-percentage-point more positive response ($p < 0.01$), consistent with agents who most actively route leads, and whose workflow is therefore most complementary to AI-assisted triage, benefiting most when the automation absorbs low-value contacts. Pre-expansion ability enters with the opposite sign but is small and only marginally significant, an approximately 0.4-percentage-point more negative response per interquartile increase ($p < 0.05$); we do not emphasize it, both because of its size and because a naive measure of prior performance is mechanically prone to mean reversion, which our leave-out construction removes. The remaining characteristics are not distinguishable from zero: daily call volume ($p = 0.78$), average log call duration ($p = 0.67$), tenure ($p = 0.85$), and gender ($p = 0.86$).

Taken together, the estimated conversion response is organized most sharply along agents' reallocation behavior, not their raw productivity, and because it is roughly symmetric around a small average, negative in the bottom third and positive in the top third, it largely offsets in the aggregate even though the underlying heterogeneity is pronounced and statistically detectable. These patterns are exploratory for three reasons: treatment varies at the calendar-date level, not independently across agents; several moderators are estimated proxies based on the same agents' historical outcomes; and we do not adjust for multiple testing across moderators. The forest also identifies heterogeneity in a within-window, selection-on-observables sense, not through the discontinuity, so we anchor the level of the effect to the regression-discontinuity estimates in the main text and read these results as suggestive patterns of heterogeneity, not independently identified subgroup causal effects. We report the graphical detail of the exploratory heterogeneity analysis summarized in Section 4.5: the doubly-robust group treatment effects across predicted-effect tertiles (Figure G1) and the best linear projection of the forest-predicted conditional average treatment effects onto agent-level moderators (Figure G2).

**Figure G1. Doubly-Robust Group Treatment Effects by Predicted-Effect Tertile**

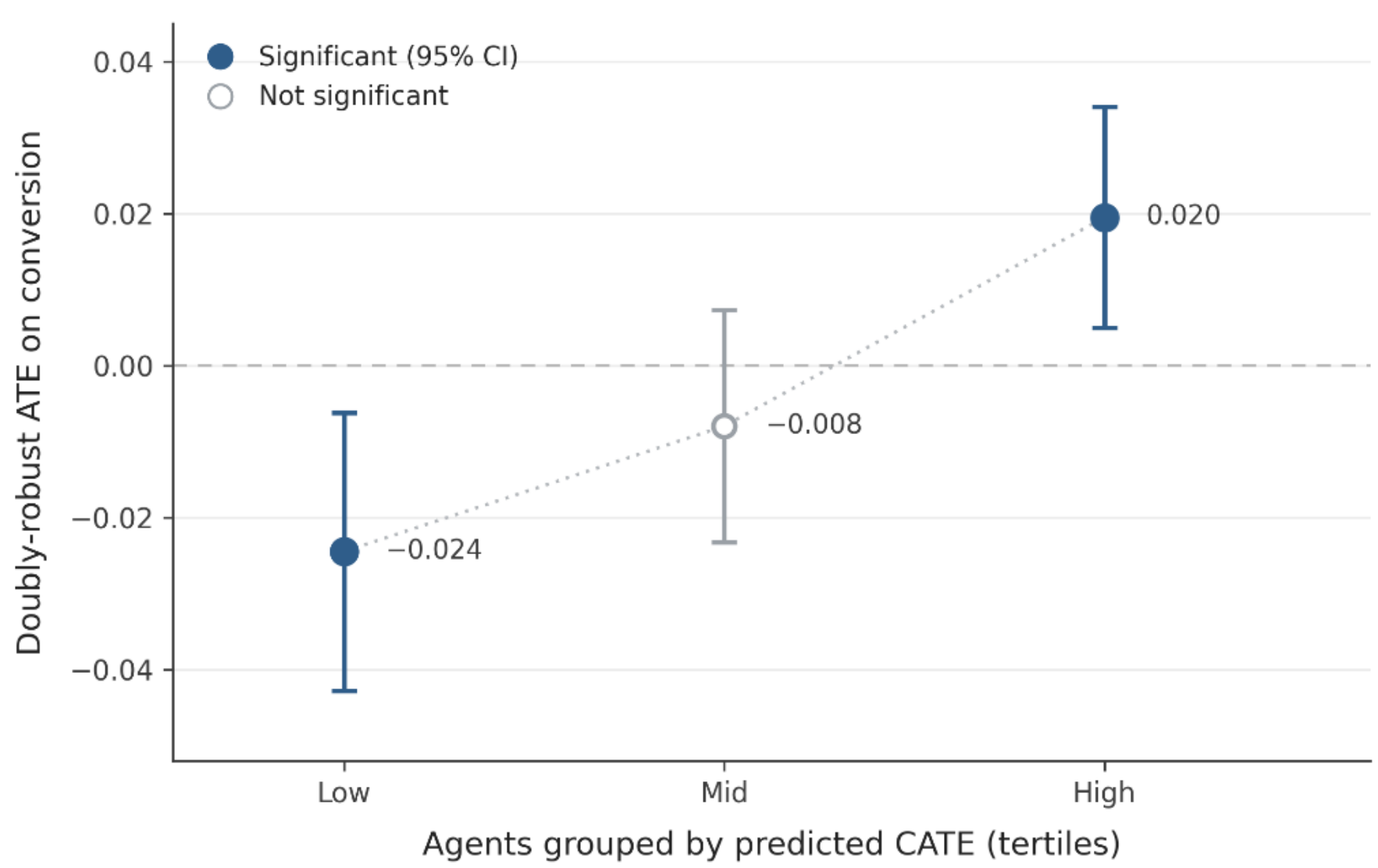

***Notes.*** Doubly-robust ATE (AIPW) of the AI expansion, by predicted-effect tertile from a causal forest (Wager and Athey 2018; Athey et al. 2019) on human-handled hot leads at positions 1–9 in the ±60-day window. Agent-day-clustered 95%

CIs. Bottom and top tertiles individually significant at 5%.

**Figure G2. Best Linear Projection of CATE on Agent-Level Moderators**

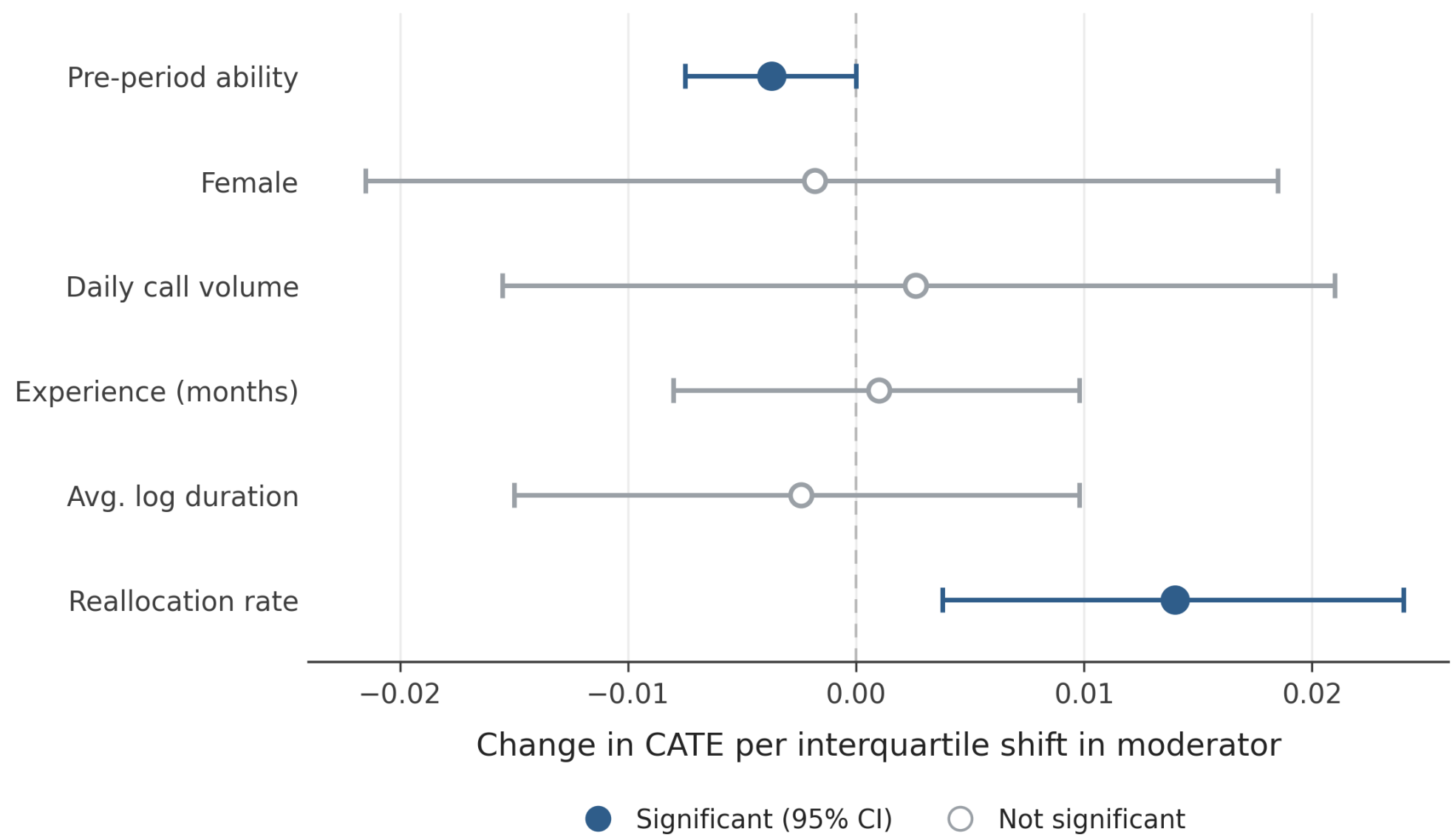

***Notes.*** Best linear projection of CATE onto agent characteristics, same forest as Figure G1. Each coefficient scaled to an interquartile shift. 95% CIs cluster- and HC-robust; filled = significant at 5%. Reallocation rate is the only robust moderator. Calibration test rejects no heterogeneity ($p<0.01$).

## Appendix H: RDiT Estimation Details

### H.1 Two-Step Estimation Procedure

A standard concern in time-based discontinuity designs is that outcomes may reflect persistent worker differences, date-specific shocks, and regular intraday patterns. Following Hausman and Rapson (2018), we therefore use a two-step residualization procedure.

In the first step, we regress each focal call-level outcome on human-agent fixed effects, calendar-date fixed effects, and hour-of-day fixed effects using the primary analytic sample:

$$Y_{ijt} = \eta_i + \delta_t + \xi_{h(jt)} + u_{ijt},$$

where $Y_{ijt}$ denotes the outcome for call j placed by agent i on calendar date t; $\eta_i$ denotes agent fixed effects; $\delta_t$ denotes calendar-date fixed effects; and $\xi_{h(jt)}$ denotes fixed effects for the hour in which the call is placed.

We define the residualized outcome as

$$\hat{u}_{ijt} = Y_{ijt} - \hat{\eta_i} - \hat{\delta_t} - \hat{\xi}_{h(jt)}.$$

Agent fixed effects absorb persistent differences across human agents. Calendar-date fixed effects absorb all factors shared by calls placed on the same date, including aggregate demand, system-wide operating conditions, daily lead quality, marketing activity, and other date-level shocks. Hour-of-day fixed effects absorb regular intraday rhythms.

For the day-level regressions reported in Appendix C Tables C1 and C2, the agent-day–level regression in Appendix C Table C3, and column (1) of Appendix Table D.5.1, we replace calendar-date fixed effects with day-of-week fixed effects. Calendar-date fixed effects cannot be included in a day-level panel because each observation corresponds to a single date; day-of-week fixed effects preserve the ability to absorb regular weekly rhythms while allowing the running variable to identify the discontinuity at the cutoff.

The residualized outcome therefore measures the outcome of a particular call relative to the overall outcome level on the same date, net of persistent agent differences and regular hour-of-day patterns.

In the second step, we estimate the following local linear specification:

$$\hat{u}_{ijt} = \beta_0 + \beta_1 1\{D_t \geq 0\} + \beta_2 D_t + \beta_3 1\{D_t \geq 0\} \times D_t + \varepsilon_{ijt},$$

where $D_t = t - c$ denotes the number of calendar days between date t and the August 28 cutoff c. The coefficient $\beta_0$ is the pre-expansion intercept, $\beta_2$ is the pre-expansion local time trend, and $\beta_3$ allows the trend to differ after the expansion. The coefficient of interest, $\beta_1$, captures the discontinuous change at the cutoff in the residualized outcome.

### H.2 Interpretation of the Estimand

Because the first-stage specification includes calendar-date fixed effects, $\beta_1$ does not estimate the absolute level change in an outcome at a particular position. Instead, it estimates how the outcome at that position changes relative to the overall outcome level observed on the same date. For example, a negative discontinuity in early-position conversion indicates that the conversion advantage associated with early calls declines relative to the daily average following the expansion. It does not necessarily imply that the absolute conversion rate at that position declines by the same magnitude.

The primary estimand is therefore a change in the shape of the within-day outcome profile. Differences in $\beta_1$ across position bins identify whether the relative distribution of conversion, reallocation, and effort across the workday changes discontinuously at the AI expansion. This estimand is well suited to the theoretical prediction that the within-day outcome profile flattens. The framework predicts that the expansion should produce larger behavioral changes at the beginning of the sequence, where future opportunities remain valuable, and weaker behavioral changes in the late tail, where those opportunities are close to exhaustion. Because the date fixed effects absorb the daily overall outcome level, the primary call-level specification does not identify the absolute aggregate daily effect of the expansion. We therefore report a complementary RDiT using raw daily averages in Appendix C, Table C1. The day-level analysis estimates the aggregate daily effect directly and is conceptually distinct from the position-specific, within-date estimand used in the primary analysis.

### H.3 Inference

Our baseline call-level standard errors are clustered at the agent-day level. This clustering accounts for dependence among repeated decisions made by the same agent within the same workday, including serial dependence in effort, work pacing, and reallocation behavior across calls.

The timing of the AI expansion, however, varies at the calendar-date level. We therefore complement the call-level estimates with an early-versus-late position contrast constructed at the date level. This analysis uses daily observations and provides a check that the position-specific results are not driven solely by the large number of calls observed within each agent-day. We do not treat the number of calls as the number of independent AI-expansion shocks. Rather, the large call-level sample allows us to measure the relative within-day profile precisely, while the date-level analysis provides complementary inference using the level at which the routing change occurs.

## Appendix References